\documentclass[12pt]{article}

\usepackage{etex}
\usepackage{tikz-cd}
\tikzcdset{
  shift left/.default=+0.7ex,
  shift right/.default=+0.7ex}

\newenvironment{tikzar}[1][]{{}\kern-4pt\begin{tikzcd}[ampersand replacement=\&,#1]}%
{\end{tikzcd}\kern-4pt{}}

\usetikzlibrary{cd}
\usetikzlibrary{decorations.pathmorphing,positioning}

\usepackage{amssymb}
\usepackage{amstext}
\usepackage{amsmath,stmaryrd}
\usepackage{txfonts}
\usepackage{bbm}
\usepackage{enumerate}
\usepackage{fullpage}
\usepackage{multicol}
\usepackage{rotating}

\newdimen\proofrulebreadth \proofrulebreadth=.05em
\newdimen\proofdotseparation \proofdotseparation=1.25ex
\newdimen\proofrulebaseline \proofrulebaseline=2ex
\newcount\proofdotnumber \proofdotnumber=3
\let\then\relax
\def\hfi{\hskip0pt plus.0001fil}
\mathchardef\squigto="3A3B
\newif\ifinsideprooftree\insideprooftreefalse
\newif\ifonleftofproofrule\onleftofproofrulefalse
\newif\ifproofdots\proofdotsfalse
\newif\ifdoubleproof\doubleprooffalse
\let\wereinproofbit\relax
\newdimen\shortenproofleft
\newdimen\shortenproofright
\newdimen\proofbelowshift
\newbox\proofabove
\newbox\proofbelow
\newbox\proofrulename
\def\shiftproofbelow{\let\next\relax\afterassignment\setshiftproofbelow\dimen0 }
\def\shiftproofbelowneg{\def\next{\multiply\dimen0 by-1 }%
\afterassignment\setshiftproofbelow\dimen0 }
\def\setshiftproofbelow{\next\proofbelowshift=\dimen0 }
\def\setproofrulebreadth{\proofrulebreadth}

\def\prooftree{
\ifnum  \lastpenalty=1
\then   \unpenalty
\else   \onleftofproofrulefalse
\fi
\ifonleftofproofrule
\else   \ifinsideprooftree
        \then   \hskip.5em plus1fil
        \fi
\fi
\bgroup
\setbox\proofbelow=\hbox{}\setbox\proofrulename=\hbox{}%
\let\justifies\proofover\let\leadsto\proofoverdots\let\Justifies\proofoverdbl
\let\using\proofusing\let\[\prooftree
\ifinsideprooftree\let\]\endprooftree\fi
\proofdotsfalse\doubleprooffalse
\let\thickness\setproofrulebreadth
\let\shiftright\shiftproofbelow \let\shift\shiftproofbelow
\let\shiftleft\shiftproofbelowneg
\let\ifwasinsideprooftree\ifinsideprooftree
\insideprooftreetrue
\setbox\proofabove=\hbox\bgroup$\displaystyle 
\let\wereinproofbit\prooftree
\shortenproofleft=0pt \shortenproofright=0pt \proofbelowshift=0pt
\onleftofproofruletrue\penalty1
}

\def\eproofbit{
\ifx    \wereinproofbit\prooftree
\then   \ifcase \lastpenalty
        \then   \shortenproofright=0pt  
        \or     \unpenalty\hfil         
        \or     \unpenalty\unskip       
        \else   \shortenproofright=0pt  
        \fi
\fi
\global\dimen0=\shortenproofleft
\global\dimen1=\shortenproofright
\global\dimen2=\proofrulebreadth
\global\dimen3=\proofbelowshift
\global\dimen4=\proofdotseparation
\global\count255=\proofdotnumber
$\egroup  
\shortenproofleft=\dimen0
\shortenproofright=\dimen1
\proofrulebreadth=\dimen2
\proofbelowshift=\dimen3
\proofdotseparation=\dimen4
\proofdotnumber=\count255
}

\def\proofover{
\eproofbit 
\setbox\proofbelow=\hbox\bgroup 
\let\wereinproofbit\proofover
$\displaystyle
}%
\def\proofoverdbl{
\eproofbit 
\doubleprooftrue
\setbox\proofbelow=\hbox\bgroup 
\let\wereinproofbit\proofoverdbl
$\displaystyle
}%
\def\proofoverdots{
\eproofbit 
\proofdotstrue
\setbox\proofbelow=\hbox\bgroup 
\let\wereinproofbit\proofoverdots
$\displaystyle
}%
\def\proofusing{
\eproofbit 
\setbox\proofrulename=\hbox\bgroup 
\let\wereinproofbit\proofusing
\kern0.3em$
}

\def\endprooftree{
\eproofbit 
  \dimen5 =0pt
\dimen0=\wd\proofabove \advance\dimen0-\shortenproofleft
\advance\dimen0-\shortenproofright
\dimen1=.5\dimen0 \advance\dimen1-.5\wd\proofbelow
\dimen4=\dimen1
\advance\dimen1\proofbelowshift \advance\dimen4-\proofbelowshift
\ifdim  \dimen1<0pt
\then   \advance\shortenproofleft\dimen1
        \advance\dimen0-\dimen1
        \dimen1=0pt
        \ifdim  \shortenproofleft<0pt
        \then   \setbox\proofabove=\hbox{%
                        \kern-\shortenproofleft\unhbox\proofabove}%
                \shortenproofleft=0pt
        \fi
\fi
\ifdim  \dimen4<0pt
\then   \advance\shortenproofright\dimen4
        \advance\dimen0-\dimen4
        \dimen4=0pt
\fi
\ifdim  \shortenproofright<\wd\proofrulename
\then   \shortenproofright=\wd\proofrulename
\fi
\dimen2=\shortenproofleft \advance\dimen2 by\dimen1
\dimen3=\shortenproofright\advance\dimen3 by\dimen4
\ifproofdots
\then
        \dimen6=\shortenproofleft \advance\dimen6 .5\dimen0
        \setbox1=\vbox to\proofdotseparation{\vss\hbox{$\cdot$}\vss}%
        \setbox0=\hbox{%
                \advance\dimen6-.5\wd1
                \kern\dimen6
                $\vcenter to\proofdotnumber\proofdotseparation
                        {\leaders\box1\vfill}$%
                \unhbox\proofrulename}%
\else   \dimen6=\fontdimen22\the\textfont2 
        \dimen7=\dimen6
        \advance\dimen6by.5\proofrulebreadth
        \advance\dimen7by-.5\proofrulebreadth
        \setbox0=\hbox{%
                \kern\shortenproofleft
                \ifdoubleproof
                \then   \hbox to\dimen0{%
                        $\mathsurround0pt\mathord=\mkern-6mu%
                        \cleaders\hbox{$\mkern-2mu=\mkern-2mu$}\hfill
                        \mkern-6mu\mathord=$}%
                \else   \vrule height\dimen6 depth-\dimen7 width\dimen0
                \fi
                \unhbox\proofrulename}%
        \ht0=\dimen6 \dp0=-\dimen7
\fi
\let\doll\relax
\ifwasinsideprooftree
\then   \let\VBOX\vbox
\else   \ifmmode\else$\let\doll=$\fi
        \let\VBOX\vcenter
\fi
\VBOX   {\baselineskip\proofrulebaseline \lineskip.2ex
        \expandafter\lineskiplimit\ifproofdots0ex\else-0.6ex\fi
        \hbox   spread\dimen5   {\hfi\unhbox\proofabove\hfi}%
        \hbox{\box0}%
        \hbox   {\kern\dimen2 \box\proofbelow}}\doll%
\global\dimen2=\dimen2
\global\dimen3=\dimen3
\egroup 
\ifonleftofproofrule
\then   \shortenproofleft=\dimen2
\fi
\shortenproofright=\dimen3
\onleftofproofrulefalse
\ifinsideprooftree
\then   \hskip.5em plus 1fil \penalty2
\fi
}

\newcommand{\To}[2]{\left({#1}\!\Rightarrow\!{#2}\right)}

\newcommand{\TTTo}[2]{\left[{#1},{#2}\right]}
\newcommand{\TTo}[1]{\left[{#1}\right]}
\newcommand{\xxp}[2]{\To{#1}{#2}}
\newcommand{\xxpp}[2]{\xxp{{#1}^+}{#2}}

\newcommand{\stream}[1]{{#1}^\omega}  
\newcommand{\thenn}{\supset}  
\renewcommand{\to}{\xrightarrow{\ \ \ }}
\newcommand{\ot}{\xleftarrow{\ \ \ }}
\newcommand{\tto}[1]{\xrightarrow{#1}}
\newcommand{\oot}[1]{\xleftarrow{#1}}
\newcommand{\mono}{\rightarrowtail}
\newcommand{\epi}{\twoheadrightarrow}

\newcommand{\inclusion}{\hookrightarrow}
\newcommand{\mmono}[1]{\stackrel{#1}\rightarrowtail}
\newcommand{\eepi}[1]{\stackrel{#1}\twoheadrightarrow}

\newcommand{\pfn}{\rightharpoonup}

\newcommand{\rel}{\longleftrightarrow}
\newcommand{\rrel}[1]{\stackrel{#1}\rel}

\newcommand{\safety}[2]{{#1}_{\subsetpluseq{#2}}}
\newcommand{\safet}[2]{{#1}_{\subseteq{#2}}}

\newcommand{\preccc}{\preccurlyeq}

\newcommand{\Precdown}{\curlyveedownarrow}

\newcommand{\id}{{\rm id}}

\newcommand{\cons}{{\, ::}}
\newcommand{\icomp}{\bullet}

\newcommand{\tail}{\lft \pi}

\newcommand{\WP}{\mbox{\large $\wp$}}
\newcommand{\KV}{\HHH_\pm}
\newcommand{\QV}{\RrR}
\newcommand{\RRrr}{\overline\RRr}
\newcommand{\RRbf}{\RRR}
\newcommand{\RRb}{\underline{\RRbf}}
\newcommand{\Rel}{\mathbf{R}}
\newcommand{\Pfn}{\mathbf{P}}
\newcommand{\TPfn}{\mathbf{tP}}
\newcommand{\TRel}{\mathbf{tR}}
\newcommand{\Set}{\mathbf{S}}
\newcommand{\fSet}{\mathbf{fS}}
\newcommand{\TSet}{\mathbf{tS}}

\newcommand{\Int}{{\sf Int}}
\newcommand{\Ord}{{\sf Ord}}
\newcommand{\SProc}{{\sf SProc}}
\newcommand{\ASProc}{{\sf ASProc}}
\newcommand{\DProc}{{\sf DProc}}
\newcommand{\dProc}{{\sf dProc}}
\newcommand{\aProc}{{\sf aProc}}

\newcommand{\Gam}{{\sf Gam}}
\newcommand{\dGam}{{\sf gam}}

\newcommand{\SFun}{{\sf SFun}}
\newcommand{\ASFun}{{\sf ASFun}}
\newcommand{\Dom}{{\rm Dom}}

\newcommand{\pp}{-}
\newcommand{\oo}{+}

\newcommand{\Tr}{{\rm Tr}}

\newcommand{\CCC}{{\cal C}}

\newcommand{\EEE}{{\cal E}}

\newcommand{\HHH}{{\cal H}}

\newcommand{\MMM}{{\cal M}}

\newcommand{\PPP}{{\cal P}}
\newcommand{\QQQ}{{\cal Q}}
\newcommand{\RRR}{{\cal R}}

\newcommand{\NNn}{{\mathbb N}}

\newcommand{\PPp}{{\mathbb P}}

\newcommand{\RRr}{{\mathbb R}}

\newcommand{\ZZz}{{\mathbb Z}}

\newcommand{\HhH}{{\mathfrak H}}

\newcommand{\LlL}{{\mathfrak L}}

\newcommand{\RrR}{{\mathfrak R}}

\newcommand{\VvV}{{\mathfrak V}}

\newcommand{\mathbold}[1]{\mbox{\boldmath $#1$}}

\newcommand{\Hhh}{{\mathbold H}}

\mathcode`\<="4268 
\mathcode`\>="5269 
\mathchardef\gt="313E 
\mathchardef\lt="313C 

 \def\pushright#1{{
    \parfillskip=0pt            
    \widowpenalty=10000         
    \displaywidowpenalty=10000  
    \finalhyphendemerits=0      
    \leavevmode                 
    \unskip                     
    \nobreak                    
    \hfil                       
    \penalty50                  
    \hskip.2em                  
    \null                       
    \hfill                      
    {#1}                        
    \par}}                      

 \def\qed{\pushright{$\square$}\penalty-700 \smallskip}

\newenvironment{prf}[1]{\begin{trivlist} \item[{\bf ~Proof}#1.]}%
{\qed\end{trivlist}}

\newcommand{\beq}{\begin{equation}}
\newcommand{\eeq}{\end{equation}}
\newcommand{\ba}[1]{\begin{array}{#1}}
\newcommand{\ea}{\end{array}}
\newcommand{\bea}{\begin{eqnarray}}
\newcommand{\eea}{\end{eqnarray}}
\newcommand{\bear}{\begin{eqnarray*}}
\newcommand{\eear}{\end{eqnarray*}}
\newcommand{\bpr}{\begin{prf}{}}
\newcommand{\epr}{\end{prf}}
\newcommand{\bprf}[1]{\begin{prf}{#1}}
\newcommand{\eprf}{\end{prf}}

\newtheorem{theorem}{Theorem}[section]
\newtheorem{proposition}[theorem]{Proposition}
\newtheorem{lemma}[theorem]{Lemma}
\newtheorem{corollary}[theorem]{Corollary}

\newtheorem{definition}[theorem]{Definition}

\newcommand{\cata}[1]{\llparenthesis {#1} \rrparenthesis}
\newcommand{\ana}[1]{\left\llbracket {#1} \right\rrbracket}

\newcommand{\uev}[1]{\left\{{#1}\right\}}

\newcommand{\Run}[1]{\left\{\!\lvert {#1} \rvert\!\right\}}

\newcommand{\Prog}{\PPp}

\newcommand{\Kar}[1]{{#1}^{\circlearrowleft}}
\newcommand{\lft}[1]{\protect\overleftarrow{#1}}

\newcommand{\Del}[1]{{#1}_\bot}
\newcommand{\may}{\Diamond}
\newcommand{\must}{\Box}

\newcommand{\iinn}{\,\epsilon\,}
\newcommand{\nnii}{\raisebox{1ex}{\hspace{1.3ex}\turnbox{180}{$\epsilon$}\hspace{.1ex}}}

\newcommand{\enco}[1]{\left\ulcorner{#1}\right\urcorner}

\newcommand{\sta}[1]{{#1}^\circ}
\newcommand{\out}[1]{{#1}^\bullet}
\newcommand{\conss}[1]{{#1}_{\scriptscriptstyle (::)}}
\newcommand{\innn}[1]{{#1}_{\scriptscriptstyle (-)}}

\newcommand{\lseq}{\,|\!\!\rightarrow\!}
\newcommand{\llseq}[1]{\,|\hspace{-0.9ex}\tto{#1}\!}
\newcommand{\seq}[1]{\Big(\, #1\, \Big)}
\newcommand{\sseq}[1]{\left(\, #1\, \right)}

\newcommand{\paar}[1]{\big<\hspace{-.66ex}\big<{#1}\big>\hspace{-.66ex}\big>}

\newcommand{\truth}{\top}
\newcommand{\para}[1]{\noindent\textbf{#1}}

\newcommand{\pplus}{\oplus}
\newcommand{\retra}{\begin{tikzar}{}
 \hspace{.1ex} \ar[thin,bend right=13,tail]{r} \&  \hspace{.1ex}  \ar[thin,bend right=13,two heads]{l}
\end{tikzar}}

\graphicspath{{PIC/}}

\title{Retracing some paths in categorical semantics:\\
From process-propositions-as-types\\ to categorified reals 
and computers}

\author{Dusko Pavlovic\thanks{Supported by NSF and AFOSR.}\\
University of Hawaii, Honolulu HI\\
{\small dusko@hawaii.edu}
}

\date{}

\begin{document}

\maketitle

\begin{abstract}
The logical parallelism of propositional connectives and type constructors extends beyond the static realm of predicates, to the dynamic realm of processes. Understanding the logical parallelism of \emph{process}\/ propositions and \emph{dynamic}\/ types was one of the central problems of the semantics of computation, albeit not always clear or explicit. It sprung into clarity through the early work of Samson Abramsky, where the central ideas of denotational semantics and process calculus were brought together and analyzed by categorical tools, e.g. in the structure of \emph{interaction categories}. While some logical structures borne of dynamics of computation immediately started to emerge, others had to wait, be it because the underlying logical principles (mainly those arising from coinduction) were not yet sufficiently well-understood, or simply because the research community was more interested in other semantical tasks. Looking back, it seems that the process logic uncovered by those early semantical efforts might still be starting to emerge and that the vast field of results that have been obtained in the meantime might be a valley on a tip of an iceberg.

In the present paper, I try to provide a logical overview of the gamut of interaction categories and to distinguish those that model computation from those that capture processes in general. The main coinductive constructions turn out to be of the latter kind, as illustrated towards the end of the paper by a compact category of all real numbers as processes, computable and uncomputable, with polarized bisimulations as morphisms. The operation of addition of the reals arises as the biproduct, real vector spaces are the enriched bicompletions, and linear algebra arises from the enriched kan extensions. At the final step, I sketch a structure that characterizes the computable fragment of categorical semantics.
\end{abstract}

\newpage
\tableofcontents
\clearpage

\section*{Personal introduction}
I first learned about Samson Abramsky's work from his invited plenary lecture at the International Category Theory Meeting in Montreal in 1991. It was the golden age of category theory, and Montreal was at the heart of it, and I got to be a postdoc there. Just a few years earlier, I was a dropout freelance programmer, but had become a mathematician, and was uninterested in computers. I was told, however, that Abramsky had constructed some categories that no one had seen before, so I came to listen to his talk. I also had a talk to give later that day myself, but for some reason, I do not recall how that went. At the end of Abramsky's plenary lecture, Saunders Mac\,Lane stood up, one of the two fathers of category theory, high up near the ceiling of the amphitheater, and spoke for a long time.  He criticized computer science in general. After that, Bill Lawvere stood up, and provided some friendly comments, suggesting directions for progress and improvement. 

Two years later, I became an \emph{"EU Human Capital Mobility"}\/ fellow within the Theory Group at the Imperial College in London, led by Samson Abramsky. I started learning computer science and spent a lot of time trying to understand Samson's \emph{interaction categories} \cite{AbramskyS:icats}. In the meantime, he had constructed more categories that no one had seen before. My fellowship ended after a year or two, and the human capital mobility turned out to be much greater than anyone could imagine, but I continued to think about interaction categories for years. Here I try to summarize some of that thinking.

\section{Introduction: On categorical logics and propositions-as-types} \label{Sec:sem}

\paragraph{The category of sets or types.}
This is a paper about categorical semantics. It is written for a collection intended for logicians. If you are reading this, then you are presumed to be interested in categorical logic, although you may not be interested in categories in general. To ease this tension, I will avoid abstract categories, and mostly stick with the category $\Set$ of sets and functions. It is presented, however, as a universe of types, by specifying which type constructors are used in each construction. Initially, we just need the cartesian products, but later we need more. The naive set theory used to be presented incrementally.  Nowadays most mathematicians think of types as sets, and most programmers think of sets as types, so it seems reasonable for logicians and computer scientists to identify the two.  To keep the naive-set-theory flavor, we usually call the type inhabitants \emph{elements}, where type theorists use the term \emph{terms}.

When a set is constructed as a type, then it can also be construed as a proposition: since its elements are constructions, they can be viewed as proofs \cite{MartinLoefP:inttt}. Such interpretations originate from logic, where the idea of propositions-as-types was first encountered along the paths of proofs-as-constructions \cite{ChurchA:types,KolmogorovA:BHK,MartinLoefP:predicative}. We retrace these paths first, and proceed throughout with propositions-as-types, types-as-sets, terms-as-elements, elements-as-morphisms \cite{LambekJ:Advances,LawvereFW:sets}.

\paragraph{Naming names.} While sets and types signal different approaches, many concepts are studied in different communities under different names even if there are no significant differences. This is useful to place narratives in their contexts or to authenticate speakers' allegiances. It is not easy to avoid such connotations when they are undesired. In some cases, I resorted to renaming. E.g., the \emph{histories}\/ from Sec.~\ref{Sec:history} onwards are known as non-empty lists, or words, or strings. There are other examples. I am not trying to reinvent them but to dissociate them from narrow contexts. I hope they will not be too distracting.

\subsection{Logics of types} 
Bertrand Russell proposed his \emph{ramified theory of types}\/ \cite{RussellB:types} as a logical framework for paradox prevention. Alonzo Church and Stephen Kleene advanced type theory into a model of computation \cite{ChurchA:types,KleeneS:integers}. Dana Scott adopted type theory as the foundation for a mathematical approach to the semantics of computation \cite{ScottD:ISWIM}. The semantics of programming languages were built  steadily upon that foundation \cite{GunterC:book,ReynoldsJC:book}. Process semantics also arose from that foundation \cite{MilnerR:processes}, but had to undergo a substantive conceptual evolution before the types could capture dynamics. I followed these developments through Samson Abramsky's work.

The propositions-as-types paradigm was discovered many times. In logic and computer science, it is attributed to Haskell Curry and William Howard \cite[Ch.~3]{curry-festschrift,GirardJY:prot}. Howard got the idea from Georg Kreisel \cite{WadlerP:curry-howard}, and Kreisel's goal was to formalize Brouwer's concept of proofs-as-constructions \cite{KreiselG:survey}. An early formalization of Brouwer's concept goes back to Kolmogorov \cite{KolmogorovA:BHK}.  

The structural reason why propositions and types obey analogous laws was offered by Lawvere \cite{LawvereFW:dialectica}, who pointed out that the propositional and the typing rules are instances of analogous categorical  \emph{adjunctions}; and that the proof constructions and the term derivations arise from the adjunction units and counits. This gave rise to the idea of categorical proof theory, pursued  by Lambek \cite{LambekJ:dedsc1,LambekJ:dedsc2,LambekJ:dedsc3}, and to the basic structures of categorical semantics, succinctly described in \cite[and the references therein]{Lambek-Scott:book}. In the preface to his seminal report \cite{ScottD:ISWIM}, Dana Scott explained that 
\begin{quote}
"
a category represents the 'algebra of types', just as abstract rings give us the algebra of polynomials, originally understood to concern only integers or rationals. One can of course think only of particular type systems, but, for a full understanding, one  needs also to take into account the general theory of types, and especially \emph{translations}\/ or \emph{interpretations}\/ of one system in another."
\end{quote}
Samson Abramsky spearheaded the efforts towards expanding the categorical semantics of program abstraction, as formalized in type theory and merge it with a categorical semantics of process abstraction and interaction, as formalized in the theory of concurrency and process calculi. This led to interaction categories \cite{AbramskyS:icats,CockettR:sproc,PavlovicD:CCPS1,PavlovicD:CCPS2}, specification structures \cite{AbramskyS:spec,PavlovicD:CTCS97}, and a step further to geometry of interaction \cite{AbramskyS:GoI} and game semantics \cite[and many other publications]{AbramskyS:interaction,AbramskyS:AJM92,AbramskyS:PCF,AbramskyS:AJM94}. As the realm of program abstraction expanded, e.g. into quantum computation and protocols \cite{Abramsky-Coecke:LICS04}, the semantical apparatus also expanded \cite{AbramskyS:LICS10,AbramskyS:Synthese-big}, the tree branched \cite{PavlovicD:QMWS,PavlovicD:Qabs12}, some branches crossed\footnote{E.g., \cite{PavlovicD:QPL09} used the methods of \cite{PavlovicD:CTCS97} to expand the models of \cite{Abramsky-Coecke:LICS04}.}. In the present paper, however, we are only concerned with the root. And even that might be overly ambitious.

\subsection{Categorical proof theory}
\paragraph{Proofs-as-constructions.} The Curry-Howard isomorphism is one of the conceptual building blocks
of type theory, built deep into the foundation of computer science and
functional programming \cite[ch.~3]{GirardJY:prot}. The fact that it
is an {\em isomorphism} means that the type constructors
on one side obey the same laws as the propositional connectives on the other side; and these laws are expressed as a bijection between the terms and the proofs.

\subsubsection{Entailments as morphisms}\label{Sec:entailment}
In categorical proof theory, logical sequents are treated as arrows in a category \cite{LambekJ:dedsc1,LambekJ:dedsc2,LambekJ:dedsc3,LawvereFW:dialectica}. The reflexivity and the transitivity of the entailment relation then correspond to the main categorical structures: the identities and the composition.
\begin{gather}\label{eq:cut}
\prooftree
\justifies
A \vdash A
\endprooftree
\qquad\qquad\qquad\qquad\qquad
\begin{tikzar}[row sep = 6ex]
1\ar{d}[description]{\enco \id}\\ \Set(A,A)
\end{tikzar}
\\[3ex] \notag
\prooftree
A\vdash B\qquad \qquad  B \vdash C
\justifies
A \vdash C
\endprooftree
\qquad\qquad
\begin{tikzar}[row sep = 6ex]
\Set(A,B) \times \Set(B,C)\ar{d}[description]{(-;-)}\\
\Set(A,C)
\end{tikzar}\\[3ex] \notag
\end{gather}
But while there is at most one sequent $A\vdash B$ for given $A$ and $B$, there can be many arrows between $A$ and $B$ in a category. Categorical semantics of the logical entailment must therefore be imposed by equations:
\begin{gather*}
\begin{tikzar}{}
\&\Set(A,B)\ar{dd}[description]{\id} \ar{dl}[description]{\left<\enco{\id}, \id\right>} \&\& \Set(A,B)\ar{dd}[description]{\id} \ar{dr}[description]{\left<\id,\enco{\id}\right>}\\
\Set(A,A)\times \Set(A,B) \ar{dr}[description]{(-;-)} \&\&\&\& \Set(A,B)\times \Set(B,B) \ar{dl}[description]{(-;-)}
\\
\& \Set(A,B) \&\&  \Set(A,B)
\end{tikzar}
\\[3ex]
\begin{tikzar}{}
\Set(A,B)\times \Set(B,C)\times \Set(C,D) \ar{dd}[description]{(-;-)\times \id} \ar{rr}[description]{\id \times (-;-)}\&\& \Set(A,B)\times \Set(B,D) \ar{dd}[description]{(-;-)}
\\
\\
\Set(A,C)\times \Set(C,D) \ar{rr}[description]{(-;-)}\&\& \Set(A,D)
\end{tikzar}
\end{gather*}

\subsubsection{Conjunction and disjunction as product and coproduct}
Algebraically, the conjunction and the disjunction are the meet and the join in the proposition lattice. Categorically, they are the product and the coproduct:
\begin{gather}\label{eq:conj}
\prooftree
X\vdash A\qquad\qquad X\vdash B
\Justifies
X\vdash A\wedge B
\endprooftree
\qquad\qquad
\begin{tikzar}{}
\Set(X,A) \times \Set(X,B)\ar[bend right]{d}[swap]{{<-,->}} \\
\Set(X, A\times B) \ar[bend right]{u}[swap]{\paar{\pi_A\circ-,\pi_B\circ-}}
\end{tikzar}
\\[3ex]
\prooftree
A\vdash X\qquad\qquad B\vdash X
\Justifies
A\vee B \vdash X
\endprooftree
\qquad\qquad
\begin{tikzar}{}
\Set(A,X) \times \Set(B,X)\ar[bend right]{d}[swap]{{[-,-]}} \\
\Set(A+ B,X) \ar[bend right]{u}[swap]{\paar{-\circ \iota_A,-\circ \iota_B}}
\end{tikzar} \label{eq:disj}
\end{gather}

\begin{definition}\label{def:cartesian} A category with the product constructor $\times$ supporting the correspondence \eqref{eq:conj} is called \emph{cartesian}. A category with the coproduct constructor $+$supporting the correspondence \eqref{eq:disj} is called \emph{cocartesian}.
\end{definition}

The difference between the algebraic and the categorical view is that in the first case there is at most one entailment $X\vdash A$, whereas in the second case there can be many arrows $X\rightarrow A$, usually labelled, and viewed as functions in the category $\Set$. The mapping in \eqref{eq:conj} on the right establishes the bijection between the proofs or functions $X\rightarrow A\times B$ and the pairs of proofs or functions $X\rightarrow A$ and $X\rightarrow B$. The proof transformations thus become function manipulations. If the elements of sets or entries of data types, witness the corresponding propositions, then the logical operations are operations on data. E.g., proof of conjunction becomes a pair of data entries. It often comes as a surprise that such simple-minded analogies can be effective tools in functional programming \cite{PavlovicD:AndreFest}. They also have far-reaching logical consequences, some of which are pursued in the present paper.

\subsubsection{Function abstraction and cartesian closed categories}\label{Sec:static-then}
The fact that the conjunction $A\wedge(-)$ is the right adjoint to the implication  $A\thenn(-)$ \cite{LawvereFW:dialectica} means that the implication introduction and elimination can be expressed as the reversible logical rule in  \eqref{eq:static-then} on the left. 
\beq\label{eq:static-then}
\prooftree
(A\wedge X) \vdash B
\Justifies
X \vdash (A\thenn B)
\using{\thenn}
\endprooftree
\qquad
\qquad\qquad\qquad
\begin{tikzar}[row sep = 5ex]
\Set(A\times X,B) \ar[bend right]{d}[swap]{\To A - \circ \eta_X} \\
\Set\big(X,\To A B\big) \ar[bend right]{u}[swap]{\varepsilon_X\circ(A\times-)}
\end{tikzar}
\eeq
The corresponding type-theoretic correspondence in  \eqref{eq:static-then} on the right was the first example of the propositions-as-types phenomenon.  This bijection between two sets of proofs-as-terms was noticed by Haskell Curry  back in the 1930s. The operation corresponding to the implication introduction, i.e. going down, is called the \emph{abstraction}. The operation corresponding to the implication elimination, i.e. going up, is called the \emph{application}. The categorical view of the resulting correspondence captures its uniformity with respect to all indexing types $X$, i.e. its \emph{polymorphism}, as the \emph{naturality}\/ with respect to the type constructors $(A\times -)$ and $\To A {-}$. A correspondence between two constructors is natural if it is preserved under all substitutions. For \eqref{eq:static-then} every $f\in \Set(X,Y)$ induces the two squares in \eqref{eq:ccc-nat}, one formed by $\eta$s, the other by $\varepsilon$s.
\beq\label{eq:ccc-nat}
\begin{tikzar}[row sep = 5ex,column sep = 10ex]
\Set(A\times Y,B) \ar[bend right]{d}[swap]{\To A - \circ \eta_Y} \ar{r}[description]{-\circ(A\times f)}
\& \Set(A\times X,B) \ar[bend right]{d}[swap]{\To A - \circ \eta_X}\\
\Set\big(Y,\To A B\big) \ar[bend right]{u}[swap]{\varepsilon_Y\circ(A\times-)} \ar{r}[description]{-\circ f}
\& \Set\big(X,\To A B\big) \ar[bend right]{u}[swap]{\varepsilon_X\circ(A\times-)}
\end{tikzar}
\eeq
The naturality of these squares means that they are commutative. The commutativity of these squares captures the type-theoretic \emph{conversion rules}\/ imposed on the abstraction operation and the application operation:
\begin{align}
\begin{tikzar}[column sep = 1ex,row sep = 3ex]
\& A\times\To A {(A\times X)} \ar{dr}[description]{\varepsilon_{A\times X}}\\
A\times X\ar{rr}[description]{\id} \ar{ur}[description]{A\times \eta_X}\&\& A\times X
\end{tikzar}\ \ \  & &
\Big(\lambda a.\  f_x(a)\Big)\cdot b & =  f_x(b)\tag{$\beta$}\\ 
\begin{tikzar}[column sep = .1ex,row sep = 3ex]
A\,\Rightarrow\, X\ar{rr}[description]{\id} \ar{dr}[description]{\eta_{\To A X}}\&\& A\,\Rightarrow\, X\\
\& A\,\Rightarrow\, {(A\times \To A X)} \ar{ur}[description]{\To A{\varepsilon_{X}}}
\end{tikzar}
& & \lambda a.\ \Big(g_x\cdot a\Big) & =  g_x \tag{$\eta$}
\end{align}
The application operation is derived from the adjunction counit $\varepsilon_X: A\times \To A X \to X$ in the form $g\cdot a =\varepsilon(a,g)$. The abstraction operation is written in type theory it is written using the variables, like in the rules $(\beta)$ and $(\eta)$ above, but can also derived from the adjunction unit $\eta_X: X\to \To A {(A\times X)}$, in the form $\lambda(f) = \To A f \circ \eta_X$. 

\begin{definition}\label{def:closed} A \emph{cartesian closed} category is a cartesian category $\Set$ with the static implication $\To{A}{B}$ for every pair of types $A,B$ and the $X$-natural (function) abstraction operation
\bea\label{eq:cartabs}
\Set(A\times X, B) & \xrightarrow[\raisebox{2ex}{$\sim$}]{\ \ \lambda^{AB}_X\ \ }& \Set(X, \To A B)
\eea
\end{definition}

\para{Remark.} Towards aligning with Def.~\ref{def:proc-closed}, note that the function abstraction $\lambda_{AB}$ is with respect to the functors  $H_{AB}, \nabla_{AB}\colon \Set^o\to \Set$ defined  
\[ H_{AB} X = \Set(A\times X, B) \qquad\qquad\qquad \nabla_{AB} X = \Set(X,\To A B\] 
The arrow parts are induced by precomposition.

\subsection{Modalities as monads and comonads}
\subsubsection{Possibility and side-effects}\label{Sec:side-effect}
A possibility modality can be introduced by the rules on the left. 
\beq\label{eq:mnd}
\prooftree \justifies A \vdash \may A
\endprooftree
\qquad\qquad
\prooftree
A\wedge B \vdash \may C
\justifies
\may A \wedge \may B \vdash \may C
\endprooftree
\qquad\qquad\qquad\qquad\qquad
\begin{tikzar}{}
\Set(A\times B,MC) \ar[bend right]{d}[swap]{\#\phi} \\
\Set\big(MA\times MB, MC\big) \ar[bend right]{u}[swap]{\eta\eta}
\end{tikzar}
\eeq
Each of the logical rules corresponds to one of the categorical transformations on the right, where $\eta\eta$ precomposes $A\times B\tto{\eta_A\times \eta_B} MA\times MB\to MC$, whereas $\#\phi$ first $\#$-lifts $A\times B\tto g MC$, and then precomposes to $MA\times MB\tto\phi M(A\times B) \tto{g^\#} MC$. The quadruple $(M, \eta, \#,\phi)$ is the structure of a \emph{commutative monad} \cite{BarrM:ttt,KockA:comm-closed,KockA:comm-dist,ManesE:algt}. The type constructor $M$, the unit $\eta: A\to MA$ and the lifting $\#$ of $A\tto f MC$ to $MA\tto{f^\#} MB$ satisfy the equations
\beq\label{eq:monad}
\eta_A^\#  = \id_{MA}
\qquad\qquad\qquad\qquad
 \ f^\# \circ \eta_{A\times B}  =  f 
\qquad\qquad\qquad\qquad
 \ (f^\#\circ t)^\#  =  f^\#\circ t^\# 
\eeq
This triple is one of the equivalent presentations of the structure of a monad \cite{ManesE:algt}. Most presentations \cite{BarrM:ttt,Lambek-Scott:book} define a monad as a triple $(M,\eta,\mu)$, where  $\mu_A:MMA\to MA$ are the (cochain) evaluators. The lifting $\#$ is derivable from the evaluators by setting $f^\# = \left( MA\tto{Mf} MMB \tto\mu MB\right)$, whereas the evaluators are derivable as the liftings of the identities, in the form $\mu_A = \left(MMA \tto{\id^\#} MA\right)$. The lifting operation $\#$ seems more convenient for programming. The last component $\phi$ of the structure $(M, \eta, \#,\phi)$ (or of the equivalent form $(M, \eta, \mu,\phi)$) is the \emph{bilinearity}\/ $\phi_{AB}:MA\times MB \to M(A\times B)$, which makes the monad commutative \cite{KockA:comm-closed,KockA:comm-dist}. This natural family satisfies
\beq\label{eq:monad-comm}
\phi_{AC}\circ (\eta_A\times \eta_C)  = \eta_{A\times C}
\qquad\qquad\qquad\qquad
\big(\phi_{BD}\circ(f \times g)\big)^\# \circ \phi_{AC} =   \phi_{BD}\circ(f^\# \times g^\#) 
\eeq
for all pairs $f:A\to  MB$ and $g:C\to MD$.  Similar equations are valid for all tuples. The equations in \eqref{eq:monad} define the identities and the composition in the category of free algebras (in the Kleisli form)
\bear
|\Set_M|&=& |\Set|\\
\Set_M(A,B) & = & \Set(A,MB)
\eear
While correspondences \eqref{eq:disj} persist in $\Set_M$, the natural bijection in \eqref{eq:conj} does not, and $\Set_M$ is not a cartesian category any more. However, the equations in \eqref{eq:monad-comm} assure that the product $\times$ from $\Set$ persists as a monoidal structure in $\Set_M$ \cite{KockA:comm-dist}. Intuitively, a function $A\to MB$ produces not just the outputs in $B$, but also some \emph{side-effects} \cite{MilnerR:processes}, represented in the type $MB$.  E.g., the fact that computations may not terminate means that they implement functions in the form $A\to \Del B$ where the monad
\bea\label{eq:Del}
\Del{(-)} :\Set&\to&\Set\\
X&\mapsto & X\cup\{\bot\}\notag
\eea
adjoins to each set a fresh element $\bot$. This is the \emph{maybe} \/ monad, corresponding to the algebraic theory with a single constant and no operations or equations. The category $\Set_\bot$ is (equivalent to) the category of sets and partial functions.

Another side-effect of interest is the \emph{nondeterminism}. Some computations may depend on the states of the computer, which may depend on the environment. Running the same program on the same inputs may therefore produce different outputs at different times, for no unobservable reason. Such computations implement functions in the form $A\to \WP B$, where the monad
\bea\label{eq:WP}
\WP:\Set&\to&\Set\\
X&\mapsto & \{V\subseteq X\}\notag
\eea
maps each set to the set of its subsets, a.k.a. its \emph{powerset}. This is the \emph{nondeterminism}\/ (or powerset) monad. It maps to every function $X\tto g Y$ the function $\WP X \tto{\wp g} \WP Y$, which takes $V\subseteq X$ to $\WP g(V) = \{g(x)\in Y\ |\ x\in V\}$.  The unit $X\tto \eta \WP X$ maps $x\in X$ to $\eta(x) =\{x\}$. The lifting maps a function $A\tto f\WP B$ to $\WP A \tto{f^\#} \WP B$ where $f^\#(V) = \cup_{v\in V} f(v)$. For reasons  discussed in Appendix A, it satisfies
\bear
\Set(A,\WP B) & \cong & \Set (B,\WP A)
\eear
which makes the category $\Set_\wp$ of nondeterminisic functions self-dual, along the natural bijection $\Set_\wp(A,B) \cong \Set_\wp(B,A)$. The idea is that, given a nondeterministic function $A\to \WP B$, knowing all possible $B$-outputs for each $A$-input allows us to extract all possible $A$-inputs for each $B$-output, which yilelds just another nondeterministic function $B\to \WP A$.  See Appendix A for more.
 
\paragraph{Notation.} Since they will be cast in leading roles, the above categories of functions with effects are abbreviated to:
\begin{itemize}
\item $\Pfn = \Set_\bot$ --- category of partial functions, and
\item $\Rel = \Set_\wp$ --- category of relations.
\end{itemize}

\paragraph{Background.} In mathematics, monads emerged as a "standard construction" of free algebras involving topologies \cite{BarrM:ttt,ManesE:algt}. The observation that the type constructors $M$ that add side-effects also carry the monad structure goes back to \cite{MoggiE:monad}. Initially proposed as a semantical tool, monads turned out to be a powerful and convenient programming tool. Nowadays, monads' popularity among programmers drives interest in semantics. Mathematically, a monad $M$ is a saturated view of an algebraic theory, presented not by operations and equations, but as a mapping from any set of generators $B$ to the free algebra $MB$. The unit $\eta$ maps each generator to its place in the free algebra. The lifting $\#$ expands the assignment $A\tto f MB$ from the generators $A$ to the algebra homomorpism $MA\tto{f^\#} MB$. Any monad corresponds to an algebraic theory, albeit with infinitary operations. The semantical assumption that all computational side-effects can be captured by algebraic operations has deep repercussions on the concept of computation.

\subsubsection{Necessity and reductions}
Dually, a necessity modality can be introduced by
\beq\label{eq:cmn}
\prooftree \justifies \must A \vdash A
\endprooftree 
\qquad\qquad
\prooftree
\must A\vdash B\vee C
\justifies
\must A \vdash \must B\vee \must C
\endprooftree
\qquad\qquad\qquad\qquad\qquad
\begin{tikzar}{}
\Set(GA,B+C) \ar[bend right]{d}[swap]{\#} \\
\Set\big(GA, GB+GC\big) \ar[bend right]{u}[swap]{-\circ\varepsilon_{B+C}}
\end{tikzar}
\eeq
This time the triple $(G,\varepsilon,\#)$ is made into a comonad by the equations:
\[
\varepsilon_A^\#  = \id_{GA}
\qquad\qquad\qquad\qquad
 \varepsilon_{B+C} \circ\ f^\#   =  f 
\qquad\qquad\qquad\qquad
 \ (f\circ t^\#)^\#  =  f^\#\circ t^\# 
\]
The third equation defines the composition in the category of coffee coalgebras, in the \emph{Kleisli}\/ form again:
\bear
|\Set_G|&=& |\Set|\\
\Set_G(A,B) & = & \Set(GA,B)
\eear
Computational interpretations of comonads are less standard, but overviews can be found in \cite{Brookes-Geva,uustalu2008comonadic}. We will need a \emph{history comonad}\/ to capture the time extension of processes in  Sec.~\ref{Sec:history}. For the moment, let us just mention the \emph{indexing}\/ comonads
\bea
A\times(-):\Set&\to&\Set\\
X&\mapsto & A\times X\notag
\eea
which exist for each $A\in \Set$, with the counits $A\times X\tto\varepsilon X$ realized by the projections, and the lifting $A\times X\tto h Y+Z$ defined to be $A\times X\tto{<\id_A, h>} A\times (Y+Z)\cong (A\times Y) + (A\times Z)$. The Kleisli category $\Set_{A\times}$ freely adjoins an indeterminate arrow $1\tto{x} A$ to $\Set$, and plays the role of the polynomial extension $\Set[x:A]$  
 \cite{Lambek-Scott:book,PavlovicD:MSCS97}. Like any Kleisli category, $\Set_{A\times}$ provides a \emph{resolution}\/ of its comonad, in the sense that it factors through the functors
\bear A\times (-) & = & \Big(\Set \tto{-\circ\varepsilon} \Set_{A\times} \tto{\#} \Set\Big)
\eear
as displayed in \eqref{eq:cmn}. While the Kleisli resolution is \emph{initial}\/ among the resolutions of the comonad $A\times(-)$, some of the constructions in this paper are built upon the fact that the resolution
\bear A\times (-) & = & \Big(\Set \tto{\Pi_A} \Set/A \tto{\Dom} \Set\Big)\eear
is \emph{final}\/ among all resolutions. Here $\Set/A$ is the category of $\Set$-morphisms into $A$, the functor $\Pi_A$ functor maps $X$ to the projection $A\times X\tto{\pi_A} A$, whereas the $\Dom$ functor $\Dom$ takes the $\Set/A$-objects, which are the $\Set$-morphisms with the codomain $A$, to their domains $\Dom(X\to A) = X$. 

\begin{lemma}\label{lemma:total} The domain functor $\Dom:\Set/A \to \Set$ is final among all functors $F:\mathbf{C}\to \Set$ which map the terminal object $1$ into $A$. 
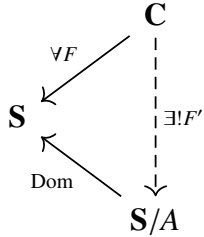
\[
\begin{tikzar}{}
\&\mathbf{C}\ar[dashed]{dd}{\exists ! F'} \ar{dl}[swap]{\forall F}\\
\Set \\
\& \Set/A \ar{ul}{\Dom}
\end{tikzar}
\]
\end{lemma}
\bpr Given $F$ with $F1 = A$, the unique $F'$ with $\Dom \circ F' = F$ is $F' X = F(X\tto ! 1)$.\epr

\subsection{Labelled sequents, commutative monads, and surjections} \label{Sec:commutativity}
In propositional logic, a sequent $X\vdash Y$ transforms proofs of $X$ into proofs of $Y$. If there are several different ways to derive one from the other, the sequent $X\vdash Y$ identifies them all. This leads to a mismatch within the propositions-as-types interpretation because it implies that there is at most one proof $X\vdash A\thenn B$, while there can be many different terms of type $X\tto t \To A B$. This mismatch is resolved by labelling the sequents, writing $X\llseq t A\thenn B$ for the former sequent. We use the symbol $\lseq$ (and not $\vdash$) for labelled sequents, to be able to write $X\, |\hspace{-1ex}\rightarrow\hspace{-0.75ex} Y$ instead of $X\llseq f Y$ when the label $f$ is irrelevant. The categorical proof theory originates from studies of labelled sequents in \cite{LambekJ:dedsc1,LambekJ:dedsc2,LambekJ:dedsc3}. A non-categorical theory of labelled sequents was developed in  \cite{GabbayD:labelled}.

For a modality $\may$, the sequents between the propositions $\may A \wedge \may B$ and $\may(A\wedge B)$ are derivable both ways, and the two are considered equivalent. The proposition $\may \top$ is also equivalent to the truth $\top$. For a monad $M$, the maps $M(A\times B) \tto{<M\pi_A, M\pi_B>} MA\times MB$ and $M1\tto ! 1$ are derivable from the cartesian structure, and the maps $MA\times MB \tto{\phi} M(A\times B)$ and $1\tto \eta M1$ are given by the monad structure.  However, these pairs of functions both ways are generally not inverse to one another. The type $M1$ is generally not final, and the type $M(A\times B)$ is generally not a product. The side-effects of type $M(A\times B)$ are different from the side-effects that arise when the outputs  are received into $MA$ and $MB$ separately. 

While this state of affairs is justified in algebra, where $M1$ is the free algebra over a single generator, it is not justified in the semantics of computation, where the trivial outputs of type $1$ should not cause nontrivial side effects contained in type $M1$. Viewing the monad $M$ as an algebraic theory shows that the nontrivial elements of $M1$ arise from the constants of the algebraic theory. This requirement is not satisfied either by the maybe monad, or by the nondeterminism monad, as the former gives the universe $\Pfn = \Set_\bot$ of sets and partial maps, the latter the universe $\Rel = \Set_\wp$ of sets and relations. The former is the category of free algebras for the theory with a single constant $\bot$, and no other operations. The latter is the category of free join semilattices, where the lattice unit is a constant again. 

Lemma~\ref{lemma:total} says that making 1 into the unit type (final object) in $\Rel = \Set_\wp$ leads to the slice category $\TRel = \Rel/1$, which boils down to 
\bea
|\TRel| & = & \coprod_{A\in |\Set|} \WP A\notag\\
\TRel(\safet S A, \safet T B) & = & \Big\{R\in \Rel(A,B)\ |\ (x\in S \iff \exists y\in T.\ xRy)\ \wedge \label{eq:TRel}\\
&& \hspace{5.3em} \wedge\ (y\in T\iff \exists x\in 
S.\ xRy)\Big\}\notag
\eea
Since the $\Rightarrow$-direction of each of the conjuncts in \eqref{eq:TRel} implies the $\Leftarrow$-direction of the other conjunct, the requirement boils down to $\forall x\in S \exists y\in T.\ xRy$ and $\forall y\in T \exists x\in 
S.\ xRy$. The category $\TRel$ is thus equivalent to the subcategory of $\Rel$ comprised of the relations that are total in both directions. Proceeding in a similar way to make 1 into the final type in the category $\Set_\bot = \Pfn$ leads to the slice category $\TPfn = \Pfn/1$, which is equivalent to the subcategory $\TSet$ of $\Set$ spanned by the surjective functions: 
\bea
|\TPfn| & = & \coprod_{A\in |\Set|} \WP A\notag\\
\TPfn(\safet S A, \safet T B) & = & \Big\{f\in \Set(S,T)\ |\ y\in T \Rightarrow \exists x\in S.\ f(x)=y\Big\}  \label{eq:TPfn}
\eea
\paragraph{Remark for the category theorist.} The forgetful functor $\TPfn \to \TSet$, where $\TSet$ is the category of sets and surjections, is an equivalence because it is surjective on the objects, and full and faithful on the morphisms. However, for each set $S\in \Set$ there is a proper class of sets $A$ such that $\safet S A \in \TPfn$ is mapped to $S\in \TSet$. Constructing the adjoint equivalence $\TSet \to \TPfn$ thus involves a choice from these proper classes of objects.

\section{Process logics}
\label{Sec:ProcLog}


\subsection{Idea of process}
The alignment of logic and type theory remains stable as long as the world is stable: the true propositions remain true, and the data types remain as given. The problems arise when processes need to be modeled, and their dynamic aspects need to be taken into account.

There are physical processes, chemical processes, mental processes, social processes. The common denominator is that they change state: a physical process changes the state of matter; a mental process changes the state of mind. Computation is also a process. Although already a local execution of a program changes the local states of a computer, it seems that the crucial aspects of processes of computation arise from their interleaving with the processes of communication, from the resulting computational interactions, and only emerge into light when the problem of concurrency is taken into account. That is why the semantics of computational processes, formalized in process calculi, initially forked off from the main branch of the semantics of programming languages. The main part of Samson Abramsky's work, which I am trying to reconstruct in logical terms, was concerned with bringing the two branches together.

\subsection{Process propositions and implications}
\subsubsection{Process sequents must be labelled}
Process logics involve modeling states. There are many different ways to model states, but within a propositions-as-types framework, state spaces occur together with the data types, subject to the same derivation rules. Although the two must be treated differently within the rules (as we shall see already in Sec.~\ref{Sec:problem-cut}), both require \emph{labelled}\/ sequents. For state spaces, this is clearly unavoidable. As mentioned in Sec.~\ref{Sec:commutativity}, an unlabelled sequent $X\vdash Y$ identifies all different proofs that $X$ entails $Y$. In particular, there is just one entailment $X\vdash X$, the trivial one. But if $X$ is a state space, then modeling state transitions requires nontrivial sequents $X \llseq \xi X$. The labels allow distinguishing the nontrivial sequents, where the states change, from the trivial one, where they do not.

\subsubsection{Machine abstraction and process-closed categories}\label{Sec:proc-implication}
A process implication $\TTTo A {B}$ asserts not just that $A$ implies $B$, but also that $A$ implies $\TTTo A {B}$. Under the propostions-as-types interpretation, the type $\TTTo A B$ thus comes with two functions

\bigskip
\begin{minipage}{.95\linewidth}
\begin{itemize}
\item $A\wedge \TTTo A {B}\llseq\varepsilon B$ \hfill $\left(\out \upsilon\right)$   
\vspace{-.5\baselineskip}
\item $A\wedge {\TTTo A {B}} \llseq \zeta {\TTTo A {B}}$\hfill $\left(\sta \upsilon\right)$ 
\end{itemize}
\end{minipage}

\bigskip
\noindent The label says that the latter is not an instance of the propositional conjunction elimination, a.k.a. projection on the types side. The sequent $\zeta$ is a \emph{coinductive}\/ clause, saying that $\TTTo A B$ is true on its own whenever it is true together with $A$ as a \emph{guard} \cite{CoquandT:coind,PavlovicD:GIFC}. This is a typical \emph{impredicative}\/ claim, of kind which is often used mathematical analysis \cite{PavlovicD:CRN,PavlovicD:CRN1,PavlovicD:LICS98}. The general idea is that, whenever a proposition $X$, guarded by a proposition $A$, entails a proposition $B$, \emph{and moreover also itself}, i.e. whenever $X$ comes with the sequents
\bigskip

\begin{minipage}{.95\linewidth}
\begin{itemize}
\item $A\wedge X \lseq B$ \hfill $\left(\out{\ana{-}}\right)$   
\vspace{-.5\baselineskip}
\item $A\wedge X \lseq X$\hfill $\left(\sta{\ana{-}}\right)$ 
\end{itemize}
\end{minipage}

\bigskip
\noindent then $X$ also entails the process implication $\TTTo A B$. Putting it all together, we get the following rules: 
\beq\label{eq:dynamic-then}
\prooftree
\hspace{3em}
\justifies
A\wedge \TTTo A B \llseq \upsilon B\wedge  \TTTo A B
\using{\upsilon}
\endprooftree
\qquad\qquad
\prooftree
A\wedge X \llseq \varphi B \wedge X
\justifies
X\llseq{\ana\varphi} \TTTo A B
\using{\scriptstyle\ana-}
\endprooftree
\qquad\qquad\qquad\qquad
\begin{tikzar}[row sep = 7ex]
 \Set(A\times X,B\times X) \ar{d}[description]{\ana-_X}\\
 \Set\left(X,\TTTo A B\right)
\end{tikzar}
\eeq
\paragraph{Terminology.} A function in the form $\xi: A\times X \to B\times X$ is often called a \emph{machine}, and the set $X$ is construed as its \emph{state space}. The induced description ${\ana\xi}:X\to\TTTo A B$  is called \emph{anamorphism}.\footnote{Anamorphisms are the coalgebra homomorphisms into final coalgebras. The name is due, I believe, to Lambert Meertens. It seems to have caught on without having been introduced in a publication. Many functional programmers call them \emph{unfolds}, generalizing the special case of lists. A machine $A\times X \to B\times X$ can be viewed as a coalgebra $X\to \To A {(B\times X)}$.}

\paragraph{Naturality.} Comparing the $\ana-$-rule with the $(\thenn)$-rule in Sec.~\ref{Sec:static-then}, shows the sense in which $\TTTo A B$ is a dynamic version of the implication $(A\thenn B)$. But note that the rule $(\thenn)$ is reversible, whereas  the rule $\ana -$ is not; and that the $X$-natural bijection in \eqref{eq:static-then} on the right boils down to an $X$-natural transformation on the right in \eqref{eq:dynamic-then}. Moreover, since $X$ occurs on both sides of the sequent $A\wedge X \llseq \varphi B \wedge X$, and thus in both covariant and contravariant position in $\Set(A\times X,B\times X)$, the naturality of $\ana -_X$ is not as simple as in \eqref{eq:ccc-nat}, and it genuinely adds to the story. This time the naturality is in the form 
\beq\label{eq:dcc-nat}
\begin{tikzar}[row sep = 13ex,column sep = 15ex]
\Set\left(A\times Y, B\times Y\right) \ar{d}[description]{\ana-_Y} \ar[leftrightarrow]{r}[description]{\Theta_{AB} f}
\& \Set\left(A\times X, B\times X\right) \ar{d}[description]{\ana-_X}\\
\Set\left(Y,\TTTo A B\right)\ar{r}[description]{(-\circ f)} 
\& \Set\left(X,\TTTo A B\right)
\end{tikzar}
\eeq
where $\Theta_{AB}$ is the functor
\bea\label{eq:Theta}
\Theta_{AB}\ :\ \Set^o& \to & \Rel\\
X&\mapsto & \Set(A\times X,B\times X)\notag
\eea
where $\Rel$ is the category of sets and relations, described in Appendix A. The arrow part of this functor transforms a function $f\in \Set(X,Y)$ into the relation $\Theta_{AB} f = (f)$ which is a subset of $\Set(A\times Y,B\times Y) \times \Set(A\times X,B\times X)$ defined by
\bea\label{eq:coalg-homom}
\zeta (f) \xi & \iff & \begin{tikzar}[row sep =7ex,column sep =7ex]
A\times Y\ar{d}[description]{\zeta}  \& A\times X \ar{d}[description]{\xi}\ar{l}[description]{A\times f}\\
B\times Y \& B\times X \ar{l}[description]{B\times f}
\end{tikzar}
\eea
The relation $(-\circ f)$ in \eqref{eq:dcc-nat}  is the arrow part of the functor 
\bea\label{eq:Xi}
\nabla_{AB}:\Set^o & \to & \Rel\\
X &\mapsto & \Set(X,\TTTo A B)\notag
\eea
where $\nabla_{AB} f = (-\circ f)$ is the subset of $\Set(Y,\TTTo A B) \times \Set(X,\TTTo A B)$ defined
\bea\label{eq:widehat}
y (- \circ f) x & \iff & y\circ f = x
\eea
$\nabla_{AB}$ is, of course, just homming into $\TTTo A B$, i.e. a functor to the category  of sets $\Set$ extended along the inclusion $\Set\inclusion \Rel$ of functions as special relations. The naturality of $\ana -:\Theta_{AB}\to \nabla_{AB}$ genuinely depends on this casting. It says that $\ana -$ must preserve the machine (i.e. coalgebra) homomorphisms specified in \eqref{eq:coalg-homom}. The concept of an $AB$-machine homomorphism has herewith been reconstructed logically, from the properties of the dynamic implication $\TTTo A B$ in \eqref{eq:dynamic-then}. 

To reconstruct the structure of final $AB$-machine, substitute $\TTTo A B$ for $Y$ in \eqref{eq:dynamic-then}, to get the outer square in
\beq\label{eq:dcc-final}
\begin{tikzar}[row sep = 2ex,column sep = 2ex]
\Set\left(A\times \TTTo A B, B\times \TTTo A B\right) \ar{ddddd}[description]{\ana- } \ar[leftrightarrow]{rrrrr}[description]{\left(\ana \varphi\right)}
\&\&\&\&\& \Set\left(A\times X, B\times X\right) \ar{ddddd}[description]{\ana-}
\\ 
\& \upsilon \ar[mapsto]{ddd} \ar[mapsto]{rrr}  \&\&\& \varphi \ar[mapsto]{ddd}
\\
\\
\\
\& \id _{\TTTo A B} \ar[mapsto]{rrr} \&\&\& \ana \varphi 
\\
\Set\left(\TTTo A B ,\TTTo A B\right)\ar{rrrrr}[description]{\left(-\circ \ana\varphi\right)} 
\&\&\&\&\& \Set\left(X,\TTTo A B\right)
\end{tikzar}
\eeq
The inner square says that, if we bind together the two left-hand rules in \eqref{eq:dynamic-then} by requiring that   
\bear
\ana \upsilon_{\TTTo A B} & = & \id_{\TTTo A B}
\eear
then the naturality requirement in \eqref{eq:dynamic-then} implies that $A\times \TTTo A B \tto \upsilon B\times \TTTo A B$ is final among all $AB$-machines. This is conveniently summarized in the next definition, intended for the readers with categorical background.

\begin{definition}\label{def:proc-closed} A \emph{process closed} category is a cartesian category $\Set$ with a process implication $\TTTo A B$ for any pair of types $A,B$ and the $X$-natural \emph{machine abstraction}\/ operaton
\bea\label{eq:procabs}
\Set(A\times X, B\times X) & \tto{\ \ \ana{-}^{AB}_X \ \ }& \Set(X, \TTTo A B)
\eea
The naturality of $\ana{-}^{AB}$ is with respect to the functors 
$\Theta_{AB}, \nabla_{AB} \colon  \Set^o \to \Rel$ defined in   (\ref{eq:Theta}--\ref{eq:widehat}).
\end{definition}

\para{Remark.} Def.~\ref{def:proc-closed} can be viewed as a lifting of Def.~\ref{def:closed} to process logics. While the latter is the categorical setting of the static propositions-as-types paradigm, the former recasts categories with final $AB$-machines in a logical form. The simple logical relation between the two structures will be spelled out in Prop.~\ref{ccc-ana}.

\subsubsection{Process propositions}\label{Sec:source}
A static proposition $B$ is equivalent with the static implication $\truth \thenn X$, where $\truth$ is the true proposition. All propositions can thus be viewed as special implications: namely the implications from the truth. A dynamic proposition $\TTo B$ can thus be defined in the form $\TTo B = \TTTo \truth B$. Since the conjunctions $\truth \wedge X$ are also equivalent with $X$, dynamic propositions can be defined by the rules
\[\prooftree
\hspace{3em}
\justifies
\TTo{B} \llseq{\upsilon} B\wedge \TTo{B}
\using{\scriptstyle\upsilon}
\endprooftree
\qquad\qquad
\prooftree
X \llseq \beta B \wedge X
\justifies
X\llseq{\ana \beta} \TTo{B}
\using{\scriptstyle\ana-}
\endprooftree
\qquad\qquad\qquad\qquad\qquad
\begin{tikzar}[row sep = 7ex]
\Set(X, B\times X) \ar{d}[description]{\ana-}\\
\Set(X,\TTo{B})
\end{tikzar}
\]
Retracing the analysis from Sec.~\ref{Sec:proc-implication} now presents a proposition $\TTo B$ with a structure map $\TTo B \tto\upsilon B\times \TTo B$, as final among all maps in the form $X\to B\times X$. The structure map  is thus a pair $\upsilon = <\out \upsilon,\sta \upsilon>$, where $\out \upsilon : \TTo B \to B$ gives an output of the process proposition, or an action,  and $\sta\upsilon :\TTo B \to \TTo B$ gives a resumption. It is thus a stream of elements in $B$.


\subsection{Relating process implications and static implications}
The static implication is defined by the rules and the correspondence in \eqref{eq:static-then}. The process implication is defined by the rules and the correspondence in \eqref{eq:dynamic-then}. How are they related? Under which conditions are both sets of rules supported? Prop.~\ref{ccc-ana} provides an answer. Sections~\ref{Sec:history} and \ref{subsec:retr}, introduce the structures involved in the answer. 

\subsubsection{History types}\label{Sec:history}
A \emph{process of $A$-histories over a state space $X$}\/ is a pair of functions $\kappa = \left<\innn\kappa, \conss\kappa\right>$ typed
\beq\label{eq:semigroup}
A\tto{\innn \kappa} X \oot{\conss\kappa} A\times X
\eeq
The idea is that,
\begin{itemize}
\item $\innn \kappa (a) \in X$ is the initial state of a process that starts with $a\in A$; 
\item $\conss\kappa (x,a) \in X$ is the next state of a process after the state $x\in X$ and event or action $a\in A$.
\end{itemize}
A history $a^n =\sseq{a_1\ a_2\cdots a_n}$ thus takes the process $\kappa$ to the state
\bear
x_n & = & \conss\kappa(a_n,\conss\kappa(a_{n-1},\ldots \conss\kappa(a_1,\innn \kappa(a_0))\cdots))
\eear
Each string of $n$ actions, construed as an $A$-history is thus mapped to a unique element of $X$. If the histories $\sseq{a_1\cdots a_n}$ are viewed as the elements of $A^n$, then the disjoint union (coproduct)
\bear
A^+ & = & \coprod_{n=1}^\infty A^n 
\eear
is the type of all $A$-histories. This is what we call a \emph{history type}. For any process of $A$-histories $\kappa$ over $X$ there is a unique \emph{banana-function}\/ (a.k.a. \emph{fold}, or \emph{catamorphism}) $A^+\tto{\cata\kappa} X$ that makes following diagram commute.
\[\begin{tikzar}{}
\& A^+ \ar[dashed]{dd}[description]{\cata\kappa}\& A\times A^+ \ar{l}[description]{(::)} \ar[dashed]{dd}[description]{A\times \cata\kappa} \\ 
A\ar{ru}[description]{(-)}\ar{rd}[description]{\innn \kappa}\\
\& X \& A\times X \ar{l}[description]{\conss \kappa} 
\end{tikzar}
\]
Hence the history type constructor, the functor
\bea
(-)^+ \ \colon \ \Set & \to & \Set\\ 
A &\mapsto & A^+\notag
\eea

\subsubsection{Retractions and idempotents}\label{subsec:retr}
A \emph{retraction}\/ is a pair of maps $A\overset{q}{\underset{i}{\rightleftarrows}} B$ such that $q\circ i = \id_B$.  The type \emph{$B$ is a retract of $A$} when there is such a pair. It is easy to see that the composite $\varphi = i\circ q$ is then idempotent, and the retraction $A\overset{q}{\underset{i}{\rightleftarrows}} B$ is its \emph{splitting}. The following diagram summarizes a retraction 
\[\begin{tikzar}{}
A \ar[two heads]{dd}[description]{q} \ar{rr}[description]{\varphi} \ar{dr}[description]{\varphi} \&\& A\\
\& A\ar{ur}[description]{\varphi}\ar[two heads]{dr}[description]{q}\\
B \ar[tail]{ur}[description]{i} \ar[equal]{rr} \&\& B \ar[tail]{uu}[description]{i}
\end{tikzar}
\]
It is easy to see that $i$ is then an equalizer of $\varphi$ and the identity; and that $q$ is a coequalizer of the same pair. Since any functor preserves retractions, they provide an exampe of an \emph{absolute limit and colimit}. A categorical construction is  \emph{absolute}\/ when it is preserved by all functors. It turns out that all absolute limits and colimits boil down to retractions \cite{PareR:absolute}. A category where all idempotents split into retractions is thus \emph{absolutely complete}. The \emph{absolute completion}\/ of a category takes its idempotents as the objects, and a morphism $f$ between the idempotents $\varphi$ and $\psi$ is required to preserve them, in the sense that $f = \psi\circ f\circ\varphi$. This is the weakest form of a categorical completion. Retractions, or idempotent splittings\footnote{While the term \emph{idempotent splitting}\/ is well-established in category theory, the term \emph{retraction}\/ is familiar in mathematics at large. They refer to the same fundamental operation \cite[Sec.~11]{PavlovicD:nucleus}.}, are thus an instance of the (co)limit operation.

The following proposition is a first step towards expanding the propositions-as-types paradigm to processes, promised in the title of this paper. 
\subsubsection{Proposition}
\label{ccc-ana}
{\it Let $\Set$ be a cartesian category. 
Then the static implications and the process implications induce each other in the presence of the history types and the retractions. More precisely,
\begin{enumerate}[a)]
\item a cartesian closed category is process closed whenever it has the history types; 
\item a process closed category is cartesian closed whenever it has the retractions.
\end{enumerate}}

The \textbf{proof} is given in the Appendix. 

\paragraph{Process abstraction is function abstraction over history types.} Prop.~\ref{ccc-ana} says that a cartesian closed category with history types has final $AB$-machines for all types $A$ and $B$, and that their state spaces $\TTTo A B = \To{A^+} B$ support rules \eqref{eq:dynamic-then}. A final $AB$-state machine can be constructed as a final coalgebras for the functor
\bear
E_{AB} : \Set & \to & \Set\\
X & \mapsto & \To A {(B\times X)}
\eear
i.e. as a limit of the tower in the form
\beq\label{eq:finlim}
\begin{tikzar}[column sep = 8ex]
1 \& \To A B  \ar{l}[swap]{!}
\&  \ar{l}[swap]{\To A {(B\times !)}} 
\To A {(B\times \To A B)} 
 \arrow[d, phantom, ""{coordinate, name=Z}]
\&\hspace{1ex}
\& \hspace{2em}
\\
\& E^n_{AB}(1)\arrow[ur,""{},dotted,rounded corners,to path= {%
-- ([xshift=-13ex]\tikztostart.east)%
|- (Z) \tikztonodes%
-| ([xshift=3ex]\tikztotarget.east) -- (\tikztotarget)}%
]
\&\ar{l}[swap]{E^{n}_{AB}(!)} 
E_{AB}^{n+1}(1)\&\& \ar[dotted]{ll} \To {A^+} {B}
\end{tikzar}
\eeq
The process implications $\TTTo A B$ are thus modeled together with the static implications $\To A B$, and both sets of rules \eqref{eq:static-then} and \eqref{eq:dynamic-then} are supported. Processes can thus be modeled as machines. This was indeed the starting idea of process semantics \cite{MilnerR:processes}. However, early on along this path, it becomes clear that many different machines implement indistinguishable processes. The problem of process equivalence arises  \cite{MilnerR:CCS82}. The input and the output types $A$ and $B$ of a process are observable, but the state space $X$ may not be. In fact, any observable behavior can be realized over many different, unobservable state spaces.


\subsection{The problem of cut in process logics}\label{Sec:problem-cut}
The fact that a process model may not support process composition is not just a conceptual shortcoming, but an obstacle to applications. Engineering tasks are in principle made tractable by decomposing the required processes into components, implementing the components, and composing the implementations. The process component models thus usually encapsulate and hide implementations, and display the interfaces. This methodology is conceptualized in \emph{full abstraction},  one of the tenets of semantics of computation ever since \cite[Sec.~4]{MilnerR:processes}. 

The first logical requirement of compositionality is that the state spaces must be factored out. This is necessary if the composition is to comply with a cut rule \eqref{eq:cut}. If the process sequents are state machines in the form $X\times A\llseq \varphi X\times B$ and $Y\times B\llseq \psi Y\times C$, then the cut rule would be something like
\beq\label{eq:comp-state}
\prooftree
X\wedge A\llseq \varphi X\wedge B\qquad \qquad Y\wedge B\llseq \psi Y\wedge C
\justifies
Z\wedge A\llseq {(\varphi;\psi)} Z\wedge C
\endprooftree
\eeq
The mismatch between the state spaces $X$ and $Y$ needs to be somehow resolved by composite state space $Z$. How should processes pass their internal states to one another?

The intuitive difference between data and states is that data are processed, whereas the states enable the processing. The structural difference is that data can be copied and sent in messages, whereas the states are internal, and may not be communicable. The problem of process composition is thus that the \emph{observable}\/ aspects of processes, that get passed in process composition from one process to another, need to be separated from the unobservable aspects, that remain hidden from the compositions. The same problem arises in applying processes as dynamic functions on sources as dynamic elements.The latter can, of course, be viewed as a special case of the former, just like process propositions are viewed as a special case of process implications. 

The idea towards a solution is that the observable aspects are presented as data types, the unobservable aspects as state spaces. Processes should thus keep their internal states for themselves, as any dynamics aspects of their interactions can be communicated using the process implications. This follows from the fact, spelled out at the end of Sec.~\ref{Sec:proc-implication}, that the process implications are the state spaces of the final state machines. Dispensing with the internal states, the process composition should thus be defined as a sequent in the form
\bea\label{eq:proccomp}
\TTTo A B \wedge \TTTo B C & \llseq{\gamma} & \TTTo A C
\eea
In the static logics, the sequents that establish the transitivity of implication are equivalent with the cut rule from \eqref{eq:cut}. In the process logics, the sequents like \eqref{eq:proccomp} solve the problem with \eqref{eq:comp-state}. The final machine and coalgebra structures carried by the process implications have been used to define composition in a variety of final-coalgebra-enriched categories \cite{AbramskyS:icats,AbramskyS:spec,PavlovicD:CTCS97,PavlovicD:FOSSACS01}. The general derivation pattern behind the composition sequents in the form \eqref{eq:proccomp} is something like this:
\beq\label{eq:fcoalg-comp}
\prooftree
\prooftree
\prooftree
\prooftree
\hspace{2em}
\justifies
A\wedge \TTTo A B \llseq \upsilon B\wedge \TTTo A B
\using{\scriptstyle\upsilon}
\endprooftree
\justifies
A\wedge \TTTo A B\wedge \TTTo B C \llseq{\alpha} B\wedge \TTTo A B \wedge \TTTo B C
\endprooftree 
\qquad
\prooftree
\prooftree
\hspace{2em}
\justifies
B\wedge \TTTo BC \llseq \upsilon C\wedge \TTTo BC
\using{\scriptstyle\upsilon}
\endprooftree
\justifies
B\wedge \TTTo A B\wedge \TTTo B C \llseq{\beta} C\wedge \TTTo A B \wedge \TTTo B C
\endprooftree 
\justifies
A\wedge \TTTo A B \wedge \TTTo B C\  \llseq{(\alpha;\beta)} \ C\wedge \TTTo A B \wedge \TTTo B C 
\endprooftree
\justifies
\TTTo A B \wedge \TTTo B C\  \llseq{\gamma\, =\, \ana{\alpha;\beta}} \  \TTTo A C
\using{\scriptstyle\ana-}
\endprooftree
\eeq
The task of composing processes thus boils down to interpreting the process implications $\TTTo A B$. The task of applying processes to sources boils down to interpreting the process propositions $\TTo A = \TTTo \truth A$.

\section{Functions extended in time}
\label{Sec:SFun}

\subsection{Dynamic elements as streams}
The outputs of a machine $a\  =\  \left(X\tto{\left<\out a, \sta a\right>} A\times X\right)$ are observable as a stream $\stream a = \sseq{a_0\ a_1  \cdots a_n\cdots}$. Starting from an initial state $x_0\in X$ the process
\begin{itemize}
\item outputs $a_0 = \out a_{x_0}$ and updates the state to $x_1 = \sta a_{x_0}$; then it
\item outputs $a_1 = \out a_{x_1}$ and updates the state to $x_2 = \sta a_{x_1}$; after $n$ steps, it
\item outputs $a_{n} = \out a_{x_n}$ and updates the state to $x_{n+1} = \sta a_{x_n}$; and so on.  
\end{itemize}
A dynamic\footnote{We use the terms \emph{"dynamic"}\/ and \emph{"extended in time"}\/ interchangeably. A distinguishing aspect, that justifies keeping both in traffic, will emerge later.} element can thus be construed as a stream of outcomes of a repeated measurement or count. Such data streams arise in science, and they are the subject of statistical inference \cite{fisher1973statistical}. If the outcomes are the truth values, then these streams are the process propositions. When the frequencies are counted, then they are the streams of random variables called \emph{sources}\/ in information theory \cite[Ch.~6]{AshR:IT}.

\subsection{Functions extended in time as deterministic channels}\label{Sec:dynfun}
 A dynamic function from $A$ to $B$ is generated by a machine in the form $f \  = \  \left(A\times X\tto{\left<\out f, \sta f\right>} B\times X\right)$. Starting from an initial state $x_0\in X$ the process consists of the following data maps and state updates:
\begin{align*}
a_0 &\mapsto b_0=\out f_{x_0}(a_0) & a_0 &\mapsto x_1 = \sta f_{x_0}(a_0)\\
a_0\ a_1 &\mapsto b_1= \out f_{x_1}(a_1) & a_0\ a_1 & \mapsto x_2 =\sta f_{x_1}(a_1)\\
&\cdots && \cdots\\
a_0\  a_1 \cdots a_{n} &\mapsto \  b_n= \out f_{x_n}(a_n) & a_0\ a_1 \ \cdots a_{n}   &\mapsto   x_{n+1} = \sta f_{x_n}(a_n)\\
&\cdots && \cdots
\end{align*}
A dynamic function can thus be viewed as a stream of functions in the form 
 \[\stream f \ \ =\ \  \seq{f_0\  f_1 \  \cdots f_n\cdots} \qquad \mbox{where}\qquad f_n = \out f_{x_n} :A^n\to B\]
The propositions-as-types interpretation of the process implication is based on such streams of functions. Streams of random functions are studied in information theory as channels. Those considered here correspond to the \emph{deterministic}\/ channels \cite[Sec.~3.2]{AshR:IT}. 

\subsection{History monad and comonad}
The construction $(-)^+ \ \colon \ \Set  \to  \Set$, described in Sec.~\ref{Sec:history}, supports the monad structure
\[\begin{tikzar}{}
A\ar{rr}{\eta} \&\& A^+\& A^+  \& B^+\ar{l}[swap]{g^\#}  \\
a\ar[mapsto]{rr}\&\& \seq a \& g(b_1)\cdot g(b_2)\cdots g(b_n)  \& \ar[mapsto]{l} \seq{a_1\ a_2\cdots a_n}
\end{tikzar}
\]
where $g$ is a function from $B$ to $A^+$, and $\cdot$ is the string concatenation. The algebras for this monad are semigroups: the set of finite $A$-sequences (words, nonempty lists) $A^+$ is the free semigroup over $A$.

For our concerns, it is more interesting that the construction $(-)^+ \ \colon \ \Set  \to  \Set$ also supports the comonad structure
\[\begin{tikzar}{}
A \&\&\& A^+ \ar{lll}[swap]{\varepsilon} \ar{r}{f^\#}  \& B^+\\
a_n\&\&\& (a_1 a_2\cdots a_n) \ar[mapsto]{lll} \ar[mapsto]{r} \& \Big(\, f(a_1)\  f(a_1 a_2)\ \cdots\  f(a_1\cdots a_n)\, \Big)
\end{tikzar}
\]
Thinking of the sequences $(a_1 a_2\cdots a_n)$ as sequences of events makes them into \emph{histories}. The cumulative functions $f^\#$ thus capture the functions extended in time. Prop.~\ref{ccc-ana} says that proofs of the process implications $\TTTo A B$ correspond to such functions. This correspondence makes process implications into hom-sets of a category.

\subsection{Category of functions extended in time}
The category of free coalgebras for the comonad $(-)^+$ is
\bear
|\Set_+| & = & |\Set|\\
\Set_+(A,B) & = & \Set(A^+,B)
\eear
The lifting $\#$ gives rise to the composition in this category:
\[\prooftree\prooftree
A^+\tto f B
\justifies
A^+ \tto{f^\#} B^+
\endprooftree
\qquad\qquad
B^+ \tto g C
\justifies
(f; g) =\left(A^+\tto{f^\#} B^+\tto g C \right)
\endprooftree
\]
The counit $A^+\tto \varepsilon A$ plays the role of the identity for this composition. The correspondence
\bear
\Set_+(A, B) & \cong & \Set(1, \TTTo A B)
\eear
means that $\Set_+$, in a sense, externalizes the process implications as functions extended in time, and makes their proofs composable. The time extension of their composition unfolds in their cumulative form. Since $A^+$ is the disjoint union of $\coprod_{n=1}^\infty A^n$, a function $f:A^+ \to B$ can be viewed as the stream $\stream f = \sseq{ f_1\  f_2\cdots f_n\cdots}$ of functions $f_n : A^n\to B$, like in Sec.~\ref{Sec:dynfun}. The corresponding cumulative function $f^\#:A^+\to B^+$ can then be viewed as the stream $f^\# = \sseq{ f^1\  f^2\cdots f^n\cdots}$ of functions $f^n : A^n\to B^n$ which commute  in the following diagram
\beq\label{eq:ladder}
\begin{tikzar}[row sep = large,column sep = large]
A\ar{d}[description]{f^1} \ar[leftarrow]{r}{\tail} \& A^2\ar{d}[description]{f^2} \ar[leftarrow]{r}{\tail}\& A^3\ar{d}[description]{f^3} \ar[leftarrow]{r}{\tail} \& A^4 \ar{d}[description]{f^4} \&  A^i \ar[dotted]{l}  \ar{d}[description]{f^i} \ar[dotted,leftarrow]{rr} \&\&\hspace{1em}\\
B \ar[leftarrow]{r}[swap]{\tail} \& B^2 \ar[leftarrow]{r}[swap]{\tail}\& B^3 \ar[leftarrow]{r}[swap]{\tail} \& B^4  \&  B^i \ar[dotted]{l} \ar[dotted,leftarrow]{rr} \&\&\hspace{1em}
\end{tikzar}\eeq
Each $\tail$ projects away the rightmost component. The components $f^n$ are: 
\[f^1 \ = \ f_1 \qquad \qquad\qquad\qquad f^{i+1} \ = \ \left<f^i\circ \tail, f_{i+1}\right>
\]

\section{Partial functions extended in time}
\label{Sec:partial}

\subsection{Output deletions and process deadlocks}
Recall from Sec.~\ref{Sec:side-effect}\eqref{eq:Del} that the partiality monad $\Del {(-)} \colon   \Set \to  \Set$ adjoins a fresh element $\bot$ to every type. A partial function $f:A\pfn B$ can be viewed as the total function $A\to \Del B$, which sends to $\bot$ the elements where $f$ is undefined. There are two logically different ways to lift this to processes:
\beq
\prooftree
A\wedge X \lseq  \Del B \wedge X
\justifies
X\lseq \TTTo A {\Del B}
\endprooftree
\qquad\qquad\qquad\qquad
\prooftree
A\wedge X \lseq  \Del{(B \wedge X)}
\justifies
X\lseq \Del{\TTTo A B}
\endprooftree
\eeq
On the left, the process may \emph{delete}\/ some of the outputs, but it always proceeds to the next state, whether if has produced the output or not. On the right, the process may \emph{deadlock}\/ and fail to produce either the output or the next state. The meanings of the two implications $\TTTo A {\Del B}$ and $\Del{\TTTo A B}$ are captured, respectively, by the final coalgebras of the two functors
\begin{align*}
D_{AB\bot}\ \colon \Set & \to \Set  &D_{A\bot B}\ \colon \Set & \to \Set\\
X &\mapsto \To A {(\Del B\times X)} & X &\mapsto \To A {\Del {(B\times X)}} 
\end{align*}
The state spaces of the final coalgebras of these two functors are then the hom-sets of the two categories of partial functions extended in time:
\begin{align}
|\Set_{+\bot}| & = |\Set| & |\Set_{\bot+}| & = |\Set|\notag\\
\Set_{+\bot}(A,B) &= \Set({A^+},{\Del B}) & \Set_{\bot+}(A,B) &= \coprod_{S\in  \Precdown A^+}\Set(S, B) \label{eq:nosafe-defs}
\end{align}
where $\Precdown A^+$ is the set of safety specifications in $A$ \cite{AbramskyS:icats,Alpern-Schneider,PavlovicD:CTCS97}
\bea\label{eq:safety}
\Precdown A^+ & = & \{S\subseteq A^+\ |\ \vec x \preccc \vec y \in S\ \implies \ \vec x\in S\}
\eea
and where the prefix relation $\vec x \preccc \vec y$ means that there is $\vec z$ such that $\vec x \vec z = \vec y$. 
An $\Set_{\bot+}$-morphism is a ladder like \eqref{eq:SFun-morphism}, but with partial functions $f_i$ as rungs. The commutativity requirement imples that $f_i(\vec s)$ must be defined whenever $f_{i+1}(\vec s a)$ is defined for some $a$. Hence $S\in \Precdown A^+$ in \eqref{eq:nosafe-defs}.

\subsection{Safety and synchronicity}
For $B=1$, the right-hand part of \eqref{eq:nosafe-defs} boils down to $\Set_{\bot+}(A,1) \cong\  \Precdown A^+$. The safety properties in $\Precdown A^+$ can thus be viewed as the objects of categories of \emph{safe}\/ dynamic functions. The morphisms may be \emph{synchronous}\/ or \emph{asynchronous}, depending on whether the outputs are always observable. They become asynchronous if some outputs may be hidden or deleted.

\subsubsection{Synchronous safe functions}\label{Sec:synfun}
The category $\SFun$  of \emph{safe dynamic functions} has all safety specifications as its objects. Combining the $\Set_+$-ladders from \eqref{eq:ladder} with the $\Set_\bot/1$-surjections from \eqref{eq:TPfn}  shows that the safe dynamic functions are ladders in the form
\beq\label{eq:SFun-morphism}
\begin{tikzar}[row sep = large,column sep = large]
S_1\ar[two heads]{d}[description]{f^1} \ar[leftarrow]{r}{\tail} 
\& S_2\ar[two heads]{d}[description]{f^2} \ar[leftarrow]{r}{\tail} \ar[phantom,"\llcorner^w",very near start]{dl}
\& S_3\ar[two heads]{d}[description]{f^3} \ar[leftarrow]{r}{\tail} \ar[phantom,"\llcorner^w",very near start]{dl}
\& S_4 \ar[two heads]{d}[description]{f^4} \ar[phantom,"\llcorner^w",very near start]{dl}
\&  S_i \ar[dotted]{l}  \ar[two heads]{d}[description]{f^i} \ar[phantom,"\llcorner^w",very near start]{dl} \ar[dotted,leftarrow]{rr} \&\&\hspace{1em}
\\
T_1 \ar[leftarrow]{r}[swap]{\tail} \& T_2 \ar[leftarrow]{r}[swap]{\tail}\& T_3 \ar[leftarrow]{r}[swap]{\tail} \& T_4  \&  T_i \ar[dotted]{l} \ar[dotted,leftarrow]{rr} \&\&\hspace{1em}
\end{tikzar}\eeq
The functions $f^i$ are not mere surjections, in the sense that for every history $\vec t\in T$ there is a history $\vec s\in S$ such that $\vec t = f^\#(\vec s)$. They are surjections  \emph{extended in time}, in the sense that the prefixes of $\vec t$ must have been the image of the prefixes of $\vec s$, i.e. $\lft \pi\left(\vec t\, \right) = f^\#\left(\lft \pi \vec s\, \right)$. Categorically, this amounts to saying that the squares in \eqref{eq:SFun-morphism} are weak pullbacks. Logically, the commutativity of \eqref{eq:SFun-morphism} uncovers a general coinductive pattern:
\bea\label{eq:syn-surj}
f^\#(\vec s) = \vec t &\iff & \forall b\in B\ \Big(  \vec t b \in T  \implies  \exists a\in A.\  \vec s a \in S\wedge f^\#(\vec s a) = \vec t b \Big)
\eea
Such coinductive surjections lie at the heart of process theory as components of \emph{bisimulations}, which we shall encounter in the next section. Before that, note that the dynamic surjections satisfying \eqref{eq:syn-surj} must be \emph{synchronous}, in the sense that they preserve the length of the histories: the time ticks steadily up the ladder. If there are silent actions, i.e. if functions may delete their outputs, this synchronicity may be breached.

\subsubsection{Asynchronous safe functions}
\label{Sec:asynfun}
The functions extended in time \emph{asynchronously}\/ inhabit the category $\Set_{\bot+}$. The element $\bot$ added to the outputs plays the role of the \emph{silent}, unobservable action \cite{Hennessy-Milner:observ,MilnerR:CCS}.   In synchronous models, the observer is assumed to have global testing capabilities  \cite{AbramskyS:observation-testing}. The asynchrony arises when some of the actions of the Environment may not be observable for the System.  Viewed as channels, the asynchronous functions extended in time become the deterministic \emph{deletion}\/ channels \cite{MitzenmacherM:deletion}.  This leads to coarser process equivalences.  Combining both of the constructions \eqref{eq:nosafe-defs} allows capturing both forms of the partiality in
\bea
|\Set_{\bot+\bot}| & =&  |\Set| \notag\\
\Set_{\bot+\bot}(A,B) &=& \coprod_{S\in \Precdown A^+}\Set(S, \Del B) \label{eq:botPlusBot}
\eea
A  function $f\in \Set(S,\Del B)$ can be viewed as a stream of functions $f = \left(f_n : S_{\leq n} \to \Del B\right)_{i=1}^\infty$, where $S_{\leq n}$ are safe histories of length up to $n$, including the empty history, i.e.
\bea S_{\leq n} &= & (S\cap A^{\leq n})+\big\{()\big\}\label{eq:leq}
\eea   
Here $A^{\leq n}$ is the disjoint union (coproduct) $\coprod_{i=0}^n A^i$. 
The cumulative form $f^\# = \left( f^{\leq n} : S_{\leq n} \to B^{\leq n}\right)_{n=1}^\infty$ is now defined by
\begin{alignat*}{5}
f^{\leq 1 }() & =  () &&\qquad\qquad& f^{\leq n+1 }() & =  ()
\\
f^{\leq 1}(a) 
& = 
\begin{cases}  () &\mbox{ if } f_{1}(a) = \bot\\
f_{1}(a) &\mbox{ otherwise }
\end{cases}
&&& 
f^{\leq n+1}(a\vec x) 
& =  
\begin{cases}  f^{\leq n}(\vec x) &\mbox{ if } f_{n+1}(a\vec x) = \bot\\
f^{\leq n}(\vec x)\cons f_{n+1}(a\vec x)  &\mbox{ otherwise }
\end{cases}
\end{alignat*}
Its components are this time the rungs of the ladder
\beq\label{eq:botPlusBot-morphism}
\begin{tikzar}[column sep =3ex, row sep = 3ex]
S_{\leq 1}\ar{dd}[description]{f^{\leq 1}} \ar[leftarrow]{rr}{\tail} \&\& S_{\leq 2}\ar{dd}[description]{f^{\leq 2}} \ar[leftarrow]{rr}{\tail}\&\& S_{\leq 3}\ar{dd}[description]{f^{\leq 3}} \ar[leftarrow]{rr}{\tail} \&\& S_{\leq 4} \ar{dd}[description]{f^{\leq 4}} \&\& \&\& S_{\leq i} \ar[dotted]{llll}  \ar{dd}[description]{f^{\leq i}} \ar[dotted,leftarrow]{rrrr} \&\&\&\&\hspace{1em}\\[1ex] 
\\
B^{\leq 1} \ar[leftarrow]{rr}[swap]{\tail} \&\& B^{\leq 2} \ar[leftarrow]{rr}[swap]{\tail}\&\& B^{\leq 3} \ar[leftarrow]{rr}[swap]{\tail} \&\& B^{\leq 4}  \&\& \&\& B^{\leq i} \ar[dotted]{llll} \ar[dotted,leftarrow]{rrrr} \&\&\&\&\hspace{1em}
\end{tikzar}\eeq
where each $\tail$ again projects away the last component. The category $\ASFun = \Set_{\bot+\bot}/1$ of asynchronous safe functions has the  safety specifications as its objects again, and a morphism $f\in \ASFun(\safety S A, \safety T B)$ is a  tower in the form
\beq\label{eq:ASFun-morphism}
\begin{tikzar}[row sep = large,column sep = large]
S_{\leq 1}\ar[two heads]{d}[description]{f^{\leq 1}} \ar[leftarrow]{r}{\tail} 
\& S_{\leq 2}\ar[two heads]{d}[description]{f^{\leq 2}} \ar[leftarrow]{r}{\tail} \ar[phantom,"\llcorner^w",very near start]{dl}
\& S_{\leq 3}\ar[two heads]{d}[description]{f^{\leq 3}} \ar[leftarrow]{r}{\tail} \ar[phantom,"\llcorner^w",very near start]{dl}
\& S_{\leq 4} \ar[two heads]{d}[description]{f^{\leq 4}} \ar[phantom,"\llcorner^w",very near start]{dl}
\&  S_{\leq i} \ar[dotted]{l}  \ar[two heads]{d}[description]{f^{\leq i}} \ar[phantom,"\llcorner^w",very near start]{dl} \ar[dotted,leftarrow]{rr} \&\&\hspace{1em}
\\
T_{\leq 1} \ar[leftarrow]{r}[swap]{\tail} \& T_{\leq 2} \ar[leftarrow]{r}[swap]{\tail}\& T_{\leq 3} \ar[leftarrow]{r}[swap]{\tail} \& T_{\leq 4}  \&  T_{\leq i} \ar[dotted]{l} \ar[dotted,leftarrow]{rr} \&\&\hspace{1em}
\end{tikzar}\eeq
This tower differs from \eqref{eq:botPlusBot-morphism} in that the squares are weak pullbacks, and the rungs of the ladder are surjective\footnote{Formally, in any regular category $\Set$, the fact that all rungs are surjective can be derived from the assumption that the starting component is a surjection, and that the squares are weak pullbacks.}. It shows that the \emph{asynchronous}\/ surjections exended in time satisfy the following condition:
\bea\label{eq:asyn-surj}
f^\#(\vec s) = \vec t &\iff & \Big( \forall b\in B.\ \ \vec t b \in T  \Rightarrow  \exists \vec{a}\in A^+.\  \vec s \vec a \in S\wedge f^\#(\vec s\vec a) = \vec t b \Big)
\eea 
This condition differs from \eqref{eq:syn-surj} in that each step up the $T$-side by $b\in B$ may be followed on the $S$-side by a string of steps $\vec a \in A^+$, rather than just a single step $a\in A$.

\section{Relations extended in time}
\label{Sec:rel}

\subsection{External and internal nondeterminism}
We saw in Sec.~\ref{Sec:side-effect} that nondeterminism is modeled using the powerset monad $\WP:\Set\to \Set$. Since a subset $U\subseteq A$ corresponds to an element $U\in \WP A$, a binary relation $R\subseteq A\times B$, viewed as a set of subsets $aR\subseteq B$ indexed over $a\in A$ corresponds to the function $\bullet R\colon A\to \WP B$. The same relation, viewed as a set of subsets $Rb \subseteq A$, indexed over $b\in B$ also corresponds to the function $R\bullet \colon B\to \WP A$. See Appendix A for more details. 

There are two ways again in which the side-effect, this time nondeterminism, may affect processes. Internal nondeterminism affects the outputs, whereas external nondeterminism may also affect the states:  
\beq
\prooftree
A\times X \tto{\xi} \Del{(\WP B \times X)}
\justifies
X\tto{\ana \xi} \TTTo A {\WP B}_\bot
\endprooftree
\qquad\qquad\qquad\qquad
\prooftree
A\times X \tto{\zeta}  \WP{(B \times X)}
\justifies
X\tto{\ana{\zeta}_\wp} {\TTTo A B_\wp}
\endprooftree
\eeq
The external nondeterminism on the right incorporates the partiality as the empty outcome $\emptyset\in \WP(B\times X)$. The partiality monad $\Del{(-)}$ is explicitly added to the internal nondeterminism on the left since they would otherwise never deadlock, which is problematic both conceptually and technically. If a process  $\xi$ on the left, e.g. involving some guessing that leads to internal nondeterminism, never deadlocks at a state $x\in X$ and on an input $a\in A$, then it determines a unique next state $\sta \xi(a,x) \in X$, and may produce an output from the set $\out \xi(a,x)\in \WP B$. For an externally nondeterministic process $\zeta$ on the right, both the outputs and the state transitions are impacted by the nondeterminism, and any pair from $\zeta(a,x) \in \WP(B\times X)$ may be produced when the input $a$ is consumed at state $x$. The intended meanings of the two process implications $\TTTo A {\WP B}_\bot$ and ${\TTTo A B}_\wp$ are captured, respectively, as the final coalgebras of the functors
\begin{align}\label{eq:PQ}
P_{AB}\ \colon \Set & \to \Set  &Q_{A B}\ \colon \Set & \to \Set\notag \\
X &\mapsto \big(A\Rightarrow {(\WP B\times X)}\big)_\bot & X &\mapsto \WP {(A\times B\times X)} 
\end{align}
The expression on the right is based on the bijection $\WP(A\times B\times X) \cong \big(A \Rightarrow {\WP(B\times X)}\big)$. The state spaces of the final coalgebras of these two functors are quite different. We consider them separately, in the next two sections.

\subsection{Internal nondeterminism}
\subsubsection{Synchronous safe relations}\label{Sec:SPROC}
The state space of the final coalgebra of the functor $P_{A\wp B}$ can be constructed within $\Set$ as a limit of the tower like \eqref{eq:finlim}
\bea\label{eq:finlimP}
\begin{tikzar}[column sep = 8ex]
1 \& \To A {\WP B}_\bot  \ar{l}[swap]{!}
\&  \ar{l}[swap]{\To A {(\wp B\times !)}} 
\left(A\Rightarrow \left(\WP B\times \To A {\WP B}_\bot\right)\right)_\bot 
 \arrow[d, phantom, ""{coordinate, name=Z}]
\&\hspace{1ex}
\& \hspace{2em}
\\
\& P^n_{AB}(1)\arrow[ur,""{},dotted,rounded corners,to path= {%
-- ([xshift=-13ex]\tikztostart.east)%
|- (Z) \tikztonodes%
-| ([xshift=3ex]\tikztotarget.east) -- (\tikztotarget)}%
]
\&\ar{l}[swap]{P^{n}_{A\wp B}(!)} 
P_{AB}^{n+1}(1)\& \ar[dotted]{l} \TTTo {A} {\WP B}_\bot
\end{tikzar}
\eea
or presented simply as
\bea\label{eq:nosafe-defs-Pow}
|\Set_{+\wp}| & = & |\Set| \\
\Set_{+\wp}(A,B) &= & \coprod_{S\in \Precdown A^+} \Set(S,{\WP B})  \notag
\eea
A morphism from $A$ to $B$ in $\Set_{+\wp}$ is thus a pair $\left<S, R\right>$, where $S$ is a safety specification, i.e. a prefix-closed set of $A$-histories from \eqref{eq:safety}, and $R$ is a stream of relations, presented as a stream of functions $\bullet R = \left(S_n\tto{\bullet R_n} \WP B\right)_{n=1}^\infty$, where $S_n = S\cap A^n$, or viewed cumulatively as
\bear
\bullet R^\# & = & \Big(S_n\tto{\bullet R^n} \left(\WP B\right)^n\Big)_{n=1}^\infty
\eear
The inductive definition is analogous to the one at the end of Sec.~\ref{Sec:SFun}.  On any input $(a_1\, a_2\, \cdots a_n)\in S$ the $n$-th component of $\bullet R^\#$ thus produces an $n$-tuple of subsets of $B$:
\bea\label{eq:relsfuns}
(a_1\, a_2\, \cdots a_n)R^n & = & \Big< a_1 R_1,\  \ (a_1 a_2)R_2,\ \ldots,\  (a_1\cdots a_{n-1})R_{n-1},\ \  (a_1\cdots a_{n-1} a_n)R_n\Big>
\eea
If each each function $S_n\tto{\bullet R^n} \left(\WP B\right)^n$ is viewed as a relation $S_n\rrel{R^n} B^n$, then \eqref{eq:relsfuns} says that they make the following tower commute
\beq\label{eq:spans}
\begin{tikzar}{}
S_1 \ar[leftarrow]{r}{\lft \pi} \& S_2 \ar[leftarrow]{r}{\lft \pi}\& S_3 \ar[leftarrow]{r}{\lft \pi} \& S_4  \& \& S_i \ar[dotted]{ll} \ar[dotted,leftarrow]{rr} \&\&\hspace{1em}\\
\& R_1 \ar{ul}
\ar{dl}
\ar[leftarrow]{r} 
\& R_2 \ar{ul}
\ar{dl}
\ar[leftarrow]{r}
\& R_3 \ar{ul}
\ar{dl}
\ar[leftarrow]{r} 
\& R_4 \ar{ul}
\ar{dl}
\& \& R_i \ar[dotted]{ll} \ar[dotted,leftarrow]{rr} \ar{ul}
\ar{dl}
\&\&\hspace{1em}
\\
B \ar[leftarrow]{r}[swap]{\lft \pi} \& B^2 \ar[leftarrow]{r}[swap]{\lft \pi}\& B^3 \ar[leftarrow]{r}[swap]{\lft \pi} \& B^4  \& \& B^i \ar[dotted]{ll} \ar[dotted,leftarrow]{rr} \&\&\hspace{1em}
\end{tikzar}
\eeq
To preclude any nontrivial side-effects of processes with trivial outputs, we slice over the trivial type 1 again, and define the category of \emph{safe synchronous relations extended in time}\/ as 
\bea
\SProc & = & \Set_{+\wp}/1
\eea
This is the original \emph{interaction category}, introduced in \cite{AbramskyS:icats}, and further studied in \cite{AbramskyS:spec,PavlovicD:CTCS97}. The descriptions were different, but it is easy to see that the objects coincide, since the morphisms $S\in \Set_{+\wp}(A,1)$ are the prefix-closed sets $S\subseteq A^+$. Reasoning like in Sec.~\ref{Sec:synfun}, a morphism $\safety S A\rrel{R} \safety T B$ in $\Set_{+\wp}/1$ is now reduced to a ladder of spans
\beq\label{eq:bisim}
\begin{tikzar}{}
S_1 \ar[leftarrow]{r}{\lft \pi} 
\& S_2 \ar[leftarrow]{r}{\lft \pi}
\& S_3 \ar[leftarrow]{r}{\lft \pi} 
\& S_4  \&\hspace{1ex} \& S_i \ar[dotted]{ll}   \ar[dotted,leftarrow]{rr} \&\&\hspace{1em}\\
\& R_1 \ar[two heads]{ul}
\ar[two heads]{dl}
\ar[leftarrow]{r} 
\& R_2 \ar[two heads]{ul}
\ar[two heads]{dl}
\ar[leftarrow]{r}
\ar[phantom,"\raisebox{-1.5ex}{$\scriptstyle \ulcorner_w$}",very near start]{ull} 
\ar[phantom,"\raisebox{.5ex}{$\scriptstyle \llcorner^w$}",very near start]{dll} 
\& R_3 \ar[two heads]{ul}
\ar[two heads]{dl}
\ar[leftarrow]{r} 
\ar[phantom,"\raisebox{-1.5ex}{$\scriptstyle \ulcorner_w$}",very near start]{ull}
\ar[phantom,"\raisebox{.5ex}{$\scriptstyle \llcorner^w$}",very near start]{dll} 
\& R_4 \ar[two heads]{ul}
\ar[two heads]{dl}
\ar[phantom,"\raisebox{-1.5ex}{$\scriptstyle \ulcorner_w$}",very near start]{ull}
\ar[phantom,"\raisebox{.5ex}{$\scriptstyle \llcorner^w$}",very near start]{dll} 
\& \& R_i \ar[dotted]{ll} \ar[dotted,leftarrow]{rr} \ar[two heads]{ul}
\ar[two heads]{dl}
\ar[phantom,"\raisebox{-1.5ex}{$\scriptstyle \ulcorner_w$}",very near start]{ull}
\ar[phantom,"\raisebox{.5ex}{$\scriptstyle \llcorner^w$}",very near start]{dll}  
\&\&\hspace{1em}
\\
T_1\ar[leftarrow]{r}[swap]{\lft \pi}  
\& T_2 \ar[leftarrow]{r}[swap]{\lft \pi}
\& T_3 \ar[leftarrow]{r}[swap]{\lft \pi} 
\& T_4  \&\hspace{1ex} \& T_i  \ar[dotted]{ll}  \ar[dotted,leftarrow]{rr} \&\&\hspace{1em}
\end{tikzar}
\eeq
Like in \eqref{eq:TRel}, we have relations that are total in both directions, which means that the projections $R\rightarrow S$ and $R\rightarrow T$ are surjective, in this case componentwise. Like in \eqref{eq:SFun-morphism}, the surjections are extended in time, in the sense that all rhombi in \eqref{eq:bisim} are weak pullbacks. Putting it all together, this tower says that $R$ satisfies
\bea
\vec s\, R\, \vec t  & \iff & \forall a\in A\ \left( \vec s a \in S \Rightarrow \exists b\in B.\  \vec t  b \in T \ \wedge\ \vec s a\,  R\, \vec t b\right) \ \ \wedge\notag \\
&& \forall b\in B\ \left( \vec t b \in T \Rightarrow \exists a\in A.\ \vec s a \in S \ \wedge\ \vec sa\, R\, \vec t b \right) \label{eq:bisimeq}
\eea
This condition means that $\safety S A\rrel{R} \safety T B$ is a \emph{strong}\/ or \emph{synchronous bisimulation}\/ relation \cite{MilnerR:CCS,ParkD:bisimulation}, as required in the original definition of $\SProc$ in \cite{AbramskyS:icats}. 

\paragraph{Bisimulations are intrinsic.} The notion of bisimulation was introduced in process theory as an imposed equivalence of the processes that are intended to be semantically indistinguishable  \cite{MilnerR:CCS82,ParkD:bisimulation}. The logical reconstruction of synchronous bisimulation from process-types-as-propositions shows that the same notion also arises as a property of morphisms in a category. The coinductive reconstructions of the whole gamut of bisimulations comprise a well-studied field of research. The present reconstruction, combining the nondeterminism monad $\WP$, the history comonad $(-)^+$, and the slicing  over 1, displays the synchronous bisimulations as a logical property of processes arising from nondeterministic choices extended in time, provided that that nontrivial side-effects only arise when there are nontrivial outputs.

\subsubsection{Asynchronous safe relations}\label{Sec:ASPROC}
Including in the model the silent, unobservable actions leads to asynchronicity, and to the notion of \emph{weak}\/ or \emph{observational}\/ bisimulation \cite{Hennessy-Milner:observ,MilnerR:CCS}. Proceeding like in Sec.~\ref{Sec:asynfun}, we consider the final coalgebras of the functors
\bea
P_{AB\bot}\ \colon \Set & \to & \Set  \\
X &\mapsto & \big(A\Rightarrow {(\WP(B_\bot)\times X)}\big)_\bot\notag
\eea
as the hom-sets of the category
\bea\label{eq:nosafe-defs-Pow-asyn}
|\Set_{+\wp\bot}| & = & |\Set| \\
\Set_{+\wp\bot}(A,B) &= & \coprod_{S\in \Precdown A^+} \Set(S,{\WP(B_\bot)})  \notag
\eea
The morphism tower is like \eqref{eq:spans}, but with each $S_n$, $R_n$ and $B^n$ replaced replaced with $S_{\leq n}$, $R_{\leq n}$ and $B^{\leq n}$, as in \eqref{eq:leq} and \eqref{eq:botPlusBot-morphism}. The category of \emph{safe asynchronous relations extanded in time}\/ is now 
\bear
\ASProc & = & \Set_{+\wp\bot} / 1
\eear
and the morphism tower is like \eqref{eq:bisim}, with the same modification of the subscripts and the superscripts. This modified tower characterizes the following  logical property of the \emph{asynchronous}\/ relation $R$ extended in time: 
\bea\label{eq:bisimeq-tau}
\vec s\, R\, \vec t  \ \  \iff \ \  \forall a\in A\ \Big(\, \vec sa\in S & \Rightarrow & \exists b\in B\ \big(\vec tb \in T \ \wedge\ \vec sa\,  R\, \vec t b\big) \ \ \vee\notag \\
&&  \exists \vec x\in A^\ast\ \big(\vec sa \vec x \in S \ \wedge\ \vec sa\vec x\,  R\, \vec t\, \big)\Big) \ \ \ \ \ \mbox{ \Large $\wedge$} \notag\\
\forall b\in B\ \Big(\, \vec t b \in T & \Rightarrow &\exists a\in A\ \big(\, \vec sa \in S \ \wedge\ \vec sa\, R\, \vec t b \big) \ \ \vee \notag \\
 && \exists \vec y\in B^\ast\ \big(\vec t b\vec y \in T \ \wedge\ \vec s\, R\, \vec t b\vec y \big)\Big) \eea
This characterizes the \emph{weak}\/ or \emph{observationsl}\/ bisimulations of \cite{Hennessy-Milner:observ,MilnerR:CCS}. The category $\ASProc$ is equivalent to the one introduced and studied under the same name in \cite{AbramskyS:icats,PavlovicD:CCPS2,PavlovicD:CTCS97}.

\subsection{External nondeterminism}
\label{Sec:Nondet-internal}
\subsubsection{Synchronous dynamic relations}\label{Sec:DPROC}
The state space of the final coalgebra of the functor $Q_{AB}$ from \eqref{eq:PQ} should again come with a tower like
\beq\label{eq:finlimPow}
\begin{tikzar}[column sep = 8ex]
1 \& \WP (A\times B)  \ar{l}[swap]{!}
\&  \ar{l}[swap]{\wp(A\times B\times !)} 
\WP(A\times B \times \WP (A\times B) 
 \arrow[d, phantom, ""{coordinate, name=Z}]
\&\hspace{1ex}
\& \hspace{2em}
\\
\& Q^n_{A B}(1)\arrow[ur,""{},dotted,rounded corners,to path= {%
-- ([xshift=-13ex]\tikztostart.east)%
|- (Z) \tikztonodes%
-| ([xshift=3ex]\tikztotarget.east) -- (\tikztotarget)}%
]
\&\ar{l}[swap]{P^{n}_{\wp AB}(!)} 
Q_{AB}^{n+1}(1)\& \ar[dotted]{l} \TTTo {A} {B}_\wp
\end{tikzar}
\eeq
The trouble is that such a tower never stabilizes within a universe of sets, since there is no set $X$ such that $X\cong \WP X$. If we take $A=B=1$, the tower boils down to
\beq\label{eq:H}
\begin{tikzar}{}
1 \& \WP 1   \ar{l}[swap]{\cup}
\& \ar{l}[swap]{\cup} 
\WP\WP 1
\& \ar[dotted]{l}[swap]{\cup}
\WP^n 1 
\&\ar{l}[swap]{\cup} 
\WP^{n+1}(1)\& \ar[dotted]{l} \TTTo 1 {1} _\PPP = \HhH
\end{tikzar}
\eeq
where the coinductive fixpoint $\HhH$ is the class of \emph{hypersets}, or \emph{non-wellfounded sets} \cite{AczelP:nwf}. It is dual to von Neumann's class of well-founded sets \cite{NeumannJ:numbers,ZermeloE:cumulative}, which arises as the inductive fixpoint $\VvV$ along the tower
\beq\label{eq:V}
\begin{tikzar}{}
\emptyset \ar[hook]{r}{\iinn}
\& \WP \emptyset = 1  \ar[hook]{r}{\iinn} 
\& \WP\WP 1 \ar[hook]{r}{\iinn}
\& \WP^n1  \ar[hook]{r}{\iinn}
\& \WP^{n+1} 1 \ar[dotted,hook]{r}{\iinn} \& \VvV
\end{tikzar}
\eeq
Von Neumann, of course, did not draw categorical diagrams like \eqref{eq:V}, but specified his construction using  the transfinite induction
\beq\label{eq:V-Ord}
 V_0 = \emptyset \qquad\qquad V_{\beta} = \bigcup_{\alpha\lt \beta}\WP(V_\alpha)\qquad\qquad \VvV = \bigcup_{\alpha \in {\Ord}} V_\alpha 
 \eeq
The class $\Ord$  of ordinals, over which the union in the third clause is indexed, is assumed to be given. The construction thus provides an \emph{inner\/} model of set theory within a given universe of sets and classes \cite{AczelP:nwf}, or equivalently within a universe with an inaccessible cardinal playing the role of the class $\Ord$ \cite{BarrM:terminal}.\footnote{A universe of sets and classes is a model of the \emph{NBG} set theory, whereas the one with an inaccessible cardinal can be interpreted in terms of the \emph{ZFC}\/ axioms \cite[Ch.~4]{MendelsonE:logic}.}. In any case, reach a fixpoint within a given universe, the constructor $\WP$ must be restricted to stay within a smaller universe. Early on, G\"odel restricted it to the subsets definable in the language of set theory, and constructed the universe $\LlL$ of constructible sets, proving the independence of the Continuum Hypothesis, and launching the whole industry of the independence proofs \cite{GoedelK:constructible}. 
Inner models of set-theory have also been constructed over topological spaces, concrete or abstract   \cite{JoyalA:algst}.

The above constructions also restrict to finite sets. The set theorists often explicitly exclude $\aleph_0$ from the definition of inaccessible cardinals, but the inequalities $2^n\lt\aleph_0$ and $\cup n \lt \aleph_0$ hold for all for all $n\lt \aleph_0$, and that makes $\aleph_0$ inaccessible from the universe $\fSet$ of finite sets. Formally, $\fSet$ can be viewed as the subcategory of $\Set$ spanned by $U\in \Set$ such that $\#U\lt\aleph_0$, where $\#U$ denotes the cardinality of $U$. Since computation is mostly concerned with finite sets, $\fSet$ is often by the computer scientists to be the universe of "small sets", and $\Set$ is interpreted as the universe of "classes". The powerset construction $\WP:\Set\to \Set$ where $\WP X = \{U\subset X\}$ is then replaced with $\PPP:\Set \to \Set$ where
\bea
\PPP X & = 
& \WP_{\lt \omega} X \ \ =\ \ \{U\subset X\ |\ \#U\lt \aleph_0\}
\eea
which restricts to $\PPP:\fSet\to \fSet$. The tower \eqref{eq:H} with $\PPP$ replacing $\WP$ thus lies in $\fSet$, and reaches a fixpoint $\HHH \cong \PPP\HHH$  in $\Set$ after countably many steps.  Since $\PPP$ does not preserve limits, the tower does not stabilize at its limit, but it turns out to stabilize at a retract of its limit \cite{Adamek-Koubek:fincoalg,BarrM:terminal,LambekJ:fixp,PavlovicD:GIFC}. The projections from the fixpoint down the tower are still jointly monic, and support inductive reasoning about the universe of finite hypersets $\HHH = \TTTo 1 1 _\PPP$, which arises in the finite version of \eqref{eq:H}, and about the finite $AB$-relations extended in time $\TTTo A B _\PPP$ which arises in the finite version of \eqref{eq:finlimPow}. Continuing with the workflow from the preceding sections, we use the process implications arising from these finite versions of \eqref{eq:finlimPow} to define the universe of sets with synchronous dynamic relations:
\bea\label{eq:PPP}
|\Set^{\PPP}| & = & |\Set| \\
\Set^\PPP(A,B) &= & \TTTo A B _\PPP \notag
\eea
Like before, we factor out any nontrivial side-effects of processes with trivial outputs by slicing over the trivial type 1 again and define the \emph{category synchronous dynamic relations} 
\bea
\DProc & = & \Set^\PPP/1
\eea

But now something new happens, and a path beyond the workflow from the previous sections opens up. When nondeterminism is internalized, the powerset constructor $\PPP$ generates types with enough structure to play the role of the labels. More precisely, the states built along the towers \eqref{eq:finlimPow} to be cumulatively stored and distinguished using the intrinsic structure, making the label sets $A, B\in \Set$ dispensable. In the constructions so far, the labels were used to identify the same action when it occurs in different processes. Now action can be identified by its history, which the type constructor, that generates the action, stores in the constructed type.

\subsubsection{Internalising the labels} 
\label{Sec:dProc}
All process universes presented up to so far have been built starting from a given universe $\Set$ of labels. The coinductive construction leading to $\DProc$ has a novel feature that it can be built starting from nothing: the role of the label sets $A\in \Set$ can be played by structures arising from the construction itself. The role of the labels $a\in A$ is to identify the same action when it occurs in different observations, or safety specifications $S$ or $T$. This is assured by modeling them as subsets $S,T\subseteq A^+$. The upshot is that there can be at most one label-preserving function $S\to T$, namely the inclusion $S\inclusion T$.

When all actions arise in a cumulative hierarchy, by iterating the constructor $\PPP$, be it inductively \eqref{eq:V} or coinductively \eqref{eq:H},  they are always given as sets with the element relation $\iinn$, which records the elements of each set, their elements, and so on. The axiom of extensionality
\bea
a = b & \iff & \left(\forall x.\ x\iinn a \iff x\iinn b\right)
\eea
says that this $\iinn$-structure completely determines the identity of each set: two sets with the same elements are the same set. In the cumulative hierarchy, the elements of sets are sets too, so the same elements are also the sets with the same elements. If such hereditary $\iinn$-relations are unfolded into trees, the extensionality axiom means that these trees must be \emph{irredundant}: they have no nontrivial automorphisms. In other words, they cannot contain isomorphic subtrees at the same level \cite{PavlovicD:CCPS1}. The $\iinn$-structures that arise from the cumulative processes in \eqref{eq:V} and \eqref{eq:H} are extensional, thus irredundant, because the powerset constructors impose $\{a,a,b,c,\ldots\} = \{a,b,c,\ldots\}$. The other way around, Mostowski's \emph{Collapse Lemma}\/ \cite{MostowskiA:collapse} says that every well-founded extensional relation corresponds to the $\iinn$-structure of a set somewhere in $\VvV$. Aczel's crucial observation in \cite{AczelP:nwf} is that the well-foundedness assumption can be dropped: any extensional relation, including non-wellfounded, can be reconstructed as the $\iinn$-relation of a hyperset, somewhere in $\HhH$, or for finite sets somewhere in $\HHH$. The upshot is that any two hypersets $S, T \in \HHH$, there is at most one $\iinn$-preserving function $S\to T$, or else nontrivial automorphisms arise. The role of the label sets can now be played by the $\iinn$-structures. 

\begin{lemma}\label{Lemma:splitting}
For every countable $A\in \Set$, i.e. such that $\#A\leq \aleph_0$, there are dynamic relations $A\tto{m} 1$ and $1\tto{e} A$ in $\Set^\PPP$ which make $A$ into a retract of $1$, i.e. their composite in $\Set^\PPP$ is
\bear
\id_A & = & \Big(A\eepi e 1\mmono{m} A \Big)
\eear
\end{lemma}

\noindent A \textbf{proof} is sketched in Appendix C. To a category theorist, Lemma~\ref{Lemma:splitting} says that the subcategory $\Set_{\leq \aleph_0}^\PPP\hookrightarrow \Set^\PPP$ spanned by the countable sets is the idempotent completion within $\Set^\PPP$  of the endomorphism monoid $\Hhh = \Set^\PPP(1,1)$. The underlying set of this monoid is the set $\HHH$ of finite hypersets. The monoid operation is the dynamic synchronous relational composition, spelled out below. For the categories
\beq
\dProc\ =\ \Hhh/1 \qquad \qquad\qquad\qquad \DProc_{\leq \aleph_0}\ =\ \Set_{\leq \aleph_0}^\PPP/1
\eeq
we have the following corollary, proved in Appendix D.

\begin{corollary}\label{Corollary:splitting}
The inclusion  
\bea\label{eq:corollary}
\dProc& \inclusion & \DProc_{\leq \aleph_0}
\eea 
is an equivalence of categories.
\end{corollary}

\paragraph{Remark.} The equivalence in the preceding corollary means that the embedding is full and faithful, and essentially surjective, i.e. that every type in $\DProc_{\leq \aleph_0}$ is isomorphic to a type in the image of $\dProc$. This notion of equivalence allows finding an adjoint functor in the opposite direction \emph{provided}\/ that the axiom of choice is given, in this case for classes. The equivalence in \eqref{eq:corollary} therefore does not provide an effective global representation of $\DProc_{\leq \aleph_0}$ in $\dProc$. Locally, however, any structure present in $\DProc$ can be found in $\dProc$, as long as we do not need uncountable sets of labels. In the rest of the paper, we elide the labels, and work in $\dProc$.

\subsubsection{Synchronous dynamic relations as hypersets}
The objects of the category $\dProc$ boil down the elements of the universe of finite $\HHH$, that arises as the coinductive fixpoint of the tower like \eqref{eq:H}, but with $\WP$ restricted to $\PPP = \WP_{\leq\aleph_0}$. Since $\HHH \cong \PPP\HHH$, an element of $\HHH$ can also be viewed as its finite subset, which unfolds it into a tower
\beq\label{eq:dProcObj}
\begin{tikzar}{}
S_1\ar[hook]{d} \ar[leftarrow]{r}{\nnii} 
\& S_2\ar[hook]{d} \ar[leftarrow]{r}{\nnii}
\& S_3\ar[hook]{d} \ar[leftarrow]{r}{\nnii} 
\& S_4 \ar[hook]{d} \& \& S_n \ar[dotted]{ll}  \ar[hook]{d}  \&\& S\ar[hook]{d} \ar[dotted]{ll}\\
\PPP 1 \ar[leftarrow]{r}{\nnii} \& \PPP^2 1 \ar[leftarrow]{r}{\nnii}\& \PPP^3 1 \ar[leftarrow]{r}{\nnii} \& \PPP^4 1  \& \& \PPP^n 1\ar[dotted]{ll}  \&\&\HHH \ar[dotted]{ll}
\end{tikzar}\eeq
where all $S_n$ and $\PPP^n 1$ are in $\fSet$. This seems like the most convenient presentation of the objects of $\dProc$.
A tower corresponding to a morphism $R\in \dProc(S,T)$ looks just like \eqref{eq:bisim} in Sec.~\ref{Sec:SPROC}, except that the projections $\lft \pi$ are replaced by the set-theoretic operation $\cup$. The bisimulation condition \eqref{eq:bisimeq} now becomes 
\bea
s\, R\, t  & \iff & \forall s' \iinn s\  \ \exists t' \iinn t. \ \ s'\,  R\, t'\ \ \ \wedge\ \ \  \forall t'\iinn t\  \ \exists s'\iinn s. \ \ s'\,  R\, t' \label{eq:bisimeq-set}
\eea

\subsubsection{Asynchronous dynamic relations}\label{Sec:adProc}
So far, the asynchrony has been modeled using a silent action $\bot$, which enabled waiting. When the actions are modeled using the element relation $\iinn$, i.e. each state transition is a choice of an element, then waiting can be enabled by allowing sets to contain and choose themselves, i.e. by making the relation $\iinn$ reflexive, satisfying $x\iinn x$ for all $x$. The objects of the category $\aProc$ of \emph{asynchronous dynamic relations}\/ are now the reflexive finite hypersets, conveniently viewed as towers of finite subsets
\beq\label{eq:aProcObj}
\begin{tikzar}{}
S_{\leq 1}\ar[hook]{d}  
\&  S_{\leq 2}\ar[hook]{d} \ar[two heads]{l}[swap]{\nnii}
\& S_{\leq 3}\ar[hook]{d} \ar[two heads]{l}[swap]{\nnii}  \& \& S_{\leq n} \ar[two heads,dotted]{ll}  \ar[hook]{d}  \&\& S\ar[hook]{d} \ar[dotted,two heads]{ll}\\
\PPP^{\leq 1} 1  \& \PPP^{\leq 2} 1 \ar[two heads]{l}[swap]{\nnii}\& \PPP^{\leq 3} 1  \ar[two heads]{l}[swap]{\nnii}   \& \& \PPP^{\leq n} 1\ar[dotted,two heads]{ll}  \&\&\Kar \HHH \ar[dotted,two heads]{ll}
\end{tikzar}\eeq
where $\PPP^{\leq n}{X} = \coprod_{i=0}^n\PPP^i X$, and $\Kar \HHH$ is the universe of reflexive finite hypersets. A morphism $R\in \aProc(S,T)$ is now a reflexive hyperset relation, satisfying the following property
\bea
s\, R\, t  & \iff & \forall s'\iinn s\  \left( \exists t'\iinn t. \ \ s'\,  R\, t'\  \vee\   \exists s''\iinn s'.\ s''R\, t\right)\ \ \wedge\notag\\
&& \forall t'\iinn t\  \left( \exists s'\iinn s. \ \ s'\,  R\, t'\ \vee\ \exists t''\iinn t'.\ s\, R\, t''\right) \label{eq:bisimeq-ord}
\eea
The computational origins of this simulation strategies were studied in \cite{GlabbeekR:branching-2,Glabbeek-Weijland}. Like before, it also arises from the mathematical structure of final coalgebras, and can be logically reconstructed from the paradigm of process-types-as-propositions.

%
%

\section{Integers, interactions, and real numbers}\label{Sec:reals}

\subsection{The common denominator of integers and interactions}
Counting generates the ordinals \cite{NeumannJ:numbers}, but the integers arise from the duality of counting up and down. Geometric and algebraic transformations generate monoids, but capturing the symmetries requires groups. Interactions between the system and the environment generate process universes, some of which we studied; but the dual interactions between the environment and the system were not captured. The duality inherent in process interactions was noted, albeit in passing, very early on in process theory:
\begin{quote}
"The \emph{whole}\/ meaning of any computing agent [would be that it is] a transducer, whose input sequence consists of enquiries by, or responses from, its environment, and whose output sequence consists of enquiries of, or responses to, its environment" \cite[p.~160]{MilnerR:processes}. 
\end{quote}
A similar vision of dual interactions between the system and the environment, as an ongoing session of a question-answer protocol, re-emerged in linear logic \cite{GirardJY:GoI}. It was formalized categorically in \cite{AbramskyS:GoI}, and retraced in \cite{AbramskyS:retracing}. 
The mathematical underpinning turned out to be the $\Int$-construction, generating free compact categories over traced monoidal categories \cite{JSV}. The name does not refer to \emph{\textbf{int}eractions} but to \emph{\textbf{int}egers}. Appying the $\Int$-construction to the additive monoid $\NNn$ of natural numbers,  viewed as a discrete monoidal category,  gives rise to the additive group $\ZZz$ of integers, viewed as a discrete compact category. The the trace structure on the monoid $\NNn$ is the cancellation property:
\bear
m+k = n+k & \implies & m = n
\eear
The set of integers is defined as the quotient 
\bear
\ZZz & = & \Int_\NNn\ \ =\ \ \NNn_\pp\times \NNn_\oo /\sim
\eear
where the equivalence relation $\sim$ is:
\bear
<m_\pp,m_\oo> \sim <n_\pp, n_\oo> & \iff & m_\pp+n_\oo = n_\pp+m_\oo
\eear
The two components of the product are annotated for convenience, e.g. as $\NNn_\pp = \{''\pp''\}\times \NNn$ and $\NNn_\oo = \{''\oo''\}\times \NNn$. The cancellation property guarantees that each $\sim$-equivalence class contains a unique canonical representative in the form $<n,0>$ or $<0,n>$. The former can be written as $\pp n$, the latter as $\oo n$.

The structural common denominator of integers and interactions, which makes the $\Int$-construction applicable to both, is the \emph{trace}\/ operation. It will also take us from relations extended in time to the reals. Towards that goal, we spell out how the trace operation arises in categories of relations. This will makes the $\Int$-construction applicable to the interaction categories.

Since the categories of relations, described in Appendix~\ref{Appendix:Rel}, are self-dual, the coproducts $+$ from the universe of sets and functions $\Set$ become biproducts $\pplus$ in the category $\Rel$ of sets and relations. As the coproduct lifts to the universes of functions extended in time, the biproducts lift to the universes $\SProc$, $\ASProc$, $\dProc$ and $\aProc$ of relations extended in time. By definition, the biproducts are both products and coproducts. Since the relation biproducts are induced by the function coproducts, their unit is the function coproduct unit 0. For every type $X$, the biproduct structure consists of
\begin{itemize}
\item a monoid $0\tto{\ \ !\ \ } X \oot{[\id,\id]} X\pplus X$, and 
\item a comonoid $0\oot{\ \ !\ \ } X\tto{<\id,\id>} X\pplus X$,
\end{itemize}
which are natural with respect to all morphisms in and out of $X$. 
The projections $X\oot{\pi} X\pplus Y \tto{\pi'} Y$ and the injections $X\tto{\iota} X\pplus Y \oot{\iota'} Y$ are derived from the comonoid counits and from the monoid units respectively. A propositions-as-types interpretation of biproducts is tenuous but a process category with the biproducts and the hom-sets $\TTTo A B$ supporting a coinductive rule
\bear
\prooftree
A\pplus X \tto\xi B\pplus X
\justifies 
X \tto{\ana\xi} \TTTo A B
\endprooftree
\eear 
comes with the trace structure $\Tr$ derived by
\bear
\prooftree
\prooftree
A\tto \iota A\pplus Y\qquad\qquad
\overline{A\pplus Y \pplus\TTTo{A\pplus Y}{B\pplus Y} \tto{\overset{\vspace{.5ex}}{\upsilon}} B\pplus Y \pplus\TTTo{A\pplus Y}{B\pplus Y} } \qquad\qquad B\pplus Y \tto\pi B
\justifies
A \pplus\TTTo{A\pplus Y}{B\pplus Y} \tto{\overset{\vspace{.5ex}}{\iota;\upsilon;\pi}} B \pplus\TTTo{A\pplus Y}{B\pplus Y}
\endprooftree
\justifies
\TTTo{A\pplus Y}{B\pplus Y} \tto{\Tr = \ana{\iota;\upsilon;\pi}} \TTTo A B
\endprooftree
\eear 
Each of the categories of relations, $\Rel$, $\SProc$, $\dProc$, etc., is easily seen to give rise to the trace structure in this way. See Appendix~\ref{Appendix:Int} for more.

\subsection{Games as labelled polarized relations extended in time}
The biproducts in $\ASProc$ are in the form
\beq\label{eq:biprod}
\begin{tikzar}{}
(S\oplus T)_{\leq 1}\ar[hook]{d} \ar[leftarrow]{r}
\& (S\oplus T)_{\leq 2}\ar[hook]{d} \ar[leftarrow]{r}
\& (S\oplus T)_{\leq 3}\ar[hook]{d}  \ar[hook]{d} \& \& (S\oplus T)_{\leq i} \ar[dotted]{ll}  \ar[hook]{d} \ar[dotted,leftarrow]{rr} \&\&\hspace{1em}\\
(A+B)^{\leq 1} \ar[leftarrow]{r}{\pi} \& (A+B)^{\leq 2} \ar[leftarrow]{r}{\pi}\& (A+B)^{\leq 3}   \& \& (A+B)^{\leq i} \ar[dotted]{ll} \ar[dotted,leftarrow]{rr} \&\&\hspace{1em}
\end{tikzar}\eeq
where $\safety {(S\oplus T)}{A+B}$ are all shuffles of $\safety S A$ and $\safety T B$.
\bear
(S\oplus T)_{\leq i} & = & \left\{\vec x\in(A+B)^{\leq i}\ |\ \vec x\restriction_A \in S\ \wedge \ \vec x\restriction_B \in T\right\}
\eear
The trace structure of categories of relations with respect to the biproducts as the monoidal structure was analyzed already in the final section of \cite{JSV}, and explained in more detail for the interaction categories in \cite{AbramskyS:retracing}. The analysis presented in that paper suggests that the \emph{AJM-games} \cite{AbramskyS:AJM92,AbramskyS:PCF,AbramskyS:AJM94} should be construed in terms of the $\Int$-construction. The AJM-games are, of course, one of the crowning achievements of the quest for fully abstract models of PCF, and a tool of many other semantical results. They appeared in many different semantical contexts  \cite{AbramskyS:interaction, AbramskyS:PCF,AbramskyS:AJM-deptypes}, with many refinements and different presentation details. A crude common denominator can be obtained by applying the $\Int$-construction from Appendix~\ref{Appendix:Int} to the category $\ASProc$, leading to
\bear
|\Gam| & = & |\ASProc|_\pp \times |\ASProc|_\oo\\ 
\Gam(S, T) & = & \ASProc( S_\pp \oplus T_\oo , T_\pp\oplus S_\oo)
\eear
Some of the crucial features of game semantics, such as the copycat strategy, and the various switching and starting conditions, arise in such reconstructions as abstract mathematical properties, like the notions of bisimulations arose before.

\subsection{Polarized dynamics}
Since $\PPP(A+B) \cong \PPP A\times \PPP B$, applying the powerset constructor on polarized sets $X_\pp +X_\oo$ leads to the functor
\bear
\QQQ\colon \Set &\to & \Set\\
X &\mapsto & \PPP_\pp X \times \PPP_\oo X 
\eear
where the subscripts are still just annotations, and we can take, e.g., $\PPP_\pp X = \{''\pp''\}\times \PPP X$ and $\PPP_\oo X = \{''\oo''\}\times \PPP X$.

\subsubsection{Synchronous case}
The universe of \emph{signed}\/ finite hypersets can be constructed just like the universe of hypersets in Sec.~\ref{Sec:Nondet-internal}, but bifurcating at each step into positive and negative hypersubsets:
\beq\label{eq:RRR}
\begin{tikzar}[column sep = 4ex]
1 \& \PPP_\pp 1 \times \PPP_\oo 1  \ar{l}[swap]{!}
\& \ar{l}[swap]{\QQQ !} 
 \PPP_\pp\left(\PPP_\pp 1 \times \PPP_\oo 1\right)\ \times \   \PPP_\oo\left(\PPP_\pp 1 \times  \PPP_\oo 1\right) \arrow[d, phantom, ""{coordinate, name=Z}]
\&\hspace{1ex}
\& \hspace{2em}
\\
\&\QQQ^n {1} 
\arrow[ur,""{},dotted,rounded corners,to path= {%
-- ([xshift=-13ex]\tikztostart.east)%
|- (Z) \tikztonodes%
-| ([xshift=3ex]\tikztotarget.east) -- (\tikztotarget)}%
]
\&\ar{l}[swap]{\QQQ^n {!}} 
 \PPP_\pp\left(\QQQ^n {1}\right)\ \times \  \PPP_\oo\left(\QQQ^n {1}\right)\& \ar[dotted]{l} \KV
\end{tikzar}
\eeq
The final coalgebra structure still maps each hyperset to its elements, but this time they can be positive or negative
\bea\label{eq:KV}
\KV 
& 
\begin{tikzar}{}
\hspace{.1ex}  \ar[bend right]{r}[swap]{\big<\nnii_\pp, \nnii_\oo\big>} \ar[phantom]{r}[description]{\cong} 
\&
\hspace{.1ex} \ar[bend right]{l}[swap]{\cup}
\end{tikzar}
&
\PPP_\pp \KV\times \PPP_\oo\KV
\eea
\paragraph{Notation.} Given $s\in  \KV$, we write $s^\pp = \nnii_\pp(s)$ for the negative part and $s^\oo = \nnii_\oo(s)$ for the positive part. We often tacitly identify $\KV$ with $\PPP_\pp \KV\, \times\,  \PPP_\oo\KV$, in which case $s\in \KV$ becomes a pair $s=<s^\pp | s^\oo>$, where $s^\pp = \nnii_\pp(s)$ and $s^\oo = \nnii_\oo(s)$. We follow \cite{ONAG} and denote a generic element of $s^\pp$ by $s_\pp$, and a generic element of $s^\oo$ by $s_\oo$, and abbreviate $s_\pp \in \nnii_\pp(s)$ and $s_\oo \in \nnii_\oo(s)$ to $s_\pp, s_\oo \iinn s$. Writing $s = \{s_\pp\ |\ s_\oo\}$ instead of $s = <s^\pp, s^\oo>$ is yet another well-established notational abuse, used to great effect by John Conway in \cite{ONAG}. E.g., instead of $\cup s = \left<\cup s^\pp, \cup s^\oo\right>$, the  polarized union operation, pointing left in \eqref{eq:KV}, can be written in the form
\bear
\cup s & = & \left\{s_{\pp\pp}, s_{\pp\oo} \ |\ s_{\oo\pp}, s_{\oo\oo}\right\}
\eear
Other coinductive definitions become even simpler, e,g.
\beq\label{eq:minus}
\ominus s \ =\  \{\ominus s_\oo\ |\ \ominus s_\pp\}\qquad\qquad\qquad\qquad s\pplus t = \{s_\pp\pplus t, s\pplus t_\pp\ |\ s_\oo\pplus t, s\pplus t_\oo\}
\eeq

\paragraph{Synchronous hypergames.} The objects of the category $\dGam$ are the signed finite hypersets from the universe $\KV$. The final coalgebra structure \eqref{eq:KV} separates their elements into a negative and a positive part. In the game semantics, this is interpreted as separating a game $s\in \KV$ into a pair $s=<s^\pp, s^\oo>$, where $s^\pp = \{s_\pp\iinn s\}\in \PPP_\pp(\KV)$ are the moves available to the player $\pp$, whereas $s^\oo=\{s_\oo \iinn s\}\in \PPP_\oo(\KV)$ are the moves available to the player $\oo$. The projections $\KV\tto{q_i} \QQQ^n1$ down the tower \eqref{eq:RRR} represent each game $s\in \KV$ as a stream $[s^1, s^2, s^3,\ldots, s^{n+1},\ldots]$, where $s^{n+1} = q_{n+1}(s) \in \QQQ^{n+1} 1 = \PPP_\pp(\QQQ^{n} 1) \times \PPP_\oo(\QQQ^{n}1)$, and thus $s^{n+1} = <s^{n+1}_\pp, s^{n+1}_\oo>$, where $s^{n+1}_\pp, s^{n+1}_\oo \subseteq \QQQ^n1$.

A morphism $R\in \dGam(s,t)$ should be a  \textbf{synchronous hyperstrategy}. It is a \emph{hyper}\/strategy because the players $\pp$ and $\oo$ play not one, but two games, $s$ and $t$, distinguished by the dual goals that the two players have in each of them:
\bea
s\, R\, t  & \iff & \forall s_\pp \iinn s\  \exists t_\pp\iinn t. \ \ s_\pp\,  R\, t_\pp\ \ \wedge \ \ \forall t_\oo\iinn t\   \exists s_\oo \iinn s. \ \ s_\oo\,  R\, t_\oo \label{eq:bisimeq-dGam}
\eea
The player $\pp$ is thus tasked with simulating every $s$-step by a $t$-step, whereas the player $\oo$ is tasked with simulating every $t$-step by an $s$-step. A hyperstrategy into a polarized version of a synchronous bisimulation \eqref{eq:bisimeq-set}. While a bisimulation relation between two processes provides a simulation relation of each of them in the other one, both ways, the polarization of a hyperstrategy separates the two simulation tasks, and each player is tasked with one.

\subsubsection{Asynchronous case}
Using the functor $\overline \QQQ:\Set\to \Set$ where $\overline \QQQ X = X +\QQQ X$, the tower in \eqref{eq:RRR} becomes
\beq\label{eq:RRr}
\begin{tikzar}[column sep = 7ex]
1 \& \QQQ^{\leq 1}{(1)}  \ar{l}[swap]{!}
\& \ar{l}[swap]{\overline \QQQ !} 
\QQQ^{\leq 2}{(1)}
\& \ar[dotted]{l}
\QQQ^{\leq n}{(1)} 
\&\ar{l}[swap]{\overline \QQQ^n !} 
\QQQ^{\leq n+1}(1)\& \ar[dotted]{l} \QV 
\end{tikzar}
\eeq
where $\QQQ^{\leq n}{(1)} = \coprod_{i=0}^n\QQQ^{i}(1)$. The final coalgebra structure is thus
\bea\label{eq:RV}
\QV 
& 
\begin{tikzar}{}
\hspace{.1ex}  \ar[bend right]{r}[swap]{\nnii} \ar[phantom]{r}[description]{\cong} 
\&
\hspace{.1ex} \ar[bend right]{l}[swap]{\cup}
\end{tikzar}
&
\QV\  +\ \ 
\PPP_\pp\QV\times \PPP_\oo\QV
\eea
The coalgebra structure $\nnii$ maps $s = <s^\pp, s^\oo>$ to $s = \nnii(s)$ if $s^\pp \in s^\pp$ and $s^\oo \in s^\oo$. Otherwise it unfolds its elements into $s^\pp = \nnii_\pp(s)$ and $s^\oo = \nnii_\oo(s)$ like before. A straightforward induction along the tower gives the following.

\begin{lemma}\label{Lemma:number}
Every $s\in \QV$ is $\iinn$-transitive, in the sense that for all $s_\pp, s_\oo \in s$ holds
\beq \label{eq:number}
s_{\pp}^\pp \subseteq s^\pp \subseteq s_\oo^\pp\qquad\qquad\qquad \qquad s_{\oo}^\oo \subseteq s^\oo \subseteq s_\pp^\oo
\eeq
\end{lemma} 

The elements of the universe $\QV$ of transitive finite signed hypersets can be thought of as \textbf{asynchronous hypergames}. They are the objects of the category $\RRbf$.  An \textbf{asynchronous hyperstrategy} $R\in \RRbf(s,t)$ resembles a branching bisimulation from \eqref{eq:bisimeq-ord}, except that the two simulation tasks are again separated, like in \eqref{eq:bisimeq-dGam}, and assigned to the two players:
\bea
s\, R \, t  & \iff & \forall s_\pp \iinn s\  \left( \exists t_\pp\iinn t. \ \ s_\pp\,  R\, t_\pp \  \vee\   \exists s_{\pp\oo}\iinn s_\pp.\ s_{\pp\oo} R\, t\right)\ \  \wedge \notag
 \\
& & \forall t_\oo\iinn t\  \left( \exists s_\oo \iinn s. \ \ s_\oo\,  R\, t_\oo\ \vee\ \exists t_{\oo\pp}\iinn t_\oo.\ s\, R\, t_{\oo\pp}\right) \label{eq:bisimeq-aGam}
\eea
Lemma~\ref{Lemma:number} makes the relations induced by the coalgebra structure on $\QV$ into hyperstrategies. Remember that $s_\pp \iinn s$ abbreviates $s_\pp \in \nnii(s) \in \PPP_\pp\QV$, whereas $s \nnii s_\oo$ abbreviates $s_\oo \in \nnii(s) \in \PPP_\oo\QV$.

\begin{lemma}\label{Lemma:order} \eqref{eq:bisimeq-aGam} holds when $s R t $ is instantiated to $s_\pp \iinn s$ and $s\nnii s_\oo$, 
for any $s\in \QV$ and $s_\pp, s_\oo \iinn s$.
\end{lemma} 

\bpr  $s_\pp^\pp\subseteq s^\pp$ implies that for every $s_{\pp\pp}$ there is $s'_\pp$ with $s_{\pp\pp}\iinn s'_\pp$. $s^\oo\subseteq s_\pp^\oo$ implies that for every $s_\oo$ there is some $s_{\pp\oo}$ with $s_{\pp\oo}\iinn s_\oo$. Hence \eqref{eq:bisimeq-aGam} for $s_\pp \iinn s$. $s^\pp\subseteq s_\oo^\pp$ implies that for every $s_\pp$ there is $s_{\oo\pp}$ with $s\pp \nnii s_{\oo\pp}$. $s_\oo^\oo \subseteq s^\oo$ implies that for every $s_{\oo\oo}$ there is some $s'_\oo$ with $s'_\oo\nnii s_{\oo\oo}$. Hence \eqref{eq:bisimeq-aGam} for $s\nnii s_\oo$.
\epr

\paragraph{Remark.} The property in \eqref{eq:bisimeq-aGam} is not self-dual under the relational converse, but under the polarity change $\ominus$ in \eqref{eq:minus}. In game semantics, the polarity change switches the roles of the player and the opponent. The game-theoretic concept of \emph{equilibrium}, where both players play their best responses, reimposes the \emph{bi}\/simulation requirement: that the \emph{same}\/ relation is a winning strategy (simulation) in \emph{both}\/ directions. The equilibrium strategies are thus fixed under two dualities: under the polarity change (switching the players $\pp$ and $\oo$), and under the relational converse (switching the component games $s$ and $t$). The two dualities are generally not independent, as there are situations when they do not commute. However, for games where they do commute, they induce a \emph{dagger-compact}\/ structure, akin to the adjunctions over the complex linear operators, which is induced by two commuting dualities: the complex conjugation and the matrix transposition. This structure also arises in many other areas of abelian and nonabelian geometry. It was not used in the game semantics, but it emerged in the Abramsky-Coecke models of quantum protocols and has been explored in other areas of the semantics of computation   \cite{Abramsky-Coecke:LICS04,PavlovicD:CQStruct,PavlovicD:QPL09}.

\subsection{A category of real numbers}
In closing this section, we encounter a remarkable and somewhat disturbing fact: that the posetal collapse of the category $\RRbf$ boils down to the ordered field $\RRr$ of the real numbers. It is disturbing because it shows that the described process logic and game semantic constructions impose on the processes no computability restrictions whatsoever since they include all real numbers. On one hand, this observation should not be surprising, since John Conway reconstructed numbers from games a long time ago \cite{ONAG}, and game semantics was inspired by his ideas and informed by his constructions \cite{AbramskyS:AJM92}. On the other hand, it should be surprising, because game semantics has been developed as the semantics of \emph{computational}\/ processes, albeit as a quotient of an undecidable term calculus \cite{Berry-Curien,LoaderR,MilnerR:fully}. 

\subsubsection{Coalgebra of reals}\label{Sec:coalg-reals}
We adapt the alternating dyadics from \cite[Sec.~3.2]{PavlovicD:CRN}\footnote{See also \cite{PavlovicD:CRN1,PavlovicD:LICS98} for a broader context.} to present the real numbers.  Consider the alphabet $\Sigma = \{\pp, \oo\}$, and denote by $\Sigma^\circledast$ the set of finite and infinite strings over it. It comes with the coalgebra structure
\bea\label{eq:circledast}
\Sigma^\circledast &\begin{tikzar}{}
\hspace{.1ex}  \ar[bend right]{r}[swap]{\chi} \ar[phantom]{r}[description]{\cong} 
\&
\hspace{.1ex} \ar[bend right]{l}[swap]{(::)}
\end{tikzar}& 1+\Sigma \times \Sigma^\circledast
\eea
where $\chi$ maps the empty string $()$ into 1 and each nonempty strings into its head symbol and the tail string. Equivalently, this coalgebra can be written in the form
\bea\label{eq:circledastt}
\Sigma^\circledast &\begin{tikzar}{}
\hspace{.1ex}  \ar[bend right]{r}[swap]{\kappa} \ar[phantom]{r}[description]{\cong} 
\&
\hspace{.1ex} \ar[bend right]{l}[swap]{[o,h_\pp, h_\oo]}
\end{tikzar}& 1 + \Sigma_\pp^\circledast +  \Sigma_\oo^\circledast
\eea
Where the product $\Sigma \times \Sigma^\circledast$, which is $\{\pp, \oo\} \times \Sigma^\circledast$ is expanded into $\{\pp\}\times \Sigma^\circledast \ +\ \{\oo\}\times \Sigma^\circledast$, and the products with the singletons are abbreviated as subscripts. The structure map $\kappa$ now maps the empty string into 1, and the strings in the form $\pm :: \vec x$ as $\vec x$ into $\Sigma^\circledast_\pm$, whereas the components $h_\pp$ and $h_\oo$ add $\pp$ and $\oo$ as the head, while $o$ maps the singleton from 1 into the empty string.

Each $\Sigma$-string encodes a unique real number. The idea is that we count the first string of $\pp$s or $\oo$s in the unary, and after that proceed in the alternating dyadics, e.g.
\bear
++---+-\, - & \mapsto & +1+1 -  \tfrac 1 2 - \tfrac 1 4 - \tfrac 1 8 +\tfrac 1 {16} -\tfrac 1{32} - \tfrac 1{64}\\
----+-+\cdots & \mapsto & -1-1-1-1+\tfrac 1 2 - \tfrac 1 4 +\tfrac 1 8 \cdots
\eear
Since the infinite strings of $\pp$s and of $\oo$s encode the two infinities, we will have a map into the extended reals $\RRrr = \RRr \cup \{\infty, -\infty\}$. The bijection $\Sigma^\circledast \cong \RRrr$ is described in Appendix~\ref{Appendix:reals}. We henceforth identify the two, and use both names interchangeably, since $\Sigma^\circledast$ refers to the encoding, and $\RRrr$ says what is encoded. 

\paragraph{Ordering.} The usual ordering of the reals in $\RRrr$ corresponds to the lexicographic ordering of $\Sigma^\circledast$. When the finite strings are padded by 0s, the symbol ordering is $-\lt 0 \lt +$.

\subsubsection{Numbers extended in time: Conway's version of Dedekind cuts}
\begin{theorem}\label{thm}
There are functors
\bea\label{eq:retraction}
\RRrr
&
\begin{tikzar}{}
\hspace{.1ex} \ar[bend right=25,tail]{rr}[swap]{\Gamma} \&\& \hspace{.1ex}  \ar[bend right=25,two heads]{ll}[swap]{\Upsilon}
\end{tikzar}
& \RRbf
\eea
which make the extended continuum $\RRrr$ into the posetal collapse of the category $\RRbf$ of asynchronous hypergames. In particular,
\begin{itemize}
\item for every real number $\varsigma\in \RRrr$ holds $\Upsilon\Gamma (\varsigma) = \varsigma$;
\item for every asynchronous hypergame $s\in \RRbf$ there are natural hyperstrategies 
\[ s\tto \eta \Gamma\Upsilon(s)\qquad \qquad \mbox{ and }\qquad \qquad \Gamma\Upsilon(s)\tto \varepsilon s\]
\end{itemize}
\end{theorem}

\bprf{ (sketch) } The functor $\Gamma$ can be obtained from the anamorphism $\ana\kappa$
\[\begin{tikzar}[row sep = 6ex]
\RRrr \ar{r}{\kappa} \ar{d}[description]{\ana\kappa}\& \RRrr\  +\  \PPP_\pp \RRrr\  \times\  \PPP_\oo\RRrr\ar{d}[description]{\ana\kappa + \PPP_\pp \ana\kappa+ \PPP_\oo \ana\kappa}\\
\QV \ar{r}[swap]{\nnii} \& \QV\  +\  \PPP_\pp \QV\  \times\  \PPP_\oo\QV
\end{tikzar}
\]
where $\kappa$ is derived from \eqref{eq:circledastt}, by mapping the empty string to the empty string, the $\Sigma$-strings in the form $\left(- ::\varsigma\right)$ to the pair $\big<\{\varsigma\}, \emptyset \big>$, and the strings in the form $\left(+ ::\varsigma\right)$ to $\big<\emptyset,\{\varsigma\} \big>$. Setting $\Gamma \varsigma = \ana\kappa\varsigma$, the functoriality of $\Gamma$ boils down to the observation that the lexicographic order $\varsigma \leq \vartheta$ on $\Sigma^\circledast$ lifts to a relation $s\leq t$ on $s = \Gamma\varsigma$ and $t=\Gamma\vartheta$ which satisfies \eqref{eq:bisimeq-aGam}, i.e.
\bea
s\leq t  & \iff & \forall s_\pp \iinn s\  \left( \exists t_\pp\iinn t. \ \ s_\pp\leq  t_\pp \  \vee\   \exists s_{\pp\oo}\iinn s_\pp.\ s_{\pp\oo}\leq t\right)\ \  \wedge \notag
 \\
& & \forall t_\oo\iinn t\  \left( \exists s_\oo \iinn s. \ \ s_\oo\leq t_\oo\ \vee\ \exists t_{\oo\pp}\iinn t_\oo.\ s\leq t_{\oo\pp}\right) \label{eq:bisimeq-aGamm}
\eea
As long as $\varsigma$ and $\vartheta$ are unpadded by 0s, their lexicographic ordering leads to  $s=\Gamma\varsigma$ and $t = \Gamma\vartheta$ satisfying the synchronous comparison clauses $s_\pp\leq t_\pp$ and $s_\oo\leq t_\oo$ of \eqref{eq:bisimeq-aGamm}. If $\vartheta$ is padded by 0s, then \eqref{eq:bisimeq-aGamm} is satisfied because the lexicographic ordering induces $s_{\pp\oo}\leq t$. If $\varsigma$ is padded by 0s, then it induces $s\leq t_{\oo\pp}$. This completes the definition of $\Gamma$.

The functor $\Upsilon$ arises from Conway's \emph{simplicity theorem} \cite[Thm. 11]{ONAG}. It picks the simplest representatives of the equivalence classes of the posetal collapse of $\RRbf$, where the simplicity is measured in \cite{ONAG} by the "birthday ordinal", which for our finite hypersets, signed or not, boils down each element's position on its coinduction tower. The simplicity theorem plays a central role in all presentations of surreal numbers, and suitable versions have been proved in detail in \cite{AllingN:surreal,GonshorH:surreal}. The arrow part of $\Upsilon$ collapses the $\RRbf$-morphisms to the lexicographic order on $\Sigma^\circledast$. Conway shortcuts his proof of the simplicity theorem by imposing the posetal collapse directly signed hypersets by
\bea \label{eq:Conword}
s\leq t & \iff & \forall s_\pp\in s\ \forall t_\oo\in t.\ \ \ t\not \leq s_\pp\ \ \wedge\ \ \ t_\oo \not \leq s\eea
Instantiating this definition to $t\leq s_\pp$ and to $t_\oo \leq s$ \eqref{eq:Conword} gives 
\bear
t \not \leq s_\pp  & \iff & \exists t_\pp \iinn t.\  s_\pp \leq t_\pp \  \ \ \vee\ \  \exists s_{\pp\oo}\iinn s_\pp.\  s_{\pp\oo} \leq t\\
t_\oo \not \leq s  & \iff & \exists t_{\oo\pp} \iinn t_\oo.\  s \leq t_{\oo\pp} \  \ \ \vee\ \  \exists s_{\oo}\iinn s.\  s_{\oo} \leq t_\oo
\eear
and shows that \eqref{eq:Conword} implies  \eqref{eq:bisimeq-aGamm}. The converse, spelled out along the lines of the proofs of the simplicity theorem that can be found in \cite{AllingN:surreal,GonshorH:surreal}, involves extensive but routinely case reasoning. The equivalence classes of the posetal quotient of $\RRbf$ are thus ordered by \eqref{eq:Conword}, which on $\Sigma^\circledast$ boils down to the lexicographic order.
\epr

\paragraph{Remarks.} Conway's proof of the simplicity theorem demonstrates coinduction in action, not only at the formal level in \eqref{eq:Conword}, but also at the meta-level. In order to define the $\RRrr$-ordering of the minimal representatives of the equivalence classes of his games, reduced to numbers, he imposes the sought ordering as a preorder on arbitrary representatives and then uses that preorder to prove the existence of the minimal representatives. Lemma~\ref{Lemma:order} also shows how the simplicity follows from the coinductive construction, as it implies $\Upsilon(s_\pp)\leq \Upsilon(s) \leq \Upsilon(s_\oo)$, and steers the coinductive descent towards the simplest representative.

\subsubsection{Real numbers as processes}
Thm.~\ref{thm} says that the real numbers can be viewed as processes; and the other way around, that the asynchronous, polarized, reflexive processes boil down to real numbers. The heart of the theorem is in the "boil down" part of the second statement. Its precise meaning is that the simulations between the asynchronous, polarized, reflexive processes implement (and are thus consistent with) the real number ordering. If these processes are thought of as the processes of observing, then the reals are the outcomes of the measurements. On the other hand, computing with the reals involves some embedding into a universe where each number is the outcome of many processes. This is a consequence of the observation, going back to Brouwer \cite{BrouwerLEJ:zahl}, that the irredundant representations of the reals, where each real number corresponds to a unique stream of digits, there are always basic arithmetical operations, and easily defined inputs, where no finite prefix suffices to determine a finite prefix of the output. Such operations are obviously not computable.

Dropping the infinite strings $-\infty = \sseq{ \pp \pp \pp\cdots}$ and $\infty = \sseq{ \oo\oo\oo\cdots}$ on the left-hand side of the retraction $\RRrr \retra \RRbf$ in  \eqref{eq:retraction}, and the signed hypersets bisimilar to $-\infty  = \{-\infty |\}$ and $\infty  = \{| \infty \}$ on the right-hand side, we get the retraction 
$\RRr \retra \RRb$. It lifts to $\RRr^n \retra \RRb^n$, i.e. it makes the real vector spaces into retracts of the discrete functor categories. A real matrix $L \in \RRr^{p\times q}$  becomes an $\RRb$-profunctor $\Lambda = \left(p\rrel{\Gamma L} q\right)$, and the linear operators $\RRr^p\tto{L}\RRr^q$ and $\RRr^q\tto{L^\ddag} \RRr^p$ become the $\RRb$-extensions of $\Lambda = \Gamma L$ along the Yoneda embeddings, in the enriched-category sense.\footnote{The reader unfamiliar with what any of this means is welcome to skip the next paragraph paragraph.}
\[\begin{tikzar}[row sep = 4em,column sep = 4em]
p \ar[leftrightarrow]{d}[description]{{\displaystyle \Lambda}} \ar{r}{\nabla} \&\RRb^p \ar[bend right = 20]{d}[swap]{\Lambda^\ast}  \ar[phantom]{d}[description]{\dashv}\\
q \ar{r}[swap]{\Delta} \&\RRb^q \ar[bend right = 20]{u}[swap]{\Lambda_\ast}
\end{tikzar}
\]
The left Kan extension $\Lambda^\ast$ maps the functor $\alpha \in \RRb^p$ into the coend, which is the colimit  along $\alpha$ of its tensors with the left transpose of $\Lambda$. The right Kan extension $\Lambda_\ast$ maps the functor $\beta \in \RRb^q$ into the end, which is the limit along $\beta$ of its cotensors with the right transpose of $\Lambda$. But since $\alpha$ and $\beta$ are discrete, the colimits boil down to coproducts, and the limits boil down to products. And since $\RRb$ is self-dual, the products and the coproducts coincide as the biproducts, which we write $\oplus$; and the tensors and the cotensors also coincide as $\otimes$. The Kan extensions thus become
\beq\label{eq:kan}
\Lambda^\ast(\alpha) = \left(\bigoplus_{i=1}^p \alpha_i\otimes \Lambda_{ij}\right)_{j=1}^q\qquad\qquad \qquad \Lambda_\ast(\beta) = \left(\bigoplus_{j=1}^q \Lambda_{ij}\otimes \beta_j\right)_{i=1}^p\eeq 
where 
\bear
s\pplus t & = & \Big\{\ s_\pp\pplus t,\  s\pplus t_\pp\ \ \big|\ \ s_\oo\pplus t,\  s\pplus t_\oo\ \Big\}\\
s\otimes t & = &  \Big\{\ (s_\pp\otimes t)\pplus (s\otimes t_\oo) \ominus (s_\pp\otimes t_\oo),\  (s_\oo\otimes t)\pplus (s\otimes t_\pp) \ominus (s_\oo\otimes t_\pp)\ \ \big|\\
&&\ \ \  (s_\pp\otimes t)\pplus (s\otimes t_\pp) \ominus (s_\pp\otimes t_\pp),\  (s_\oo\otimes t)\pplus (s\otimes t_\oo) \ominus (s_\oo\otimes t_\oo)\ \Big\}
\eear
correspond respectively to Conway's addition and multiplication operations \cite{ONAG}. Formally, this correspondence means that the retraction $\Upsilon$ satisfies 
\[ \Upsilon(s\oplus t)  \ = \  \Upsilon s + \Upsilon t\qquad\qquad\qquad\qquad \qquad\Upsilon(s\otimes t)\   =\   \Upsilon s \cdot \Upsilon t\]
The usual matrix operations are thus "rediscovered" as the $\Upsilon$-image of the Kan extensions in \eqref{eq:kan} of $\RRb$-profunctors (corresponding to the real matrices) along the Yoneda embeddings of the bases into their $\RRb$-completions (corresponding to the real vector spaces). 

\subsection{Where is computation?}
As exciting as it is to see the real numbers arising from the categorical structure of processes, it also suggests that we lost the computation from sight somewhere along the way, while retracing the paths of the categorical semantics of computation. The process universe $\RRb$ contains a representative $\Gamma \varsigma$ of every real number $\varsigma$ from the field $\RRr$. Whatever can be computed on such process representatives of the reals in $\RRb$ can be projected back into $\RRr$ along $\Upsilon$. Any real number can $\varsigma$ can be obtained in that way, since $\Upsilon\Gamma \varsigma = \varsigma$. But most real numbers are not computable. Arbitrarily long prefixes of uncomputable reals can be defined by enumerating all computations and avoiding all computable reals by a diagonal argument. This idea has been refined in many directions, showing that almost all real numbers are uncomputable, whichever way we quantify them \cite{ChaitinG:1969,GaczP:reducible,Martin-LoefP:random}. And they all live in $\RRb$. Everything that any oracle can tell any computer is already there. Somewhere on the path from propositions-as-types, through process-propositions-as-types-extended-in-time, to dynamic interactions, the idea of process-computability-as-programmability got lost, and we got all processes. 

In the final section, we retrace the path back to one of the original questions of categorical semantics: \emph{How can \textbf{intensional} computation be characterized semantically?}


\section{Categorical semantics as a programming language}
\label{Sec:intensional}

\subsection{Computability-as-programmability}
A process is computable if it is programmable.\footnote{Network processes are sometimes also called computations, although they are not globally controllable, and thus not programmable. They can be steered by interacting programs and protocols, but that is a different story. The notion of computability was originally defined as computability by computers, and the term is still used in that sense.} In a universe of processes, types are used to specify requirements and to impose constraints. In a universe of \emph{computable}\/ processes, there is also a  type $\Prog$ of \emph{programs}. Since any Turing-complete language can encode its own interpreter, any model of a Turing-complete language must contain\footnote{The tacit assumption is that a model of a programming language contains all types recognizable in that language.} the type $\Prog$ of programs in that language.

A model of computable processes is \emph{extensional}\/ if it only describes the extensions of  computations, i.e. their input-output functions, and does not say anything about the process of computation. Each computable function is thus assigned a unique "program". Type-theoretically, this unique "program"  is captured by the  (cartesian) abstraction operation, which fold a function ${f_x(a)}:A\times X \to B$ with parameters from $X$ to the $X$-indexed family of abstract functions ${\lambda a. f_x(a)}:X\to\To A B$. The application operation applies an abstraction to its inputs and recovers the corresponding function. The bijection between the abstractions and their applications was displayed in \eqref{eq:static-then} and formalized in Def.~\ref{def:closed} using the structure of \emph{cartesian closed}\/ categories. If "programs" do not specify some input-output mappings, but also how they change during computation, then the $X$-indexing becomes a state dependency, and the computations are presented as state machines $\xi:A\times X\to B\times X$, producing the outputs by $ \out \xi:A\times X\to B$ and updating the states by $\sta \xi:A\times X\to X$. The bijection between the parametrized functions and their abstractions \eqref{eq:static-then} becomes a mapping \eqref{eq:dynamic-then} of machines $\xi:A\times X\to B\times X$ to the anamorphisms ${\ana \xi}:X\to \TTTo A B$ assigning to each state in $X$ a dynamic function as the induced computational behavior. This \emph{machine abstraction} was formalized in  Def.~\ref{def:proc-closed}  using the structure of \emph{process closed}\/ categories. The machine abstraction is not injective because many different machines realize the same behaviors; and it is not surjective because some dynamic functions are not implementable by machines. A categorical structure capturing how actual computable functions are specified by actual programs (without the quotation marks) is formalized in Def.~\ref{def:catcom}. It characterizes computable functions using the language of Definitions  \ref{def:closed} and \ref{def:proc-closed}, but not in terms of an abstraction operation, since program abstraction is not an operation.

The conceptual distinction between the static view of the function abstraction in \eqref{eq:static-then}, and the dynamic view of the process abstraction in \eqref{eq:dynamic-then} is echoed to some extent by the technical distinction between the \emph{denotational}\/  and the \emph{operational}\/ semantics of computation \cite{AbramskyS:interaction,AbramskyS:definability,Montanari:book}. Overarching all such distinction is the logical distinction between the \emph{extensional}\/ and the \emph{intensional}\/ models of meaning, going back to Frege, Carnap, Church and Martin-L\"of \cite{FittingM:intensional}. All models of computation that capture abstraction \emph{as an operation} fall squarely on the extensional side. The intuitive reason is that abstraction as an operation readily produces a "program" to each computation; but programming is not such an easy operation. It is a process that involves programmers and evolves other processes. 

In contrast with the denotational models of the $\lambda$-abstraction of functions, and with the operational models of the $\ana{-}$-abstraction of processes, the intensional models of computations are based on the operations for evaluating programs and executing computations. There are many programs for each computation, but there is no operation that transforms computations into programs.

\subsection{Categorical semantics of intensional computation} 
The logical schema of intensional computation is dual to \eqref{eq:dynamic-then}:
\beq\label{eq:intensional-then}
\prooftree
X\llseq{p} \Prog
\justifies
A\wedge X \llseq {\Run p} \may(B \wedge X)
\using{\scriptstyle\Run- \may}
\endprooftree
\qquad\qquad\qquad\qquad
\begin{tikzar}[row sep = 7ex]
 \Set\left(X,\Prog\right)  \ar[two heads]{d}[description]{\displaystyle \Run -}\\
 \Set_M(A\times X,B\times X)
\end{tikzar}
\eeq
The idea of computability-as-programmability is expressed by the requirement that the maps $\Run-$ are surjective: for any computation $A\times X\tto g M(B\times X)$ there is a program  $\rho$ such that $\Run{\rho} = g$. Computations are presented as state machines to help capturing the dynamics of computation. Prop.~\ref{Prop:MonCom} shows that this view of computation is equivalent to the standard view in terms of acceptable enumerations.

The naturality of the program executions $\Run -$ in \eqref{eq:intensional-then} can be described, \emph{mutatis mutandis}, in a similar way like the naturality of $\ana -$ in \eqref{eq:dynamic-then}. An $X$-indexed family of functions $\Run -^{AB} _X:\Set(X,\Prog) \to  \Set_M(A\times X , B\times X)$ constitutes a natural transformation $\Run -^{AB} : \nabla_\Prog \to \Theta_{AB}$ between the functors  
\begin{align}\label{eq:widehat-theta}
\nabla_ \Prog :\Set^o & \to \Rel & \Theta_{AB}: \Set^o &\to  \Rel\\
X & \mapsto \Set(X,\Prog) & X & \mapsto \Set_M(A\times X ,B\times X)\notag
\end{align}
See \eqref{eq:coalg-homom} and \eqref{eq:widehat} in Sec.~\ref{Sec:proc-implication} for the arrow parts of these functors. The naturality requirement is dual to \eqref{eq:dcc-nat}. It implies that the diagram here on the left  commutes for every $p\in \Set(X,\Prog)$.
\beq\label{eq:catcom}
\begin{tikzar}[row sep = 1.5ex,column sep = 1.5ex]
\Set\left(\Prog, \Prog\right) \ar{ddddd}[description]{\Run-^{AB}_\Prog } \ar{rrrrr}[description]{\left(- \circ p\right)}
\&\&\&\&\& \Set\left(X, \Prog\right) \ar{ddddd}[description]{\Run-^{AB}_X}
\\ 
\\
\\
\\
\\
\Set_M\left(A\times \Prog, B\times\Prog \right)\ar[leftrightarrow]{rrrrr}[description]{(p)} 
\&\&\&\&\& \Set_M\left(A\times X,B\times X\right)
\end{tikzar}
\qquad\quad
\begin{tikzar}[column sep = 5ex, row sep = 7.5ex]
A \times X \ar{r}{\Run{\rho}}\ar{d}[description]{A\times \rho} \&M(B\times X)\ar{d}[description]{M(B\times \rho)}\\
A\times \Prog \ar{r}[swap]{\Run{\id}} \& M(B\times \Prog)
\end{tikzar}
\eeq
The diagram on the right arises by chasing $\id \in \Set(\Prog,\Prog)$ around the diagram on the left. The left-hand diagram says that $\Run{\id}^{AB}_\Prog$ and $\Run{p}^{AB}_X$ are related under $\Theta_{AB}p$, which by the definition in \eqref{eq:coalg-homom} means that the right-hand square commutes. Since the naturality implies that  
\[\Run{p f}_Y\  =\  \Run{p}_X (A\times f) \ =\  \Run{\id}_\Prog (A\times p f)
\] 
holds  for all $f\in \Set(Y,X)$ and $p\in \Set(X,\Prog)$, droping the subcripts $X$ from $\Run -_X$ seldom causes confusion. The other way around, by the surjectivity of $\Run-$, for every computation $g\in \Set_M(A\times X, B\times X)$ there is an $X$-indexed program $\rho \in \Set(X,\Prog)$ such that $\Run{\rho}^{AB}_X = g$, making the right-hand square in \eqref{eq:catcom} commute. Since this is true for all $A$ and $B$, the claim is thus that $\Prog$ is the state space of a weakly\footnote{The word "weakly" refers to the fact that the programs $\rho_g$ are not unique: each machine $g$ can be represented by many of them; in fact infinitely many.} final $AB$-machine $\Run{\id}^{AB}_\Prog \in \Set_M(A\times \Prog, B\times \Prog)$ --- for \emph{all types}\/ $A$ and $B$ in $\Set$. The categories of computable-as-programmable functions, induced by \eqref{eq:catcom}, are thus process-closed in a suitable intensional sense that is both weaker and stronger than the extensional process-closed structure \eqref{eq:dynamic-then}. It is weaker in the sense that the abstractions are not unique, but it is stronger in the sense that all abstractions, over all types, are of the same type $\Prog$. They are the programs. Hence the intensional cousin of the cartesian-closed and the process-closed categories defined in \ref{def:closed} and \ref{def:proc-closed}:

\begin{definition}\label{def:catcom} A \emph{categorical computer} is a cartesian category $\Set$ with a commutative monad  $M:\Set\to \Set$, a fixed type $\Prog$ of \emph{programs}, and for any pair of types $A,B$ an $X$-natural family of surjections, called \emph{program executions}:
\bea\label{eq:exec}
\Set(X,\Prog) &\begin{tikzar}{}
\hspace{.1em} \ar[two heads]{r}{\Run{-}^{AB}_X} \&\hspace{.1em}
\end{tikzar} & \Set_M(A\times X, B\times X)
\eea
The naturality of the program executions $\Run{-}^{AB}$ is with respect to the functors 
$\nabla_\Prog,  \Theta_{AB} :\Set^o  \to \Rel$ from   \eqref{eq:widehat-theta}.
\end{definition}

\begin{proposition}\label{Prop:MonCom} Let $\Set$ be a cartesian category, $\Prog\in \Set$ a fixed type, and $M:\Set\to \Set$ a commutative monad. Specifying the the program executions $\Run -$ in \eqref{eq:exec}, and establishing $\Set$ as a categorical computer,  is equivalent to specifying the following data for all types $A,B, X$:
\begin{enumerate}[a)]
\item a \emph{universal evaluator}\/ (or \emph{interpreter}) $\varphi^{AB} \in \Set_M(A\times \Prog ,B)$ and 
\item a \emph{partial evaluator}\/ (or \emph{specializer}) $\sigma^{X} \in \Set(X\times \Prog, \Prog)$
\end{enumerate}
such that for any $f\in \Set_M(A,B)$ there is $p\in \Set(1,\Prog)$ with
\beq\label{eq:eqs}\begin{split} f \ = \ \varphi^{AB}\circ\Big(A\times p\Big)\ \ \ \ \\
 \varphi^{(AX)B} \ = \ \varphi^{AB}\circ\left(A \times \sigma^{X}\right)
 \end{split}
 \qquad \qquad \qquad
 \begin{tikzar}[row sep = large,column sep = 10ex]
A\ar{r}{f} \ar{d}[swap]{A\times p} \& B\\
A\times \Prog \ar{ur}[description]{\varphi^{AB}} \& A\times X\times \Prog \ar{u}[description]{\varphi^{(AX)B}} \ar{l}[description]{A\times \sigma^{X}}
\end{tikzar}
\eeq
\end{proposition}

\bprf{ (sketch)} Given a categorical computer, the interpreters are $\varphi^{AB} = \pi_B\circ \Run{\id}^{AB}_\Prog$ and the specializers $\sigma^X$ are chosen using the surjectivity of $\Run{-}^{AB}_{(X\Prog)}$. Showing that the same $\sigma^X$ can be chosen for all $A$ and $B$ is the only part which requires work\footnote{Most computability theory goes through with non-uniform specializers, which may vary with the context $A,B$.}. Towards he converse, setting $\Run{p}^{AB}_X = \varphi^{A(BX)}\circ(A\times p)$ defines a natural transformation. To show that its components are surjective, for an arbitrary computation $A\times X\tto g M(B\times X)$, set $\rho(x) =  \sigma^{X\Prog}(x,r,r)$ using in the following diagram.
\beq\label{eq:kleenorm}\begin{tikzar}[row sep = 4em,column sep = 4em]
A\times X \ar[dashed,bend right=55]{ddd}[description]{A\times \rho} \ar{rr}{g} \ar{d}{A\times X \times r} \&\& M(B\times X) \ar[dashed,bend left=55]{ddd}[description]{B\times \rho} \ar{d}[swap]{M(B\times X \times r)}\\
A\times X\times \Prog \ar{r}{g\times \eta} \ar[dashed]{d}{A\times X \times \Prog \times  r} \& M(B\times X)\times M\Prog\ar{r}{\phi}\& M(B\times X\times \Prog) \ar{d}[swap]{M(B\times X\times \Delta)}\\
A\times X\times \Prog\times \Prog \ar{rrd}{\varphi^{(AX\Prog)(B\Prog)}} \ar{d}{A\times \sigma^{X\Prog}} \&\& M(B\times X\times \Prog\times \Prog) \ar{d}[swap]{M(B\times \sigma^{X\Prog})}\\
A\times \Prog \ar{rr}[swap]{=\ \Run{\id}_\Prog^{AB}}{\varphi^{A(B\Prog)}} \&\& M(B\times \Prog) 
\end{tikzar}
\eeq
The program $r$ is defined by the commutative trapezoid in the middle. It encodes the computation where the state output $A\times X\tto{\sta g} MX$ is fed into the function $X\times \Prog\tto{X\times \Delta} X\times \Prog\times \Prog \tto{\sigma^{XP}}\Prog$ where $\sigma^{X\Prog}$ partially evaluates any program on itself. This computation is the composite of the arrows going from $A\times X\times \Prog$ right along the top and down along the right side of the trapezoid. Some programs $r$ that make the trapezoid commute when substituted as the left dashed side exist by Prop.~\ref{Prop:MonCom}(a). The top rectangle is obtained by feeding some such $r$ as the input to to the partial evaluator, to evaluate it on itself. The triangle at the bottom commutes by Prop.~\ref{Prop:MonCom}(b). The commutativity of the whole diagram gives $\Run\rho^{AB}_X = g$.
\epr

\paragraph{Historic background.} Prop.~\ref{Prop:MonCom} says that the structure of categorical computer is a categorical version of the standard concept of \emph{acceptable enumeration} \cite{Hartley:book}.  In the standard  notation, the enumeration would be a sequence $\sseq{\varphi^n_x}^{n\in\NNn}_{x\in \Prog}$, where $x$ is the program index, and $n$ is the arity of the computable function $\varphi_x$. While the computable functions are usually modeled over natural numbers, and the arity $n$ means that the function takes the inputs of type $\NNn^n$, and always produces a single output of type $\NNn$, the categorical treatment is over abstract types, so we write $\varphi^{AB}$ to specify the input type $A$ and the output type $B$. 

\paragraph{Programming background.} The construction in the proof of Prop.~\ref{Prop:MonCom} is easily seen to be a version of Kleene's construction of the fixpoint in his Second Recursion Theorem \cite[Ch.~11]{Hartley:book}. The partial evaluator evaluating all programs on themselves plays the central role. This capability of self-evaluation lies at the heart of many computational constructions \cite{MoschovakisY:Kleene}. While the diagram chase above elides many equations, the string diagrammatic versions do not just abridge the constructions but display the geometric patterns behind many of them.  They support a diagrammatic programming language with convenient implementations of computable logic and arithmetic, program schemas, abstract metaprogramming concepts like compilation, supercompilation, synthesis, and to derive static, dynamic, and algorithmic complexity measures \cite{PavlovicD:MonCom,PavlovicD:MonCom3}. 

The $\lambda$-calculus and the underlying type theories have been used as abstract programming languages in the semantics of computation from the outset \cite{ScottD:ISWIM}, and remained at the heart of the semantical investigations \cite{AbramskyS:PCF,Hyland-Ong:PCF}. Programming in abstract programming languages has also been pursued since early on \cite{PlotkinG:LCF}. It led to functional programming, which now permeates programming practices beyond the realm of. However, the mere presence of the abstraction operations makes the underlying type systems essentially extensional. Dropping the extensional $\lambda$-conversions allows that multiple programs may correspond to a single computation, but still provides a canonical choice among them, maintains a canonical extensional core of the type system \cite{HayashiS:semifunctors}. This has been the main obstacle to studying genuinely intensional algorithmic phenomena, such as complexity, within  the semantics of computation. 

\subsection{Computability as an intrinsic property}
A poset may be a monoid in many different ways: e.g., the reals are a monoid for addition, for multiplication, and for many other operations. But a poset may be a lattice (an idempotent monoid) in at most one way: the joins are the least upper bounds, the meets are the greatest lower bounds, and if they exist, they are uniquely determined by the order. A category can be monoidal in many different ways, but it can be cartesian in at most one way because the cartesian products are uniquely determined.  The lattice structure of a poset and the cartesian structure of a category are unique, and they are therefore the \emph{properties}\/ of their carriers.  When the meets in a poset have the right adjoints, the implications that arise are also unique,  and the structure of a Heyting algebra in is also a property. For the same reason, the cartesian-closed structure from Def.~\ref{def:closed} is a property of a category, as is the process-closed structure from Def.~\ref{def:proc-closed}. 

The structure of a categorical computer from Def.~Def.~\ref{def:catcom} is also essentially unique and thus a property of the category that carries it. Proving this requires a little more work than the simple arguments above, but not much more. It boils down to a categorical encoding of the theorem that all parametrized interpreters isomorphically interpret one another \cite{Hartley:book}. A categorical computer thus displays computability as a categorical property: that all of its morphisms are programmable functions. 

On the other hand, it was explained in \cite[Sec.~1.2.3]{AbramskyS:definability} that the notion of computability, \emph{as defined in the standard Church-Turing approach}, is \emph{extrinsic}, in the sense that a particular computable function is recognized as such only by referring to a particular external model of computation, say a Turing machine or a definitional schema. The invoked model then describes a particular process of computing the function, which is not recorded or recognizable on the function itself. It was thus argued in \cite{AbramskyS:definability} that the standard definitions do not specify computability as an intrinsic structure, even less a property of a function. In contrast, \eqref{eq:intensional-then} expresses the idea of \emph{com\-pu\-ta\-bi\-li\-ty-as-pro\-gram\-ma\-bi\-li\-ty} as a logical structure; and by the virtue of uniqueness of that structure, as a logical property. Whatever programming language $\Prog$ might be used to encode programs, they are always assigned semantics along some program executions  $\Set(X,\Prog)\epi \Set_M(A\times X, B\times X)$, or along some equivalent mappings. The Rogers' isomorphism theorem says that all programming languages are isomorphic along semantics-preserving computable functions. Whichever Church-Turing model of computation might be used to define computability, the underlying execution model will map its process descriptions to the corresponding computational processes, and this mapping will make it into a categorical computer. This structure provides a \emph{"canonical form witnessing computability"}, sought in \cite[Sec.~1.2.3]{AbramskyS:definability}. 

Many languages of logic claim universality and establish their universality on their own terms. The set theory proves that it is the foundation of all mathematics, first-order logic is the language of predicates, category theory is the language of structures. The statement that logic is tasked with discovering the universal laws of logic is a tautology, in a logic of logic. A universal law should not be misunderstood as the last word about anything, but as the first word about something else. The idea that \emph{computability-as-programmability}\/ is a model-invariant, syntax-independent, device-free concept, and a property intrinsic to all computable objects and processes, is broader than any particular structure, categorical or otherwise, in which it may be expressed. The idea  of computability-as-programmability lurks behind  Kolmogorov's invariance theorem \cite[Sec.~2.1]{Vitanyi:book}. While recognizing a particular function as computable depends on encodings in a particular model, the invariance theorem is built upon the fact that the encodings and their transformations are programmable, and that the programs are of constant lengths. Kolmogorov's invariance theorem can be construed as a quantitative counterpart of Rogers' isomorphism theorem \cite[Thm.~2.4.14]{CaludeC:complx}. Both theorems characterize computability as an intrinsic property.  Computability-as-programmability is not just testable by any of the equivalent models of computation, as claimed by the Church-Turing thesis, but it is also \emph{quantifiable}, in Kolmogorov's formulation by the length of programs. Kolmogorov's \emph{algorithmic complexity}\/ is thus the quantitative view of the intrinsic property of computability-as-programmability. By displaying programmability as a structure, categorical semantics provides the qualitative view of this property.

It should be noted that the qualitative and the quantitative views of computability as an intrinsic property of processes come about in disguise in many arenas of science.  Although the search for a program that makes a process computable is generally not a computable process, its average algorithmic complexity is an intrinsic quantity again: the Shannon entropy \cite{MuchnikA:shannon-kolmo,ZurekW:shannon-kolmo}. Information theory as the theory of information processing has been viewed as a theory of computation in microsystems, averaged out in thermodynamics. Domain theory has been viewed as a theory of computability-as-approximation in suitable topologies \cite[Sec.~5.1]{AbramskyS:APAL91,SmythM:topology}. A natural task for categorical semantics is to bring such conceptual threads together. That is the message that I got from Samson Abramsky's categories that no one had seen before.

\section{Summary}
\label{Sec:conclusion}

In the propositions-as-types view, the extensional operations of  abstraction and application, \emph{viz}\/ the structure of cartesian closed categories, correspond to the introduction and the elimination of the propositional implication:
\[
\prooftree
(A\wedge X) \vdash B
\Justifies
X \vdash (A\thenn B)
\using{\thenn}
\endprooftree
\qquad
\qquad\qquad\qquad
\begin{tikzar}[row sep = 5ex]
\Set(A\times X,B) \ar[bend right]{d}[swap]{\To A - \circ \eta_X} \\
\Set\big(X,\To A B\big) \ar[bend right]{u}[swap]{\varepsilon_X\circ(A\times-)}
\end{tikzar}
\]
In process logics, the process implication introduction rule corresponds to the coinductive interpretation of arbitrary states as process behaviors, captured in the final machine:
\[
\prooftree
A\wedge X \llseq \varphi \may(B \wedge X)
\justifies
X\llseq{\ana\varphi} \TTTo A B_\may
\using{\scriptstyle\ana-}
\endprooftree
\qquad\qquad\qquad\qquad
\begin{tikzar}[row sep = 7ex]
 \Set_M(A\times X,B\times X) \ar{d}[description]{\ana-_X}\\
 \Set\left(X,\TTTo A B_M\right)
\end{tikzar}
\]
In terms of dynamic types, computation corresponds to program execution. In terms of process propositions, computability-as-programmability is thus an elimination rule, mapping programs, as intensional proofs of the universal proposition, the programming language, into computations as their extensions:
\[\prooftree
X\llseq{p} \PPp
\justifies
A\wedge X \llseq {\uev p} \may(B \wedge X)
\using{\scriptstyle\uev-}
\endprooftree
\qquad\qquad\qquad\qquad
\begin{tikzar}[row sep = 7ex]
 \Set\left(X,\PPp\right)  \ar[two heads]{d}[description]{\uev -}\\
 \Set_M(A\times X,B\times X)
\end{tikzar}\]
Categorical semantics provides convenient and sometimes effective tools for reasoning about types and processes. Samson Abramsky led many of us through its vast landscape. I followed him to the best of my ability. The present paper is an attempt at a travel report. But the territory is largely uncharted, and there were times when I lost sight of Samson, probably somewhere far ahead. It is thus likely that the travel report is not just about what I learned from Samson, but also about what I misunderstood by getting lost, and maybe most of all about what I did not learn at all. Categorical semantics of computational processes is a computational process itself, and it is the nature of such processes that they may terminate, or not.


%

\bibliographystyle{plain}
\bibliography{specgames-ref,PavlovicD,semantics,AbramskyS,IT,logic,CT}

\appendix
\section*{Appendices}
\addcontentsline{toc}{section}{Appendices}

\section{Category $\Rel$ of sets and relations}\label{Appendix:Rel}
Relations $A\ot R\to B$ arise in two ways:
\begin{enumerate}[a)]
\item as subsets $R\mmono{r} A\times B$, so that
\bear
aRb & \iff & \exists x\in X.\ a = r_A(x) \wedge r_B(x) = b
\eear
\item as a nondeterministic functions $A\tto \varrho \WP B$ and $B\tto{\varrho^o} \WP A$, so that
\[
aRb\ \  \iff \ \  \varrho(a)\ni b\ \ \iff\ \ a\in \varrho^o(b)
\]
where $\WP:\Set\to\Set$ is the powerset monad. 
\end{enumerate}
The equivalence between the two views lies at the heart of the elementary structure of topos \cite{BarrM:ttt,Freyd-Scedrov:book,Lambek-Scott:book}, which can be defined in terms of the correspondece between the subsets $R\mono A\times B$ and the elements $\chi_R\in \WP(A\times B)$, and the natural bijections
\beq\label{eq:pow}
\Set(X\times A,\WP B)\  \cong\  \Set(X,\WP(A\times B))\  \cong\  \Set(X\times B,\WP A)
\eeq
A relational calculus can, however, be developed entirely in terms of subobjects $R\mono A\times B$, in type universes without the powerset monad. Process relations are presented from this angle. The universe $\Set$ only needs to be \emph{regular} \cite{BarrM:regular-book,PavlovicD:mapsI}. In addition to the cartesian structure, it is thus also assumed to have the equalizers (i.e., the subsets characterized by equations), which induce the pullback squares. The final assumption, crucial for the relational calculus, is that every function $f:A\to B$ has an epi-mono (surjective-injective) factorization: it can be decomposed in the form $f = \left(A\eepi{e_f} A' \mmono{m_f} B\right)$, where $e_f\in \EEE$ and $m_f\in \MMM$. The family $\EEE$ can be thought of as the quotient maps (coequalizers), whereas $\MMM$ are all monics. The family $\EEE$ is required to be stable under the pullbacks. The category of relations in $\Set$ is then defined to be
\bea\label{eq:Rel}
|\Rel | & = & |\Set|\\
\Rel(A,B) & = & \MMM_\cong\big/(A\times B)\notag
\eea 
where $\MMM_\cong$ is the set of the equivalence classes modulo the relation
\bear
m\cong m' & \iff & \begin{tikzar}[row sep = large,column sep = 1.5ex]
R\ar[bend left=20,tail]{rr} \ar[phantom]{rr}[description]{\cong} \ar[tail]{dr}[description]{m}\&\& R' \ar[tail]{dl}[description]{m'}\ar[bend left=20,tail]{ll}\\
\& X
\end{tikzar}
\eear
Without this quotienting, $\Rel(A,B)$ would in general be a proper class. The composition of relations $A\rrel R B$ and $B\rrel S C$, viewed as the $\MMM$-monics $R\mono A\times B$ and $S\mono B\times C$, is defined using the pullback $R\underset{B}{\times} S$ and the factorization in the following diagram. 
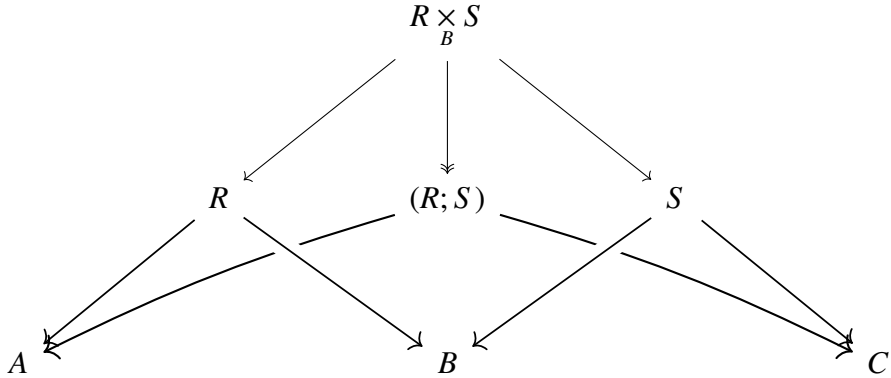
\[
\begin{tikzar}{}
\&\&\&\& R\underset{B}{\times} S \ar[thin]{ddll}\ar[thin]{ddrr} \ar[two heads,thin]{dd}
\\
\\
\&\& R \ar{ddll} \&\&(R;S)\ar[bend right=5,thick]{ddllll} \ar[bend left=5,thick]{ddrrrr}\&\& S\ar{ddrr}
\\
\\
A \&\&\&\& \ar[leftarrow,crossing over]{uull} \ar[leftarrow,crossing over]{uurr} B\&\&\&\& C
\end{tikzar}
\]
The identity $A\rel A$ in $\Rel_\Set$ is the diagonal $A\to A\times A$ in $\Set$. More general categories of relations can be defined in more general situations using technically different but conceptually similar constructions \cite{PavlovicD:mapsI,PavlovicD:mapsII}. If $\Set$ has the coproducts $+$, they become biproducts in $\Rel$. The products $\times$ from $\Set$ induce a canonical monoidal structure in $\Rel$, with the compact structure $\eta:1\leftrightarrow A \leftrightarrow A\times A$ and $\varepsilon : A\times A\leftrightarrow A\leftrightarrow 1$ on every $A$ \cite{Kelly-Laplaza}. 

\section{Proof of  Prop.~\ref{ccc-ana}}\label{Appendix:ccc-ana} 
\textbf{a)} Suppose that $\Set$ is a cartesian closed category with the static implication $\xxp A B$, and with the process of $A$-histories $A\tto{(-)} A^+ \oot{(::)} A\times A^+$ for every $A$. Then $\TTTo A B = \xxpp A B$ is the state space of the final $AB$-machine with the structure map
\bea\label{eq:finmach}
A\times \xxpp A B & \tto{\upsilon\ =\ <\out  \upsilon, \sta\upsilon>} & B\times \xxpp A B 
\eea
where the components are derived by evaluating along the components of the $A$-history process 
\begin{gather*}\prooftree
A\times \xxpp A B \tto{\big(A\times {(-)}\Rightarrow B\big)} A\times \xxp A B \tto{\ \varepsilon\ } B
\justifies
A\times \xxpp A B \tto{\out\upsilon}  B
\endprooftree
\\[2.5ex] 
\prooftree
A^+  \times A \times  \xxpp A B\ \cong\ A \times A^+ \times \xxpp A B  \tto{(::)\times \xxpp A B} A^+  \times  \xxpp A B \tto{\ \varepsilon\ } B
\justifies
A \times \xxpp A B \tto{\sta \upsilon} \xxpp A B
\endprooftree
\end{gather*}
To show that \eqref{eq:finmach} is a final machine, first note that every $AB$-machine $A\times X \tto{\xi \ =\ <\out \xi,\sta \xi >} B\times X$ induces an $A$-history process 
\beq\label{eq:kappa}
A\tto{\innn\kappa}\xxp X B\oot{\conss\kappa} A\times \xxp X B\eeq 
with the components
\begin{gather*}
\prooftree
A\times X \tto{\out \xi} B
\justifies
A \tto{\innn\kappa}  \xxp X B
\endprooftree
\qquad\qquad
\prooftree
X  \times A \times  \xxp X B\ \cong\ A \times X\times \xxp X B  \tto{\sta \xi \times \xxp X B} X  \times  \xxp X B \tto{\ \varepsilon\ } B
\justifies
A \times \xxp X B \tto{\conss \kappa} \xxp X B
\endprooftree
\end{gather*}
%
%
%
By Sec.~\ref{Sec:history}, the $A$-history process $\kappa$ induces the catamorphism (i.e. fold, banana-function) $\cata \kappa$
\beq\label{eq:cataeq}\begin{tikzar}{}
\& A^+ \ar[dashed]{dd}[description]{\cata\kappa}\& A\times A^+ \ar{l}[description]{(::)} \ar[dashed]{dd}[description]{A\times \cata\kappa} \\ 
A\ar{ru}[description]{(-)}\ar{rd}[description]{\innn \kappa}\\
\& \xxp X B \& A\times \xxp X B \ar{l}[description]{\conss \kappa} 
\end{tikzar}
\eeq
On the other hand, the transposition
\[\prooftree
A^+ \tto{\cata \kappa} \xxp X B
\justifies
X \tto{\ana \xi} \xxpp A B
\endprooftree\]
induces the anamorphism (unfold, lens-function) $\ana \xi$
\beq\label{eq:anaeq}
\begin{tikzar}[row sep=4em,column sep=2em]
X\times A \ar{r}{\xi} \ar[dashed]{d}[swap]{\ana \xi \times A} 
\& X\times B \ar[dashed]{d}{\ana \xi \times B}
\\
\xxpp A B \times A \ar{r}[swap]{\upsilon} 
\& \xxpp A B \times B
\end{tikzar}\eeq
which shows that $\upsilon$ makes $\TTTo A B$ into the process implication as in Sec.~\ref{Sec:proc-implication}. The diagram chase showing that the catamorphism \eqref{eq:cataeq} commutes if and only if \eqref{eq:anaeq} commutes is an instructive exercise.

\paragraph{b)} The assumption is that $\Set$ has final $AB$-machines
\bear
A\times \TTTo A B & \tto{\upsilon\ =\ <\out \upsilon, \sta \upsilon>} & B\times \TTTo A B 
\eear
Replacing the second component by the projection gives the machine which induces the anamorphism $\ana{\out\upsilon,\pi_1}$, which makes the outer square in the following diagram commute.
\beq\label{eq:expdef}
\begin{tikzar}{}
A\times \TTTo A B  \ar{rrr}{<\out\upsilon, \pi_1>} \ar[dashed]{dd}[swap]{A\times \ana{\out\upsilon, \pi_1}} \ar[two heads]{dr}[swap]{A\times q}
 \&\&\& B\times \TTTo A B \ar[two heads]{dl}{B\times q}
\ar[dashed]{dd}{B\times \ana{\out\upsilon, \pi_1}}
\\
\& A\times \xxp A B \ar[dashed]{r}{<\mathbf{\varepsilon},\pi_1>} \ar[tail]{dl}[swap]{A\times m} \& B\times \xxp A B \ar[tail]{dr}{B\times m}\\
A\times \TTTo A B  \ar{rrr}[swap]{<\out\upsilon, \sta\upsilon>} \&\&\& B\times \TTTo A B
\end{tikzar}
\eeq
Since $\ana{\out\upsilon,\pi_1}$ is also endomorphism on the $AB$-machine  $A\times\TTTo A B \tto{<\out\upsilon, \pi_1>}  B\times \TTTo A B$, the uniqueness of $\ana{\out\upsilon,\pi_1}$ as an $AB$-machine homomorphism from $<\out\upsilon, \pi_1>$ to $<\out\upsilon, \sta\upsilon>$ implies that it is an idempotent:  
\bear
\ana{\out\upsilon,\pi_1}\circ \ana{\out\upsilon,\pi_1} & = & \ana{\out\upsilon,\pi_1}
\eear 
Here we use the assumption that the idempotents split in $\Set$, and define $\xxp A B$ as the splitting 
\bear
\ana{\out\upsilon,\pi_1} & = & \left(\TTTo A B\eepi q \xxp A B \mmono m \TTTo A B\right)
\eear 
also displayed in \eqref{eq:expdef}. The component $A\times \xxp A B \tto{\  \varepsilon\ } B$ of the factoring defined there is the counit of the adjunction $A\times (-) \dashv \xxp A -$, defined
\bear
\Set(X\times A, B) & \tto{\ \lambda\  } & \Set(X , \xxp A B)\\
f & \longmapsto & \lambda f = q\circ \ana{f,\pi_1}
\eear
To show that $\varepsilon \circ (\lambda A\times f) \ = \ f$, chase the following diagram:
\[\begin{tikzar}{}
A\times X \ar{rrr}{<f,\pi_1>} \ar{dr}{A\times \ana{f,\pi_1}} \ar{dd}[swap]{A\times \lambda f }\& \&\& B\times X \ar{dl}[swap]{B\times \ana{f,\pi_1}} \ar{dd}{B\times \lambda f}\\
\& A\times \TTTo A B \ar{r}{<\out \upsilon,\pi_1>} \ar[two heads]{dl}{A\times q}\& B\times \TTTo A B \ar[two heads]{dr}[swap]{B\times q}\\
A\times \xxp A B  \ar{rrr}[swap]{<\varepsilon,\pi_1>}\&\&\& B\times \xxp A B \end{tikzar}
\]
\hfill $\Box$

\section{Proof sketch for Lemma~\ref{Lemma:splitting}}\label{Appendix:splitting}
Since $\#A\leq \aleph_0$, there is an ordinal number $\kappa \leq \omega$ large enough to support an retraction $\PPP(A\times A)\mono \PPP^\kappa(A)\epi \PPP(A\times A)$, and thus also $Q_{AA} 1\mono Q^\kappa_{A1} 1\epi Q_{AA} 1$ for the functor $Q$ defined in \eqref{eq:PQ}. Hence the tower of retractions:
\[
\begin{tikzar}[column sep =2.5ex, row sep = 3ex]
1\ar[equals]{dd} \&\&\&\& Q_{AA} 1\ar{llll}[swap]{!}\ar[bend left,tail]{dd}  \&\&\&\& Q_{AA}^2 1\ar[bend left,tail]{dd} \ar{llll}[swap]{Q_{AA}!} \&\&\&\& Q^n_{AA} 1 \ar[dotted]{llll}  \ar[bend left,tail]{dd}  \&\&\&\& \TTTo A A _\PPP \ar[dotted]{llll} \ar[bend left,tail,dashed]{dd}{m_0}
\\[1ex] 
\\
1\&\&\&\&Q_{A1}^\kappa 1 \ar[bend left,two heads]{uu} \ar{llll}[swap]{!} \&\&\&\&Q_{A1}^{\kappa 2} 1 \ar[bend left,two heads]{uu} \ar{llll}[swap]{Q^\kappa_{A1}!}\&\&\&\& Q^{\kappa n}_{A1} 1 \ar[bend left,two heads]{uu} \ar[dotted]{llll} \ar[bend left,two heads,dashed]{uu}  \&\&\&\& \TTTo A 1 _\PPP \ar[dotted]{llll} \ar[bend left,two heads,dashed]{uu}{e_0}
\end{tikzar}
\]
The symmetry $A\times 1 \cong 1\times A$ lifts to a smilar retraction
\[ \TTTo A A _\PPP\mmono{m_1} \TTTo 1 A_\PPP \eepi{e_1} \TTTo A A _\PPP\]
With these retractions, the proof boils down to showing the commutativity of the following diagram
\[
\begin{tikzar}[column sep =3ex, row sep = 9ex]
1
\ar{dd}[description]{\ana\id} \ar{rr}[description]{\ana\id}  \&\& \TTTo A A_\PPP \ar{dd}[description]{<\id,\id>} \ar[tail]{dl}[description]{<m_0, m_1>}  
\\
\& \TTTo A 1_\PPP\times \TTTo 1 A_\PPP \ar{dl}[description]{\ana{-;-}} \ar[two heads]{dr}[description]{<e_0, e_1>} 
\\
\TTTo A A _\PPP \&\& \TTTo A A_\PPP \times \TTTo A A _\PPP \ar{ll}[description]{\ana{-;-}}
\end{tikzar}
\]
where ${\ana{-;-}}$ are the enriched compositions, constructed like in \eqref{eq:fcoalg-comp} (or see \cite{PavlovicD:FOSSACS01} for more details), whereas $\ana{\id}$ is the enriched identity, constructed as the anamorphism (final coalgebra homomorphism) from the identity machine $A \times 1 \tto{\eta} \PPP(A\times 1)$, where $\eta$ is the unit of the monad $\PPP$. This diagram says that $j = m_0\ana{\id} \in \Set^\PPP(A,1)$ and $r = m_1\ana{\id} \in \Set^\PPP(1,A)$ display $A$ as a retract of $1$ in $\Set^\PPP$, i.e. that they compose to
\bea\label{eq:jr}
\id_A & = & \Big(A \tto{j =m_0\ana{\id}} 1 \tto{r = m_1\ana{\id}} A \Big)
\eea
\hfill $\Box$

\section{Proof of Corollary~\ref{Corollary:splitting}}\label{Appendix:corollary}
Since the embedding \eqref{eq:corollary} is full and faithful by definition, we only need to prove that it is essentially surjective: for an arbitrary object $S\in \DProc_{\leq \aleph_0}$ we must find $S'\in \dProc$ such that $S\cong S'$ in $\DProc_{\leq \aleph_0}$. An object of $\DProc_{\leq \aleph_0}$ is a dynamic relation $A\rrel S 1$ in $\Set^\PPP$, where $\#A\leq \aleph_0$. An object of $\dProc$ is a hyperset $S'$, viewed as a dynamic relation $1\rrel{S'} 1$ in $\Set^\PPP$.  By Lemma~\ref{Lemma:splitting}, there are the relations $j  \in \Set^\PPP(A,1)$ and $r \in \Set^\PPP(1,A)$ such that $(j;r) = \id_A$. Setting 
\bear
S' & = & \Big(1\rrel r A \rrel S 1\Big) 
\eear
assures that the inner triangle in the following diagram commutes.
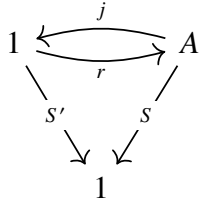
\[
\begin{tikzar}[row sep = 7ex,column sep = 3ex]
1 \ar{dr}[description]{S'} \ar[bend right=15]{rr}[swap]{r} \&\& A\ar{dl}[description]{S} \ar[bend right=15]{ll}[swap]{j}\\
\& 1
\end{tikzar}
\]
The outer triangle commutes because $S' \circ j = S\circ r \circ j = S$ by \eqref{eq:jr}. So we have the morphisms $r\in \DProc_{\leq \aleph_0}(S', S)$ and $j\in \DProc_{\leq \aleph_0}(S, S')$. They form an isomorphism because $r\circ j = \id_S$ by \eqref{eq:jr} again, and $j\circ r \in \DProc_{\leq \aleph_0}(S', S')$ must be an identity because $S'$ is a subobject of the terminal object in $\DProc_{\leq \aleph_0}$. \hfill $\Box$   

\section{Traces and the $\Int$-construction}
\label{Appendix:Int}
The \emph{trace}\/ operation on a symmetric (or braided) monoidal category $(\CCC,\otimes, I)$ is typed by the rule
\[\prooftree
A\otimes Y \tto f B\otimes Y
\justifies
A\tto{\Tr_Y(f)} B
\endprooftree
\]
The equations for this operation, with some examples and explanations can be found in \cite{AbramskyS:trace,JSV,PavlovicD:CMCS12}. The free compact category over any traced monoidal $\CCC$ 
\bea\label{eq:Int}
|\Int_\CCC | & = & |\CCC|_\pp\times  |\CCC|_\oo\\
\Int_\CCC(A,B) & = & \CCC\left(A_\pp \otimes B_\oo, B_\pp\otimes A_\oo\right)\notag
\eea 
where $X_\pp = \{\pp\}\times X$ and $X_\oo = \{\oo\}\times X$. The composition of $\Int_\CCC(A,B)\times \Int_\CCC(B,C) \tto\icomp \CCC(A,C)$ is defined by
\[\prooftree
\prooftree
A_\pp \otimes B_\oo\tto{\ f\ } B_\pp\otimes A_\oo \qquad\qquad \qquad B_\pp \otimes C_\oo \tto{\ g\ } C_\pp\otimes B_\oo
\justifies
A_\pp \otimes C_\oo \otimes B_\pp \otimes B_\oo\  \stackrel\sigma \cong\  A_\pp \otimes B_\oo \otimes B_\pp \otimes C_\oo\  \tto{\ f\otimes g\ }\  B_\pp\otimes A_\oo  \otimes  C_\pp\otimes B_\oo\   \stackrel\sigma \cong \ C_\pp\otimes A_\oo  \otimes   B_\pp \otimes B_\oo  
\endprooftree
\justifies
g\icomp f \ \ =\ \ \Big(A_\pp \otimes C_\oo \tto{ \Tr_{B_\pp \otimes B_\oo}\left(\sigma\circ(g\otimes f)\circ \sigma\right)}
\ C_\pp\otimes A_\oo\Big) 
\endprooftree
\]
%
%

\section{The extended reals as alternating dyadics}\label{Appendix:reals}
Recall from Sec.~\ref{Sec:coalg-reals} that $\RRrr = \RRr \cup \{\infty, -\infty\}$ is the extended real continuum, and that $\Sigma^\circledast = \coprod_{i=0}^{\omega + 1} \Sigma^i$ is the set of finite or infinite (countable) strings of symbols from $\Sigma = \{\pp,\oo\}$, which are treated in \eqref{eq:Phi} as $\{-1, 1\}$.

Define the value of the function $\Phi:\Sigma^\circledast \to \RRrr$ on an arbitrary string $ \varsigma = \sseq{\varsigma_0 \, \varsigma_1\, \varsigma_2,\, \ldots}$ to be
\bea\label{eq:Phi}
\Phi(\varsigma) & = & z\cdot \varsigma_0 + \sum_{i = z+1}^\infty \frac{\varsigma_i}{2^{i-z}}
\eea
where $z = \mu n.\ \varsigma_n\neq \varsigma_{n+1}$ is the length of the initial segment before the sign flips. If $\varsigma$ is the infinite string of either one sign or the other, then $z$ is infinite, and the value of $\Phi(\varsigma)$ is either $\infty$ or $-\infty$. Leaving the two infinities aside, $\Phi$ establishes a bijection between the remaining $\Sigma$-stings, where the sign eventually flips, and the finite real numbers from $\RRr$. For an arbitrary $x\in \RRr$, the string $\upsilon \in \Sigma^\circledast$ such that $x = \Phi(\upsilon)$ can be constructed as follows:
\begin{itemize}
\item Decompose the real line as the disjoint union of the closed-open and open-closed intervals
\bear
\RRr & = & \coprod_{n=1}^\infty [-n, -n+1) + \{0\} + \coprod_{n= 1}^\infty (n-1,n]
\eear
leaving the 0 on its own. Then there are 3 cases: 
\item (0) If $x = 0$ then $\upsilon$ is the empty string $()$.
\item (-) If $x\in [-n_0, -n_0+1)$, then $\upsilon$ begins with $\underbrace{\pp \pp\cdots \pp}_{n_0}$.
\item (+) If $x\in [n_0-1, n_0)$, then $\upsilon$  begins with $\underbrace{\oo \oo\cdots \oo}_{n_0}$. 
\item In case (-), find
\begin{itemize}
\item  the smallest $n_1$ such that $x\leq -n_0 +\sum_{i=1}^{n_1} \frac 1 {2^i}$ and append\  $\underbrace{\oo \cdots \oo}_{n_1}$ to $\upsilon$;
\item the smallest $n_2$ such that $x\geq -n_0 +\sum_{i=1}^{n_1} \frac 1 {2^i} - \sum_{i=1}^{n_2} \frac 1 {2^{n_1+i}}$ and append\  $\underbrace{\pp \cdots \pp}_{n_2}$ to $\upsilon$;
\item  the smallest $n_3$ such that $x\leq \cdots$, etc.
\end{itemize}
\item In case (+), find
\begin{itemize}
\item  the smallest $n_1$ such that $x\geq n_0 -\sum_{i=1}^{n_1} \frac 1 {2^i}$ and append\  $\underbrace{\pp \cdots \pp}_{n_1}$ to $\upsilon$;
\item the smallest $n_2$ such that $x\leq\cdots$, etc.
\end{itemize}
\item If you ever reach a sum equal to $x$, then halt and leave $\upsilon$ finite. Otherwise $\upsilon$ is infinite.
\end{itemize}
In any case, it is easy to see that $\Phi(\upsilon) = x$ and that $\Phi(\upsilon) = \Phi(\zeta)$ implies $\upsilon = \zeta$. So $\Phi$ is an injection. And we have just shown that it is a surjection by constructing for an arbitrary $x\in \RRrr$ a $\upsilon\in \Sigma^\circledast$ such that $x=\Phi(\upsilon)$. The function $\Phi$ defined by \eqref{eq:Phi} is thus the claimed bijection.

\end{document}